\documentclass[12pt,a4paper]{amsart}
\usepackage[T1]{fontenc}
\usepackage[utf8]{inputenc}
\usepackage{lmodern,microtype}
\usepackage[margin=2.7cm]{geometry}
\usepackage{amsmath,amssymb,bm,amsthm}
\usepackage{booktabs,tabularx,array}
\usepackage{graphicx}
\usepackage{hyperref,setspace}
\usepackage{tikz}
\usetikzlibrary{arrows.meta,positioning,shapes.geometric,fit,calc,decorations.pathreplacing}
\hypersetup{colorlinks=false,pdfborder={0 0 0}}
\newtheorem{theorem}{Theorem}
\newtheorem{proposition}[theorem]{Proposition}

\newtheorem{corollary}[theorem]{Corollary}
\theoremstyle{definition}\newtheorem{definition}[theorem]{Definition}
\theoremstyle{remark}\newtheorem{remark}[theorem]{Remark}
\newcommand{\ind}{\mathbf{1}}
\newcommand{\R}{\mathbb{R}}
\newcommand{\E}{\mathbb{E}}
\newcommand{\doi}[1]{\href{https://doi.org/#1}{doi:#1}}
\newcommand{\lingnote}[1]{\par\smallskip\noindent\textit{Linguistic reading.}\ #1\par\smallskip}
\title[Identifiability of Neologism Dynamics]{Identifiability of Neologism Dynamics from Aggregate Corpus Counts}
\author[Area]{Iv\'an Area}
\address[Area]{IFCAE, E.E. Aeron\'autica e do Espazo, Universidade de Vigo, Campus de Ourense, 32004 Ourense, Spain}
\email{area@uvigo.gal}
\author[Moreno-Sandoval]{Antonio Moreno-Sandoval}
\address[Moreno-Sandoval]{Dept. Linguistics, Universidad Aut\'onoma de Madrid, Spain}
\email{antonio.msandoval@uam.es}
\urladdr{https://orcid.org/0000-0002-9029-2216}
\author[Porta Zamorano]{Jordi Porta-Zamorano}
\address[Porta Zamorano]{Dept. of Computer Engineering, Universidad Aut\'onoma de Madrid, Spain}
\email{jordi.porta@uam.es}
\subjclass[2020]{91D30, 60J28, 05C82, 62M09}
\keywords{neologism, lexical change, multilayer network, continuous-time Markov chain, complex contagion, observation model, identifiability}
\date{\today}
\begin{document}
\begin{abstract}
Aggregate lexical frequencies reflect both what speakers do and how a corpus records it. We develop a stochastic framework for lexical adoption, abandonment and reactivation on a temporal multilayer network, together with a count observation model that separates speaker prevalence, use intensity and detectability, and we ask what such data leave identified. Three results organise the analysis. We give a necessary and sufficient condition for the never-active and dormant speaker states to be merged exactly, so that the three-state chain collapses to a two-state system, and show that it is a condition on the exposures a given network can actually deliver rather than on the response function as a whole. We separate the exact observational invariances, under which no count design can distinguish use intensity from detectability, from the merely practical loss of information in sparse count series, and both from the exact parameter equivalence established next. And we prove that in a low-prevalence community approximation the rates enter only through the net growth $\beta-\mu$ and the coupling product $\beta\kappa$, so that a one-parameter family of rate triples has identical trajectories, with the separating information carried entirely by a term quadratic in prevalence whose size depends on the emission scale. Simulations characterise parameter recovery, prediction and sensitivity to misspecification, and locate where each degrades. An exploratory analysis of $144$ candidate lexical forms in the CREA and CORPES XXI press subcorpora then illustrates the practical consequences: under the chosen emission scale and a symmetric contact graph, substantially different rate triples produce nearly identical fitted trajectories, and in a leave-one-country-out comparison using training-only proxies and seeds within a retrospectively selected cohort a constant-rate count baseline outperforms the evaluated dynamic specifications on $84$ to $94\%$ of forms. These findings neither establish nor refute contact-driven diffusion. They delimit what the fitted model and this corpus panel can support, and identify the calibration, the geographical contrast and the repeated speaker observations that would make the question answerable.
\end{abstract}

\maketitle

\section{Introduction}\label{sec:intro}
Lexical innovations are continually produced, but only a small and non-random subset becomes established. Aggregate token frequency alone cannot reveal who adopts a form, through which ties, or whether apparent growth reflects broad uptake, repeated use by a few speakers, or changing documentary coverage. Labov's distinction between transmission within a speech community and diffusion across communities provides the linguistic motivation for keeping these mechanisms separate \cite{labov1994,labov2001,labov2007,labov2010}.
Network studies of lexical innovation show that forms with similar token totals can have markedly different social distributions \cite{wuerschinger2021}. Online studies additionally report associations between network structure and lexical retention, and evidence that some linguistic innovations behave as complex contagions \cite{goel2016,zhu2021,laitinen2020}. A parallel corpus-based study of Anglicisms drawn from Twitter in U.S.\ Spanish illustrates both the promise and the labour of manual validation, because a very large tweet collection yielded a comparatively small manually validated set of candidate Anglicisms \cite{morenosandoval2018}. These findings motivate a generative model with four linked components: emergence, speaker-level state transitions, cross-community diffusion, and imperfect observation.
A fundamental distinction must be drawn between the birth (or emergence) of an innovation and its subsequent spatio-temporal consolidation.  Our empirical design does not claim to track the initial coinage of neologisms.

Our contribution is an analysis of the relationship between lexical mechanisms and the observations used to infer them. The framework (i) defines a finite-state stochastic process on a temporal multilayer network; (ii) states exactly which reduced system each classical formula belongs to, and under what conditions the reduction is exact; (iii) distinguishes local consolidation from cross-community reach; and (iv) supplies an observation layer that separates latent prevalence, use intensity, and detectability. The network, epidemic, and factor-model ingredients are established, and no new theory of any of them is claimed. What the paper adds is their disciplined integration for lexical innovation, and a precise account of which conclusions the resulting observations support. We distinguish observational scale ambiguities, which are exact, from parameter equivalence in a linear reduction, which is also exact, and from practical loss of information in sparse count series, which is a matter of degree. The empirical analysis illustrates these distinctions on a national press panel and evaluates the proposed dynamics against simpler count models. The resulting diagnosis concerns the stated observation design and model restrictions; it does not imply that diffusion is absent, or that every aggregate corpus design is uninformative.

Two things follow for the reader. First, the theoretical sections are deliberately conservative: every threshold or S-curve statement is attached to a named approximation, and Proposition~\ref{prop:lump} says precisely when that approximation is exact. Second, the empirical sections are organised so that each one answers a stated question. Section~\ref{sec:insilico} characterises the estimator on data of known structure and measures where it is precise and where it is not; Section~\ref{sec:protocol} sets out the prospective protocol a confirmatory study would follow, and is explicitly a proposal rather than a completed analysis; Section~\ref{sec:diagnosis} carries out the fit that the available press corpora permit and reports what it does and does not determine; Section~\ref{sec:inference} quantifies what pooling across a cohort buys; and Section~\ref{sec:limits} and Section~\ref{sec:conclusion} state the scope of the whole.

\section{Related statistical models of lexical change}\label{sec:related}
Quantitative models of how words spread already exist, and the contribution here is defined by what they leave unaddressed rather than by their absence. Five strands are directly relevant.

\emph{S-curve mechanism models.} The sigmoidal trajectory that motivates much of this literature has been formalised and, importantly, shown to be mechanism-underdetermined: Blythe and Croft argue that an aggregate S-curve is compatible with several distinct propagation processes, so curve shape alone does not identify the underlying dynamics \cite{blythe2012,ghanbarnejad2014}. Related work derives self-actuating S-curves from replicator-type dynamics \cite{stadler2016}. Our Proposition~\ref{prop:logistic} is consistent with this view: we obtain the logistic curve only as a controlled reduction and do not treat it as an empirical law.

\emph{Evolutionary (Wright--Fisher) models.} The principal statistical alternative to an epidemic formulation treats frequency time series as a Wright--Fisher process and tests whether an observed change reflects directional selection or neutral drift \cite{newberry2017}. The standard one-population implementation is powerful for aggregate diachronic frequencies but does not itself encode a social network, a distinction between within-community and cross-community spread, or a separate documentary-detection process. Structured and observation-augmented evolutionary models are possible, so the contrast is between the implementations considered here rather than between mutually exclusive model classes.

\emph{Network-position and fate models.} Empirically, the closest antecedents relate the social-network position of early users to whether an innovation consolidates or fades. Studies on Twitter report that forms with similar token totals diverge in social distribution \cite{wuerschinger2021}, that network structure modulates the rate of change \cite{zhu2021,laitinen2020,goel2016}, and that features derived from adopter centrality predict the survival-versus-decay outcome \cite{tarrade2024}. These analyses are typically regression-based or classifier-based on a single digital channel, without a generative latent-state process or a corpus-observation model.

\emph{Compartmental language-competition models.} A parallel tradition models not a single lexical item but whole languages in competition, through compartmental systems of
ordinary differential equations. Area and Jornet \cite{areajornet2025} give a recent and unusually general instance: $J$ local tongues competing with two ``prestigious'' languages, with intermediate compartments for speakers who use a local and a prestigious tongue equally, giving a system of $3J+2$ equations for which they establish existence, uniqueness and positivity of solutions on $[0,\infty)$, extinction of the local tongues under explicit sign conditions on the demographic parameters, and local asymptotic stability of the surviving equilibrium through the eigenvalues of a triangular Jacobian. Two features of that work bear directly on the present one. Its compartments are defined by the language a speaker \emph{prefers in social interaction} rather than by what a speaker knows, which is the same measurement decision that Section~\ref{subsec:window} must make for $U$, $A$ and $D$, and it is forced for the same reason: an aggregate model can carry only a partition that is decidable from observable behaviour. And the theory is qualitative and illustrated by simulation rather than fitted, which the authors state
plainly, remarking of an oscillation produced by one parameter set that it is unclear whether such behaviour occurs empirically or is an artefact of the Lotka--Volterra structure. Settling a question of that form is what an observation model is for, and it is the gap this paper works in.

\emph{Norm change and critical mass.} Convention-replacement models exhibit tipping-point behaviour with a critical fraction of committed users, a prediction tested directly in controlled experiments on social convention \cite{amato2018,centola2018}, and they provide useful comparators for nonlinear reinforcement; controlled experiments on behavioural spread in online networks provide the wider empirical backdrop \cite{centola2010}. A Hill response with $r>1$ is not, by itself, a committed-minority model and does not automatically imply a population-level tipping point; it only makes the local response flat at zero (Remark~\ref{rem:hill}). A contemporaneous study applies the same epidemic and complex-contagion machinery to a different socio-cognitive phenomenon, the spread of dependence on large language models, using an uncoupled/coupled/dependent three-compartment model with cooperative reinforcement and saddle-node tipping points \cite{sole2026}. The structural parallel is close, and the stance is compatible: that study also presents tipping behaviour conditionally, as something the interplay of transmission, recovery and reinforcement \emph{can} generate once a threshold is crossed, and it identifies conditions under which the transition is reversible. The difference is one of object rather than of attitude. Our concern is a corpus-observation layer that separates latent prevalence from what is recorded, and the question of what that layer leaves identified does not arise in the same form there.

\emph{The two closest generative predecessors.} Regression and classifier studies are not the relevant comparison for a generative latent-state model. Eisenstein et al.\ \cite{eisenstein2014} combine latent regional lexical dynamics, a count observation layer, global word effects, and dynamics shared across words; Ananthasubramaniam et al.\ \cite{ananthasubramaniam2024} fit a generative agent model with repeated usage, network exposure, identity, and attention decay. Table~\ref{tab:compare} places the present framework against these two directly. The incremental features are the three-state continuous-time speaker chain with an explicit dormant compartment and reactivation, and the explicit prevalence/intensity/detectability factorisation of the observed count. Table~\ref{tab:compare} also makes the evidential asymmetry plain: both predecessors work from geolocated streams in which individual users can be counted, whereas the present study works from aggregate national press counts, and that difference in observation unit is precisely what the identifiability analysis below is about. A benchmark against \cite{eisenstein2014} on matched data would be the natural next step, and Section~\ref{sec:diagnosis} explains why it cannot be carried out on the panel available here: speaker-count and token-count observation units differ, requiring a matched observation design. Weak parameter identification does not itself preclude a predictive benchmark, which remains unperformed.

\begin{table}[t]
\centering\footnotesize
\renewcommand{\arraystretch}{1.2}
\begin{tabularx}{\linewidth}{@{}p{3.25cm}>{\raggedright\arraybackslash}X>{\raggedright\arraybackslash}X>{\raggedright\arraybackslash}X@{}}
\toprule
& Eisenstein et al.\ \cite{eisenstein2014} & Ananthasubramaniam et al.\ \cite{ananthasubramaniam2024} & present framework\\
\midrule
latent state per unit & continuous regional lexical propensity & agent adoption with attention decay & three-state speaker chain $U/A/D$, continuous time\\
observation unit & counts of distinct users per area, through a binomial layer & simulated adoption events & token counts per community, through a negative-binomial layer\\
units and their coupling & metropolitan areas, with influence between areas inferred by a latent vector autoregression rather than observed & simulated agents with inferred and demographic ties & communities, with a contact matrix assumed rather than observed in the corpus study\\
count observation layer & yes & no separate layer & yes, with $\pi$, $u$, $q$ factored\\
identifiability treated & implicitly & implicitly & as the main object of study\\
empirical evaluation & latent regional inference and held-out tuning & simulation calibrated to geolocated data & national press counts, evaluated against a count baseline\\
\bottomrule
\end{tabularx}
\caption{The present framework against its two closest generative predecessors. The incremental contribution is structural: an explicit dormant compartment with reactivation, and a count layer in which prevalence, use intensity and detectability are factored separately. The row that matters for what follows is the observation unit, since the predecessors count identified users and this study counts tokens in a national press panel, and Section~\ref{sec:diagnosis} works out what that difference costs.}
\label{tab:compare}
\end{table}

\section{How to read the model: notation and linguistic interpretation}\label{sec:reading}
\begin{figure}[t]
\centering
\resizebox{\linewidth}{!}{%
\begin{tikzpicture}[
  >=Latex,
  stage/.style={draw,rounded corners,thick,minimum height=15mm,text width=26mm,align=center,font=\small},
  opt/.style={draw,rounded corners,thick,dashed,minimum height=15mm,text width=26mm,align=center,font=\small},
  note/.style={font=\scriptsize,align=center,text width=27mm},
  arrow/.style={->,thick}
]
\node[stage] (a) {\textbf{Emergence}\\a speaker coins a form};
\node[opt,right=9mm of a] (b) {\textbf{Transmission}\\(not implemented)\\faithful intergenerational acquisition};
\node[stage,right=9mm of b] (c) {\textbf{Diffusion}\\contact-driven spread across community boundaries};
\node[stage,right=12mm of c] (d) {\textbf{Detection}\\trace left in the corpora we read};
\path[arrow] (a)--(b); \path[arrow] (b)--(c);
\node[note,below=3mm of a] {intensity $\nu_g(t)$ (Sec.~\ref{sec:emergence})};
\node[note,below=3mm of b] {age-structured acquisition; an extension, see Sec.~\ref{sec:community}};
\node[note,below=3mm of c] {cross-community edges; bridging speakers};
\node[note,below=3mm of d] {prevalence $\pi$, intensity $u$, detection $q$ (Sec.~\ref{sec:observation})};
\node[note,below=20mm of c] {dormancy and revival ($D$) occur within the latent speaker process};
\draw[->,thick,dotted] (a.north) to[out=40,in=140] node[above,font=\scriptsize]{detection can occur at any stage} (d.north);
\end{tikzpicture}}
\caption{The life cycle of one lexical innovation, as the framework decomposes it. What we ultimately observe (rightmost box) is a filtered trace of what happens among speakers; keeping these stages separate is the point of the framework. The diagram is not a one-to-one picture of the implemented components: the transmission box is drawn dashed because faithful intergenerational acquisition is an extension rather than an implemented stage (Section~\ref{sec:community}), and the speaker-level chain of Section~\ref{sec:ctmc} operates inside the diffusion stage rather than beside it. The dotted arrow records that documentary detection operates throughout the life cycle and does not follow diffusion; the note below diffusion locates dormancy and revival inside the latent speaker process, without implying feedback from corpus detection.}
\label{fig:pipeline}
\end{figure}
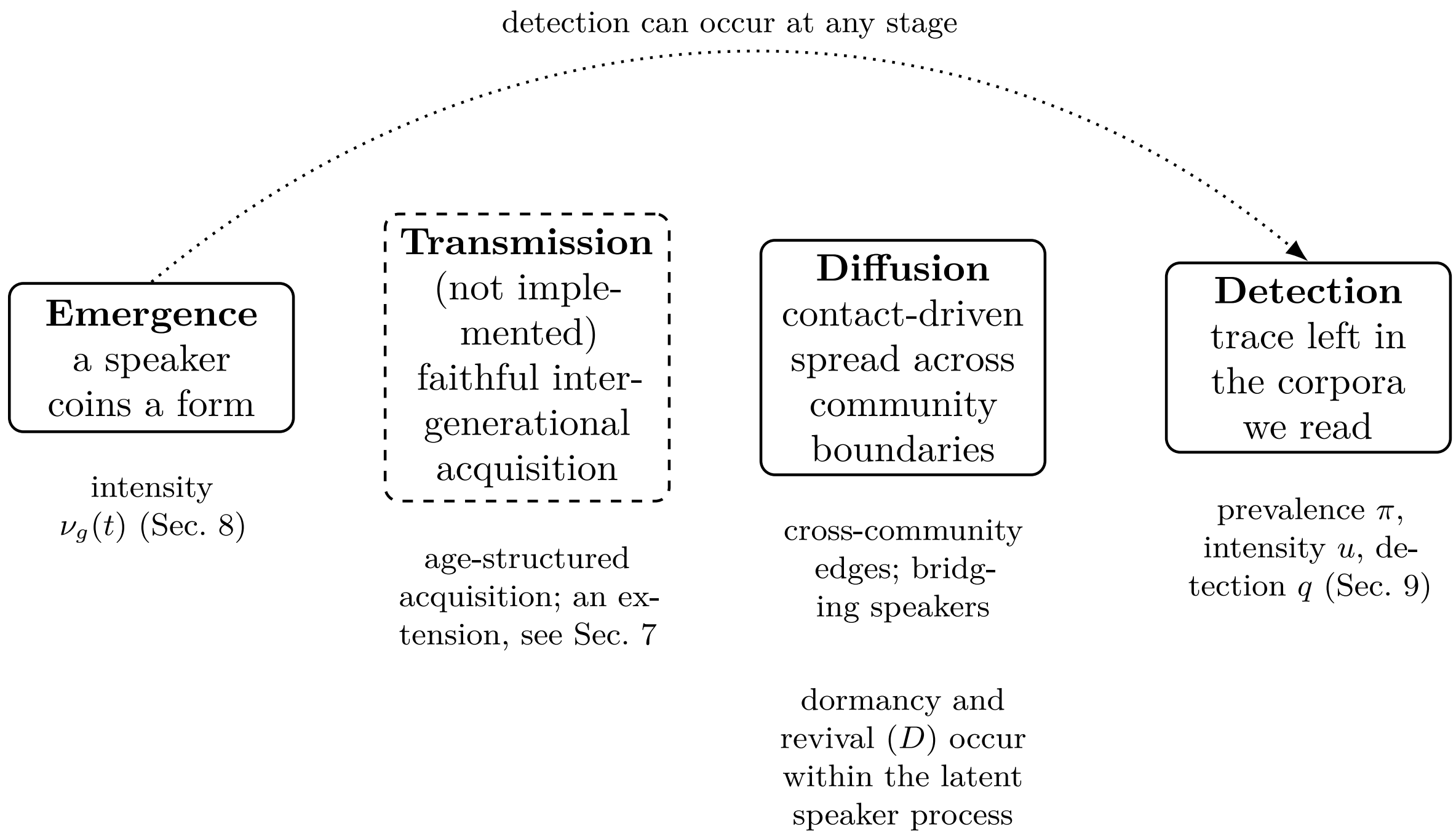

This section is an on-ramp for readers whose background is linguistic rather than mathematical. The whole model can be read as a four-stage story about the life of a single new word, shown in Figure~\ref{fig:pipeline}; Table~\ref{tab:glossary} then translates the core mathematical symbols into their linguistic meaning. A reader may consult these two items and skip the derivations without losing the argument.

\begin{table}[t]
\centering
\footnotesize
\renewcommand{\arraystretch}{1.12}
\begin{tabularx}{\linewidth}{@{}p{2.5cm}>{\raggedright\arraybackslash}X@{}}
\toprule
\textbf{Symbol} & \textbf{Linguistic meaning}\\
\midrule
$w$ & the single new word (neologism) under study\\
$N$, $V$ & the number of speakers and the set of speakers\\
$g(i)$, $G$ & the speech community of speaker $i$ and the number of communities; country, region, and medium are metadata, not automatically speech communities\\
$A^{(\ell)}(t)$ & a network layer $\ell$ of contacts (family, work, social media) at time $t$\\
$W_{ij}(t)$ & overall strength of exposure from speaker $j$ to speaker $i$ (all layers combined)\\
$M_i(t)$ & exposure to $w$ from broadcast or institutional sources (media, school)\\
$U,\ A,\ D$ & speaker states: never-active, active user, dormant or former user\\
$\Lambda_i(t,x)$ & how much of the word speaker $i$ is currently hearing from active peers plus media\\
$q_i^{UA},\ q_i^{AD},$ $q_i^{DA}$ & rates of adopting, abandoning, and reviving the word\\
$H_{r,\theta}$ & dose--response curve; $r>1$ produces nonlinear reinforcement, but repeated or socially independent exposures are represented only if $\Lambda_i$ is constructed to encode them\\
$r$ & steepness exponent of the dose--response curve: $r=1$ simple, $r>1$ nonlinear\\
$\theta$ & the exposure level at which the dose--response curve reaches one half; sets the scale against which exposure is measured\\
$\theta_D$ & the same quantity for the reactivation response of dormant speakers\\
$\alpha_i$ & exogenous adoption not represented by peer exposure or by $M_i$; independent coinage is one possible source. Media-only uptake belongs in $M_i$, not in $\alpha_i$\\
$\beta,\ \mu,\ \rho,\ \varphi$ & baseline strengths of contagion, abandonment, spontaneous revival, contact-driven revival\\
$P_g(t)$ & proportion of active users within community $g$ (speaker prevalence)\\
$\nu_g(t)$ & rate at which new candidate forms are coined in community $g$\\
$\pi,\ u,\ q$ & latent speaker prevalence, emission rate of the form per token an active user produces, chance of being recorded\\
$\pi^{\mathrm{tok}}$ & the share of the recorded text that is produced by active users; equal to $\pi$ only under the sampling assumption of Section~\ref{sec:observation}\\
$E_{gtc}$ & size or coverage of the corpus for community $g$, period $t$, channel $c$\\
$Y_{wgtc}$ & the token count we actually observe\\
$\psi_c$ & dispersion of the count model in channel $c$: how much more variable counts are than a Poisson process would be\\
$B$, $\kappa$ & the community-contact matrix used in the estimable specialisation, and the strength of cross-community coupling relative to its fixed normalisation\\
$\phi_c$ & the emission scale $u_cq_c$ of channel $c$, fixed rather than estimated (Section~\ref{sec:observation})\\
$\rho(W)$, $\mathcal R_{\mathrm{lex}}$ & connectivity of the contact network; invasion index (Sec.~\ref{sec:reductions})\\
\bottomrule
\end{tabularx}
\caption{Dictionary from mathematical notation to linguistic meaning. The core quantities of the paper appear here in plain terms. Three pairs of symbols are close enough to be worth separating explicitly: $\rho$ (spontaneous revival rate) and $\rho(\cdot)$ (spectral radius of a matrix); $\varphi$ (contact-driven revival rate of the speaker chain) and $\phi_c$ (emission scale of channel $c$); and the transition rates $q_i^{UA},q_i^{AD},q_i^{DA}$ and the scalar detectability $q$. The letter $W$ denotes the contact matrix throughout Sections~\ref{sec:network}--\ref{sec:reductions} and the number of forms in a cohort in Section~\ref{sec:diagnosis}; the two never occur together.}
\label{tab:glossary}
\end{table}

\section{State space and temporal multilayer network}\label{sec:network}
Fix one lexical innovation $w$ and a finite set $V=\{1,\ldots,N\}$ of speakers. Each speaker belongs to one primary community $g(i)\in\{1,\ldots,G\}$. For the mathematics this is a partition; linguistically it is a strong simplification because speakers participate in overlapping communities of practice and interactional networks. The empirical analysis must state an operational community definition before fitting. Overlapping memberships can be introduced later, but are not needed for the results below. For layers $\ell=1,\ldots,L$, let $A^{(\ell)}(t)=(A^{(\ell)}_{ij}(t))$ be a nonnegative, measurable contact matrix with zero diagonal. Define
\begin{equation}\label{eq:effective}
 W_{ij}(t)=\sum_{\ell=1}^L \omega_\ell A^{(\ell)}_{ij}(t),\qquad \omega_\ell\ge0.
\end{equation}
The matrices may be directed. To avoid double counting, $W$ represents interpersonal exposure, whereas a separate nonnegative term $M_i(t)$ represents broadcast or institutional exposure.

Speaker $i$ has state $X_i(t)\in\{U,A,D\}$, where $U$ means never yet an active user of $w$, $A$ means currently an active user, and $D$ means a former user who is currently dormant. This is a latent classification of the speaker, not a statement about what has been observed: both $U$ and $D$ speakers produce no tokens of $w$, and the corpus cannot separate them. What distinguishes $D$ from $U$ is prior activity, which is why the two carry different transition rates into $A$; how the latent states relate to what is recorded is the subject of Section~\ref{sec:observation}. Neither $U$ nor $D$ asserts anything about passive knowledge: a speaker in $U$ may well have heard the form, and a separate awareness state would be required to model recognition without use. Write $X(t)=(X_1(t),\ldots,X_N(t))$ and
\begin{equation}\label{eq:exposure}
 \Lambda_i(t,x)=\sum_{j=1}^N W_{ij}(t)\ind\{x_j=A\}+M_i(t).
\end{equation}
Cross-community contact is encoded in the corresponding entries of $W$; a separate community coupling term should be added only when $W$ omits those contacts.
\lingnote{A layer is one kind of social tie: face-to-face contact, a workplace, a social-media following. Stacking layers into $W$ says only that a speaker can hear a new word through several channels at once, and that some channels weigh more ($\omega_\ell$) than others. The term $M_i(t)$ keeps mass media separate from person-to-person contact, because a word heard on television is not the same evidence as a word heard from a friend.}

\subsection{Units, normalisation, and what is identified}\label{subsec:scale}
Exposure has no natural unit, so the model is invariant under rescalings that must be fixed by convention before anything is estimated. The invariances are different in the nonlinear and the linear specifications, and it matters which is being used.

\begin{proposition}[Scale invariances]\label{prop:scale}
Let $c>0$ and let $H_{r,\theta}(z)=z^r/(\theta^r+z^r)$.
\begin{enumerate}
\item[(i)] \emph{(Hill specification.)} The map $(W,M,\theta,\theta_D)\mapsto(cW,cM,c\theta,c\theta_D)$ leaves every transition intensity of Section~\ref{sec:ctmc} unchanged, with $\beta,\varphi$ and all other parameters fixed. Consequently the law of the process is unchanged.
\item[(ii)] \emph{(Linear mass-action specification, $H$ replaced by the identity.)} The map
\[(\beta,\varphi,W,M)\mapsto(\beta/c,\ \varphi/c,\ cW,\ cM)\]
leaves every transition intensity unchanged.
\item[(iii)] In (ii) the broadcast term must be rescaled together with $W$. Under $(\beta,\varphi,W)\mapsto(\beta/c,\varphi/c,cW)$ with $M$ fixed and $c\ne1$, all intensities are preserved if and only if $\beta M_i(t)=\varphi M_i(t)=0$ for every $i$ and almost every $t$. In particular, $M\equiv0$ is sufficient.
\end{enumerate}
\end{proposition}
\begin{proof}
(i) $H_{r,c\theta}(c z)=(cz)^r/\big((c\theta)^r+(cz)^r\big)=z^r/(\theta^r+z^r)=H_{r,\theta}(z)$, and $\Lambda_i$ in \eqref{eq:exposure} is positively homogeneous of degree one in $(W,M)$, so $\Lambda_i\mapsto c\Lambda_i$ and each $H$ term is unchanged. (ii) With $H$ the identity, the exposure-driven terms are $\beta\Lambda_i$ and $\varphi\Lambda_i$, and $\Lambda_i\mapsto c\Lambda_i$ compensates $\beta\mapsto\beta/c$ exactly. (iii) Under $(\beta,\varphi,W)\mapsto(\beta/c,\varphi/c,cW)$ the exposure becomes $c\sum_jW_{ij}\ind\{x_j=A\}+M_i$, so $\beta\Lambda_i$ becomes $\beta\sum_jW_{ij}\ind\{x_j=A\}+\beta M_i/c$, which differs from $\beta\Lambda_i$ whenever $\beta M_i\ne0$ and $c\ne1$. The reactivation term has the analogous difference $\varphi M_i(1/c-1)$, proving the stated equivalence.
\end{proof}

Thus in the linear specification the exposure effects enter through the scale-invariant products $\beta\omega_\ell A^{(\ell)}$, $\beta M$, and their reactivation counterparts; the factors cannot be separately identified without an external scale convention, and a normalisation such as $\rho(W)=1$, fixed layer scales, or an externally calibrated contact unit is required. In the Hill specification the free scale sits in $(\theta,\theta_D)$ rather than in $\beta$, so fixing $\theta$ removes this exposure-scale invariance. Removing an invariance is necessary for a parameter to have a determinate meaning, but it is not sufficient for identification: uninformative exposure histories, confounded covariates, unobserved contacts and the observation process can each still leave $\beta$ undetermined, and Sections~\ref{subsec:identobs} and \ref{subsec:identreal} show that in this application they do. Reporting a value of $\beta$ without stating which specification and which normalisation produced it is therefore not meaningful.

\section{Speaker-level stochastic dynamics}\label{sec:ctmc}
\begin{figure}[t]
\centering
\begin{tikzpicture}[
  >=Latex,
  state/.style={circle,draw,thick,minimum size=12mm,font=\small},
  lab/.style={font=\scriptsize,align=center},
  every edge/.style={draw,thick,->}
]
\node[state] (U) {$U$};
\node[state,right=35mm of U] (A) {$A$};
\node[state,below=23mm of A] (D) {$D$};
\path (U) edge[bend left=13] node[lab,above] {$q_i^{UA}(t,x)$\\adoption} (A);
\path (A) edge node[lab,right] {$q_i^{AD}(t,x)$\\abandonment} (D);
\path (D) edge[bend left=13] node[lab,below left] {$q_i^{DA}(t,x)$\\reactivation} (A);
\node[lab,below=8mm of U,text width=31mm] {never active};
\node[lab,above=7mm of A,text width=32mm] {active user};
\node[lab,below=8mm of D,text width=33mm] {dormant or former user};
\node[draw,dashed,rounded corners,fit=(U)(A)(D),inner sep=6mm,label={[lab]above left:latent speaker process}] {};
\end{tikzpicture}
\caption{Speaker-level state transitions for one lexical innovation. Interpersonal and broadcast exposure enter the adoption and reactivation intensities through $\Lambda_i(t,x)$. All three states are latent: the corpus distinguishes only whether tokens were recorded.}
\label{fig:states}
\end{figure}
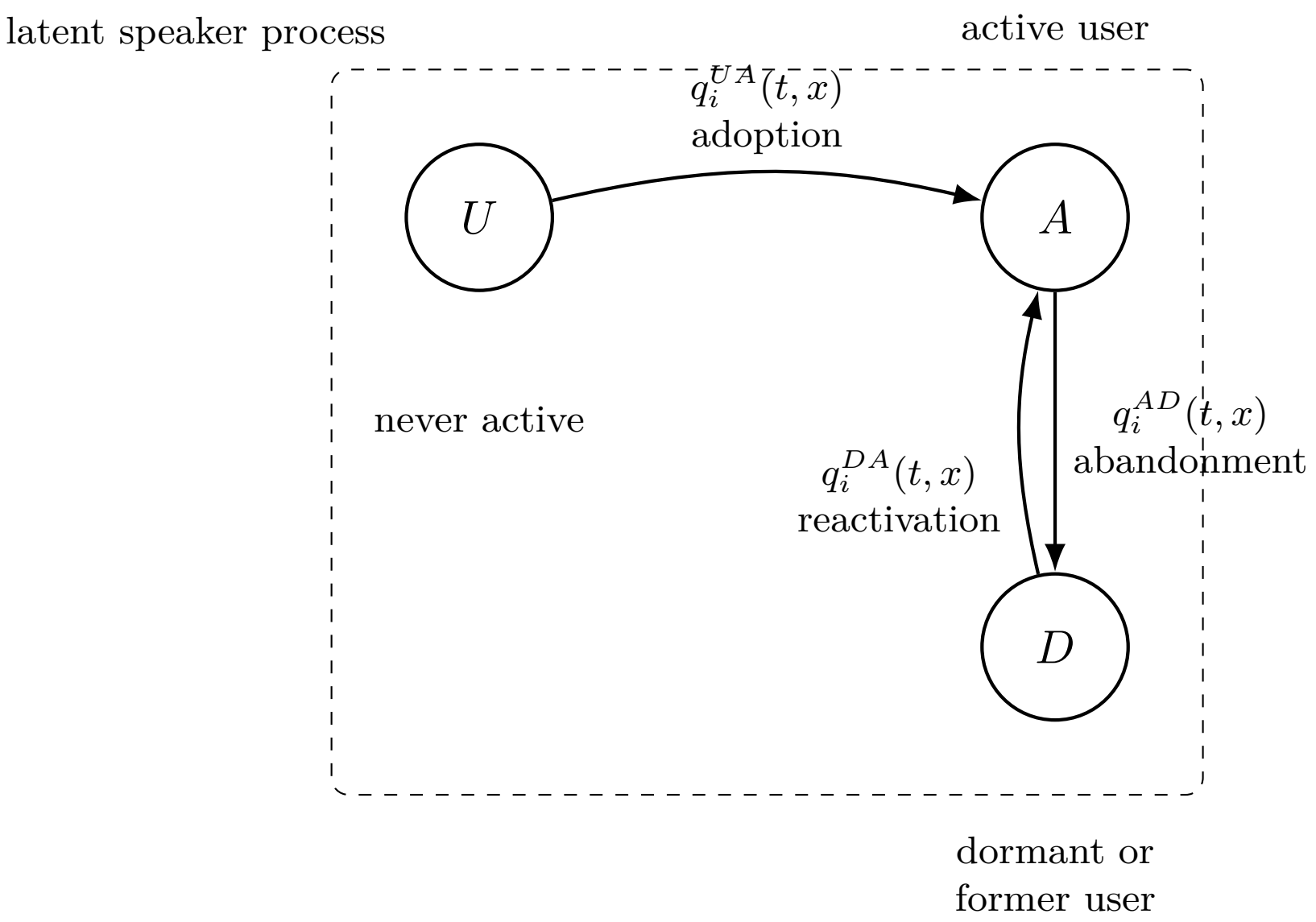

For $r\ge1$ and $\theta>0$, set $H_{r,\theta}(z)=z^r/(\theta^r+z^r)$ for $z\ge0$. Figure~\ref{fig:states} shows the three states and the transitions between them. Conditional on $X(t^-)=x$, the allowed transition intensities are
\begin{align}
 q_i^{UA}(t,x)&=\alpha_i(t)+\beta H_{r,\theta}(\Lambda_i(t,x))\exp(z_i^\top\gamma+c_{g(i)}(t)^\top\delta),\label{eq:adopt}\\
 q_i^{AD}(t,x)&=\mu\exp(-z_i^\top\eta-c_{g(i)}(t)^\top\xi),\label{eq:abandon}\\
 q_i^{DA}(t,x)&=\rho+\varphi H_{1,\theta_D}(\Lambda_i(t,x)),\label{eq:reactivate}
\end{align}
with $\theta_D>0$ and nonnegative baseline parameters $\alpha_i(t),\beta,\mu,\rho,\varphi$. Covariates $z_i$ and cultural states $c_g(t)$ are assumed bounded on every finite time interval. The use of a bounded Hill response in \eqref{eq:reactivate} prevents an unbounded intensity when contact weights are not normalised.

\subsection{What the states mean as a measurement}\label{subsec:window}
The three states are latent and the process \eqref{eq:adopt}--\eqref{eq:reactivate} is Markov in them by construction: an $A$ speaker leaves for $D$ at a rate that does not depend on how long the speaker has been active, and a $D$ speaker returns at a rate that does not depend on how long it has been dormant. Both are modelling assumptions, and they are strong ones. Real abandonment is likely to show duration dependence, and a fixed look-back window (``active means at least one use in the last $\Delta$ years'') is emphatically \emph{not} equivalent to them: a state defined by a window is a function of the recent trajectory, so the resulting label process is not Markov on $\{U,A,D\}$ alone.

We therefore keep the two notions apart throughout. The window is an \emph{operational rule for classifying observed behaviour}, used only in Section~\ref{sec:observation} and in the empirical definition of operational extinction; the latent chain has the specified instantaneous hazards and no observation window. Its individual-state sojourns are generally not exponentially distributed: time-varying inputs and changes in neighbours can change the hazard during a sojourn. When a study needs the window semantics inside the latent process, the state must be augmented with elapsed time since last use, giving a piecewise-deterministic Markov process, or the sojourn distributions must be replaced by phase-type laws; a finite phase-type augmentation can use the same finite-state argument under bounded rates, whereas an elapsed-time extension requires a separate existence and non-explosion argument for the resulting piecewise-deterministic process. We do not pursue either here, and no result below depends on the window.

\lingnote{Each speaker is in one of three latent conditions: has never used the word ($U$), is currently using it ($A$), or used it before and has stopped ($D$). A two-state ``adopt or not'' model can also fall silent and later resume; what the third state adds is the distinction between a speaker who has never used the form and one who has abandoned it, and the freedom to let the second take the form up again at a different rate. That is a substantive linguistic claim (re-adoption is not the same event as first adoption) and Proposition~\ref{prop:lump} says exactly when it makes no difference.}

\begin{theorem}[Well-posedness]\label{thm:wellposed}
Assume that the entries of $W(t)$, $M_i(t)$, $\alpha_i(t)$, and $c_g(t)$ are deterministic, measurable and locally bounded. Alternatively, draw exogenous input paths that satisfy these properties almost surely and then construct the chain conditionally on those paths using the displayed intensities and a specified conditional initial distribution. Then, for every initial distribution on $\{U,A,D\}^N$, intensities \eqref{eq:adopt}--\eqref{eq:reactivate} define a unique non-explosive time-inhomogeneous Markov jump process in distribution, conditionally on the inputs in the stochastic-input construction.
\end{theorem}
\begin{proof}
For fixed $N$, the state space has exactly $3^N$ elements. On any interval $[0,T]$, local boundedness of the inputs and boundedness of $H_{r,\theta}$ imply a finite common upper bound for every individual transition intensity. Since exactly one outgoing transition type is allowed per speaker state and hence at most $N$ individual transitions are enabled from any configuration, the total exit rate is bounded on $[0,T]$. A standard thinning construction from a dominating Poisson random measure therefore constructs the process and rules out infinitely many jumps before $T$. For every starting time $s$, the transition matrix solves the finite linear system $\partial_t P(s,t)=P(s,t)Q(t)$, $P(s,s)=I$, with locally integrable coefficients. Uniqueness of these transition matrices, together with the initial law and the Markov construction, determines all finite-dimensional distributions and hence the law of the jump process.
\end{proof}

\begin{remark}[The Markov property is conditional]\label{rem:markov}
If $W$, $M$ or $c_g$ are themselves stochastic and endogenous (for example if contact intensifies because the word is spreading) then $X$ alone is not Markov, and the joint process $(X,W,M,c)$ must be modelled. Conditioning an endogenous input on its entire realised path need not preserve the displayed intensities. All conditional statements below therefore use the exogenous pathwise construction of Theorem~\ref{thm:wellposed}; an endogenous extension requires a separately specified joint model.
\end{remark}

\begin{remark}[When is the nonactive set absorbing?]\label{rem:extinction}
Let $\mathcal N=\{x: x_i\ne A\ \forall i\}$. On $\mathcal N$ we have $\Lambda_i(t,x)=M_i(t)$, so the total rate of leaving $\mathcal N$ is $\sum_i\big[\alpha_i(t)+\beta H_{r,\theta}(M_i(t))e^{z_i^\top\gamma+c^\top\delta}\big]$ from states $x_i=U$ plus $\sum_i\big[\rho+\varphi H_{1,\theta_D}(M_i(t))\big]$ from states $x_i=D$. Hence $\mathcal N$ is absorbing if and only if, for almost every $t$ and every $i$, $\alpha_i(t)=0$, $\rho=0$, and either $M_i(t)=0$ or $\beta=\varphi=0$. Positive external exposure alone does not prevent absorption: it does so only when the corresponding response coefficient is positive. We accordingly distinguish \emph{mathematical extinction}, defined only under these conditions, from \emph{operational extinction}, defined empirically as no detected active use during a prespecified observation window. The latter is an observation event, not a state-space theorem.
\end{remark}

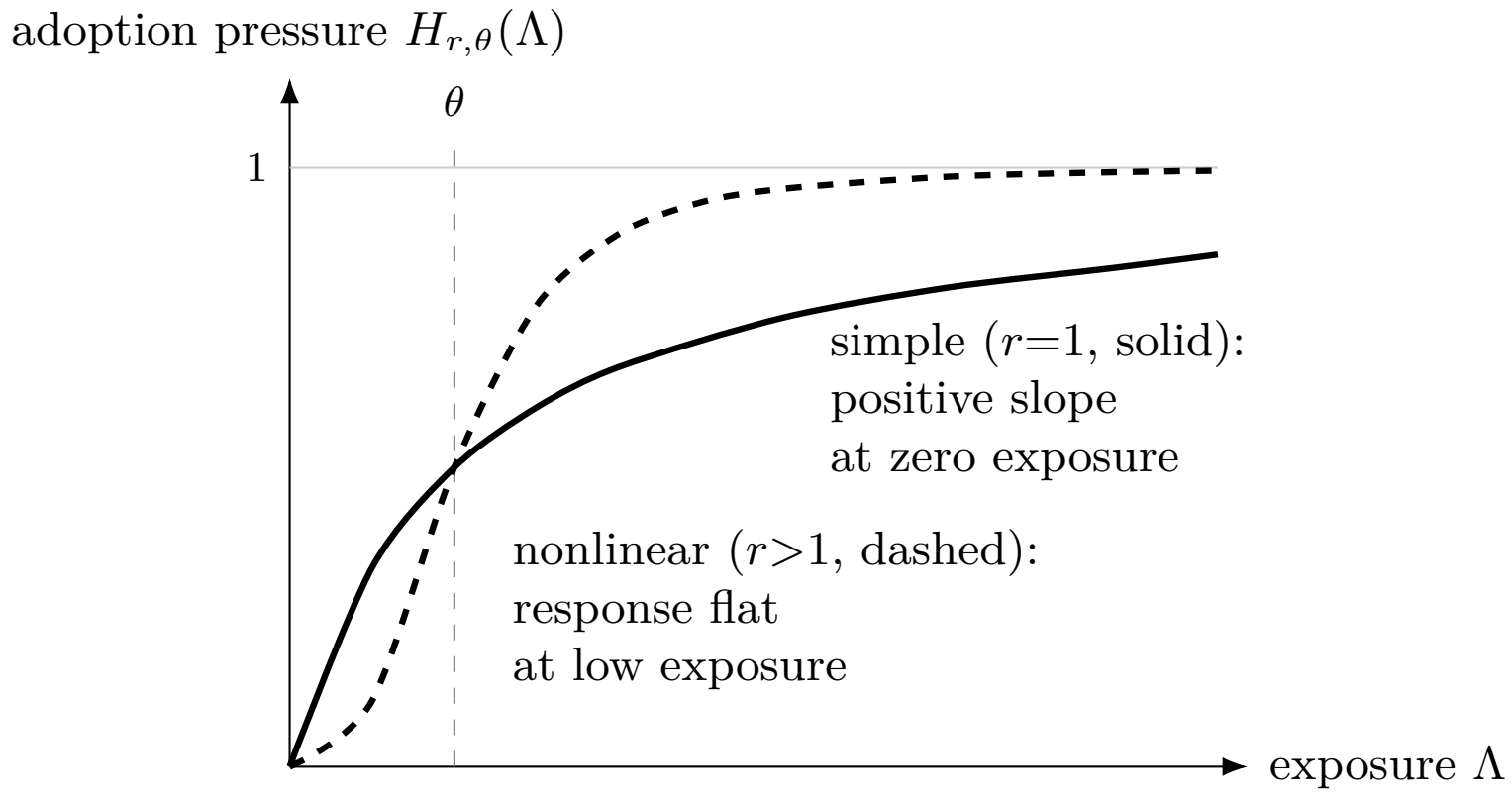
\begin{figure}[t]
\centering
\begin{tikzpicture}[x=1cm,y=1cm,font=\scriptsize,>=Latex]
\draw[->] (0,0) -- (6.4,0) node[right]{exposure $\Lambda$};
\draw[->] (0,0) -- (0,4.6) node[above]{adoption pressure $H_{r,\theta}(\Lambda)$};
\draw[dashed,gray] (1.1,0) -- (1.1,4.2) node[black,above,font=\tiny]{$\theta$};
\node[left,font=\tiny] at (0,4) {1};
\draw[gray!40] (0,4)--(6.2,4);
\draw[very thick] plot[smooth] coordinates {(0,0)(0.55,1.33)(1.1,2.0)(1.65,2.4)(2.2,2.67)(3.3,3.0)(4.4,3.2)(5.5,3.33)(6.2,3.42)};
\draw[very thick,dashed] plot[smooth] coordinates {(0,0)(0.55,0.44)(1.1,2.0)(1.65,3.08)(2.2,3.56)(2.75,3.77)(3.3,3.86)(4.4,3.94)(5.5,3.97)(6.2,3.98)};
\node[align=left] at (5.0,2.45) {simple ($r{=}1$, solid):\\ positive slope\\ at zero exposure};
\node[align=left] at (3.25,1.05) {nonlinear ($r{>}1$, dashed):\\ response flat\\ at low exposure};
\end{tikzpicture}
\caption{Simple versus nonlinear dose response, the two shapes of $H_{r,\theta}$. With $r=1$ the derivative at zero is positive. With $r>1$ the derivative at zero vanishes and low exposure has a higher-order effect. This is nonlinear reinforcement in the aggregate exposure $\Lambda$. It is not by itself a model of repeated independent contacts: repeated exposure is represented only if $\Lambda_i$ is constructed to count distinct contacts, and $r>1$ implies neither bistability nor a population-level critical mass (Remark~\ref{rem:hill}).}
\label{fig:contagion}
\end{figure}

\section{Two controlled reductions and the exact lumping condition}\label{sec:reductions}
The full process does not generally imply a logistic trajectory, and the classical threshold formulas do not apply to it as it stands. This section states precisely which system each formula belongs to, and gives a necessary and sufficient condition for the reduction from three states to two to be exact rather than merely a substitution.

\begin{definition}[Two-state replacement]\label{def:replacement}
The \emph{two-state replacement} of the process of Section~\ref{sec:ctmc} is the Markov jump process on $\{S,I\}^N$ (non-active, active) obtained by (i) setting $\alpha_i\equiv0$, $M_i\equiv0$, $\rho=0$; (ii) merging $U$ and $D$ into a single non-active state $S$; (iii) giving the $S\to I$ transition the intensity $\beta H_{r,\theta}(\Lambda_i)$ and the $I\to S$ transition the intensity $\mu$. When in addition $H_{r,\theta}$ is replaced by the identity, we call it the \emph{linear} two-state replacement; this is the standard network SIS process on $W$.
\end{definition}

Definition~\ref{def:replacement} is a change of model, not an approximation with a small parameter: it deletes the distinction between never-active and formerly active speakers, and with it the possibility that reactivation is easier or harder than first adoption. The next proposition says when nothing is lost.

\begin{definition}[Attainable exposure set]\label{def:attain}
For speaker $i$ at time $t$, the exposures the network can actually deliver form the finite set
\begin{equation}\label{eq:attain}
 \mathcal L_i(t)=\Big\{\,M_i(t)+\sum_{j\ne i}W_{ij}(t)\,z_j \;:\; z_j\in\{0,1\}\,\Big\}.
\end{equation}
\end{definition}

\begin{proposition}[Exact lumpability]\label{prop:lump}
Let $W$ have zero diagonal, so that $\Lambda_i$ does not depend on $x_i$, and write
$b_i(t)=\beta\exp(z_i^\top\gamma+c_{g(i)}(t)^\top\delta)$. The partition of $\{U,A,D\}^N$ induced by merging $U$ and $D$ speaker-wise is strongly lumpable (that is, the aggregated process is Markov with well-defined rates with the same aggregated transition family for every initial law and every starting time) if and only if, for every speaker $i$, for almost every $t$, and for every attainable exposure $\lambda\in\mathcal L_i(t)$,
\begin{equation}\label{eq:lumpcond}
 \alpha_i(t)+b_i(t)\,H_{r,\theta}(\lambda)\;=\;\rho+\varphi\,H_{1,\theta_D}(\lambda).
\end{equation}
When \eqref{eq:lumpcond} holds, the lumped process is the Markov jump process on $\{S,I\}^N$ with activation intensity $\rho+\varphi H_{1,\theta_D}(\Lambda_i)$ and deactivation intensity $\mu\exp(-z_i^\top\eta-c_{g(i)}(t)^\top\xi)$.
\end{proposition}
\begin{proof}
The standard generator criterion for strong lumpability requires that, for any two states in the same block, the total rate into each other block coincide \cite{ball1993}. For locally bounded time-dependent rates, apply this criterion almost everywhere: necessity follows from short-time transition probabilities at Lebesgue points for every starting time, and sufficiency from uniqueness of the aggregated finite forward equations. Fix a configuration $x$ and a speaker $i$ with $x_i=U$, and let $y$ agree with $x$ except that $y_i=D$; then $x$ and $y$ lie in the same block. Because $W_{ii}=0$ we have $\Lambda_i(t,x)=\Lambda_i(t,y)=:\lambda$, and this common value is of the form \eqref{eq:attain}, so $\lambda\in\mathcal L_i(t)$; conversely every element of $\mathcal L_i(t)$ arises from some such pair. For $j\ne i$ we have $\Lambda_j(t,x)=\Lambda_j(t,y)$, since $x_i,y_i\ne A$. The only transitions whose target block depends on the identity of $x_i$ are $i$'s own moves into $A$, with rates $q_i^{UA}(t,x)$ from $x$ and $q_i^{DA}(t,y)$ from $y$; every other transition has equal rate and the same image block. Equality of these two rates for all such pairs is exactly \eqref{eq:lumpcond} on $\mathcal L_i(t)$, and it is also sufficient, since then all states in a block have identical aggregated rates. The stated form of the lumped intensities follows by substitution.
\end{proof}

\begin{remark}[The condition is about attainable exposures, not about all of $[0,\infty)$]\label{rem:attain}
The restriction to $\mathcal L_i(t)$ is not a technicality. Take two speakers with unit contact weights in both directions, no broadcast exposure, no exogenous adoption and no spontaneous revival, no covariate effects, and
\[
 \beta=\varphi=1,\qquad \theta=\theta_D=1,\qquad r=2 .
\]
Each speaker can then experience only the exposures $0$ and $1$, and
\[
 H_{2,1}(0)=H_{1,1}(0)=0,\qquad H_{2,1}(1)=H_{1,1}(1)=\tfrac12,
\]
so \eqref{eq:lumpcond} holds on $\mathcal L_i(t)=\{0,1\}$ and the chain is strongly lumpable even though the reinforcement exponent is not one. The example has strictly positive contact effects, so this is not a degenerate edge case. On a fixed finite network, lumpability is a statement about the finitely many exposures that network can produce.
\end{remark}

\begin{corollary}[Functional form of the condition]\label{cor:lumpfun}
Suppose instead that \eqref{eq:lumpcond} is required to hold as an identity in $\lambda$ on all of $[0,\infty)$. For a fixed speaker and almost every fixed time, this also follows if the same response coefficients are required to satisfy the condition over a network class whose attainable exposures have an accumulation point in $(0,\infty)$. Then exactly one of the following holds:
\begin{enumerate}
\item[(i)] $\alpha_i(t)=\rho$, $r=1$, $\theta=\theta_D$ and $b_i(t)=\varphi>0$ for every $i$ and almost every $t$; or
\item[(ii)] $\alpha_i(t)=\rho$ and $b_i(t)=\varphi=0$ for every $i$ and almost every $t$, with $r$, $\theta$ and $\theta_D$ unconstrained.
\end{enumerate}
\end{corollary}
\begin{proof}
For the network-class formulation, the difference of the two response functions is real analytic on $(0,\infty)$, including for noninteger $r\ge1$. Zeros accumulating at a positive point imply that it vanishes on that connected interval; continuity extends the identity to zero. The coefficients must be held fixed in this argument: pooling exposures across times with different coefficients would not suffice. Evaluating the identity at $\lambda=0$ gives $\alpha_i(t)=\rho$, since $H_{r,\theta}(0)=H_{1,\theta_D}(0)=0$. Letting $\lambda\to\infty$, which is now legitimate, gives $b_i(t)=\varphi$. If $\varphi=0$ we are in case (ii) and both sides are constant. If $\varphi>0$, the identity reduces to $H_{r,\theta}\equiv H_{1,\theta_D}$ on $[0,\infty)$; since $H_{r,\theta}(\lambda)=\tfrac12$ exactly at $\lambda=\theta$ we obtain $\theta=\theta_D$, and then $\lambda^r/(\theta^r+\lambda^r)=\lambda/(\theta+\lambda)$ for all $\lambda>0$ forces $r=1$.
\end{proof}

\begin{remark}[Reading Proposition~\ref{prop:lump}]\label{rem:lumpread}
Two points deserve emphasis. First, the lumped process of Proposition~\ref{prop:lump} is not in general the linear network SIS process: it retains spontaneous activation $\rho$, a saturating response in $\Lambda_i$, and heterogeneous deactivation. It becomes the two-state Hill process of Definition~\ref{def:replacement} only under its additional restrictions. The linear SIS process further replaces the Hill response by the identity; that change does not follow from lumpability. Second, the conditions of Corollary~\ref{cor:lumpfun}(i) are, linguistically, the assumption that re-adopting a form one has used before is exactly the same event as adopting it for the first time. Whenever reactivation differs from adoption, which is the reason for having a dormant state at all, the two-state system is a different model, and results proved about it (Proposition~\ref{prop:logistic} and Theorem~\ref{thm:threshold} below) transfer to the three-state chain only as heuristics. The useful content of Proposition~\ref{prop:lump} is therefore sharp: the dormant distinction disappears precisely when its transition rates cannot be told apart over the exposures the network actually supplies.
\end{remark}

\subsection{Homogeneous mean-field reduction}
Take the linear two-state replacement of Definition~\ref{def:replacement}, assume homogeneous mixing, and let $p(t)$ denote the active proportion. Adoption occurs at rate $\beta p(t)$ per non-user and abandonment at rate $\mu$ per active user. Then
\begin{equation}\label{eq:logistic}
 p'(t)=\beta p(t)(1-p(t))-\mu p(t).
\end{equation}
\begin{proposition}[Logistic special case]\label{prop:logistic}
Let $\beta>\mu>0$, put $K=1-\mu/\beta\in(0,1)$, and let $p(0)=p_0\in(0,1]$. Then the solution of \eqref{eq:logistic} is
\begin{equation}\label{eq:logistic-solution}
 p(t)=\frac{K}{1+(K/p_0-1)e^{-(\beta-\mu)t}},
\end{equation}
and $p(t)\to K$ as $t\to\infty$. Moreover:
\begin{enumerate}
\item[(i)] $p$ is strictly increasing if $p_0<K$, constant if $p_0=K$, and strictly decreasing if $p_0>K$;
\item[(ii)] $p$ has an inflection point at a positive time if and only if $p_0<K/2$, at the time $t^\star=(\beta-\mu)^{-1}\log(K/p_0-1)$; if $K/2\le p_0<K$ the trajectory is increasing and concave throughout, and exhibits no S-shape.
\end{enumerate}
If $\mu\ge\beta$ and $\mu>0$ then $p(t)$ decreases to zero. If $\beta=\mu=0$ then $p'\equiv0$ and $p$ is constant, so decay to zero requires $\mu>0$.
\end{proposition}
\begin{proof}
Equation \eqref{eq:logistic} equals $\beta p(K-p)$ with $K=1-\mu/\beta$. Separation of variables yields \eqref{eq:logistic-solution} when $\beta>\mu$, and the sign of $p'=\beta p(K-p)$ on $(0,1]$ gives (i). Differentiating, $p''=\beta p'(K-2p)$, which, on a nonconstant trajectory, vanishes exactly when $p=K/2$; since $p$ moves monotonically from $p_0$ towards $K$, the value $K/2$ is attained at a positive time if and only if $p_0<K/2$, and solving $p(t^\star)=K/2$ in \eqref{eq:logistic-solution} gives $t^\star$. If $\beta\le\mu$ then $p'\le-(\mu-\beta)p\le0$, with strict decrease for $p>0$ except in the boundary case $\beta=\mu$, where $p'=-\beta p^2$. The last statement is immediate.
\end{proof}

\begin{figure}[t]
\centering
\begin{tikzpicture}[x=1cm,y=1cm,font=\scriptsize,>=Latex]
\draw[->] (0,0) -- (9,0) node[right]{time $t$};
\draw[->] (0,0) -- (0,4.4) node[above]{active proportion $p(t)$};
\draw[dashed,gray] (0,3.2) -- (8.6,3.2);
\node[left,font=\tiny] at (0,3.2) {$K$};
\draw[gray!35] (2.5,0)--(2.5,3.9);
\draw[gray!35] (5.5,0)--(5.5,3.9);
\draw[very thick] plot[smooth] coordinates {(0,0.12)(1,0.24)(2,0.48)(3,1.0)(4,1.8)(5,2.48)(6,2.88)(7,3.08)(8,3.16)(8.6,3.18)};
\node[align=center,font=\tiny] at (1.25,3.7) {slow start};
\node[align=center,font=\tiny] at (4.0,3.7) {steep phase};
\node[align=center,font=\tiny] at (7.0,3.7) {plateau};
\end{tikzpicture}
\caption{The logistic curve \eqref{eq:logistic-solution} in the small-seed case $p_0<K/2$, which is the only case that produces the familiar S-shape (Proposition~\ref{prop:logistic}(ii)). The plateau $K=1-\mu/\beta$ is not full adoption: even a successful form settles below one, because abandonment never stops. Calling the plateau ``fixation'' would be incorrect, and the phase labels are descriptive; they are not attributed to any particular author. The curve is a consequence of one set of assumptions about one reduced system, not a law imposed on the data.}
\label{fig:scurve}
\end{figure}
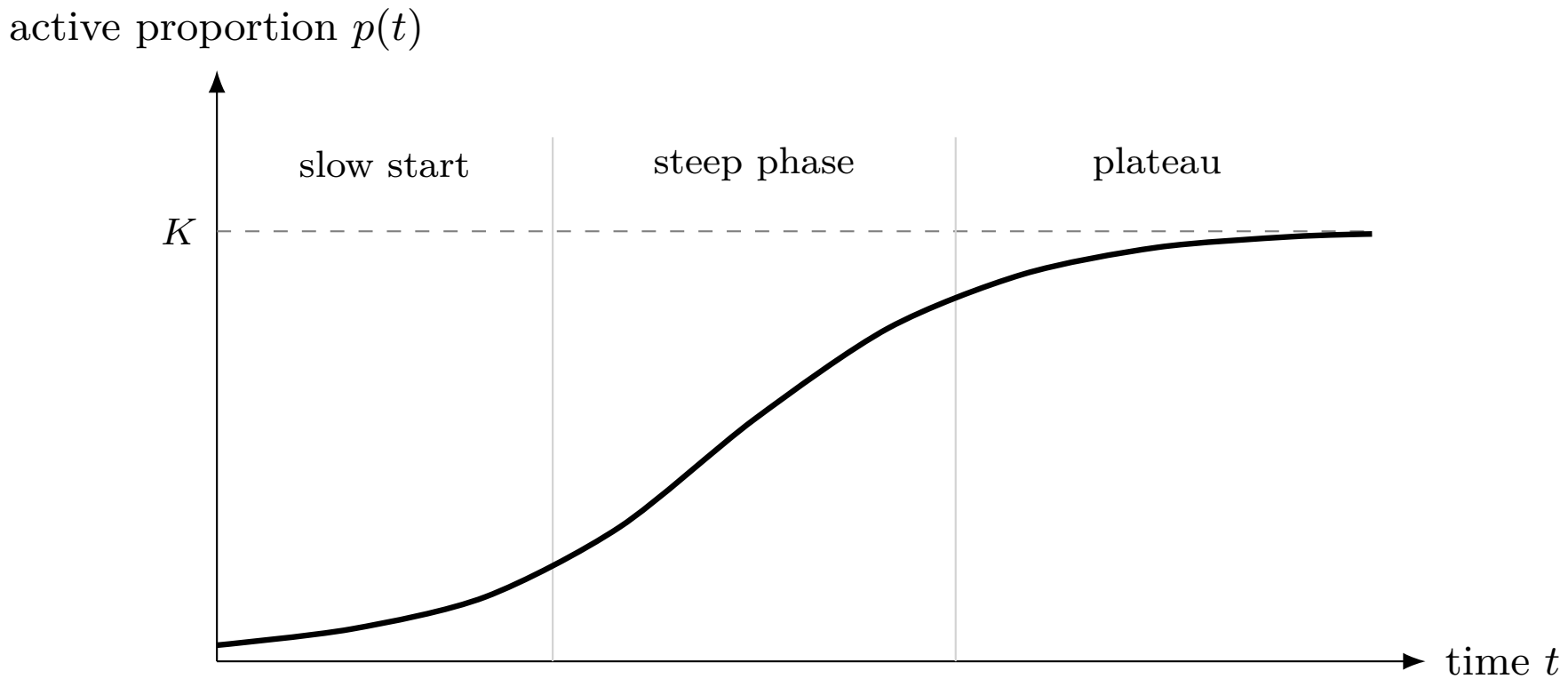

Figure~\ref{fig:scurve} shows the curve and marks the small-seed regime in which it is sigmoidal. Thus an S-curve is a special-case consequence of one reduction under a small-seed condition, not an imposed empirical law, and it belongs to the linear two-state replacement rather than to the chain of Section~\ref{sec:ctmc}. Modularity, nonlinear reinforcement, time variation, or observation bias can produce plateaus, multiple waves, or stepwise aggregate curves.

\subsection{Static-network invasion approximation}
For a threshold calculation only, take the \emph{linear} two-state replacement with $W$ static, and close the exact first-moment hierarchy by a mean-field assumption, giving the network SIS system
\begin{equation}\label{eq:nis}
 p_i'=\beta(1-p_i)\sum_j W_{ij}p_j-\mu p_i,
 \qquad i=1,\ldots,N.
\end{equation}
This is a deterministic approximation for marginal activation probabilities, not the exact first-moment equation of the Markov chain, because pair correlations are closed by a mean-field assumption \cite{lajmanovich1976,kiss2017}. Two approximations therefore separate \eqref{eq:nis} from Section~\ref{sec:ctmc}: the replacement of Definition~\ref{def:replacement}, which requires both Proposition~\ref{prop:lump} and the additional removal of forcing and heterogeneity; replacing a Hill response by the identity is a further model change. The remaining approximation is the moment closure.

\begin{theorem}[Spectral invasion criterion for \eqref{eq:nis}]\label{thm:threshold}
Let $W\ge0$ be constant and irreducible, and let $\beta,\mu>0$. The inactive equilibrium $p=0$ of \eqref{eq:nis} is locally asymptotically stable if
\begin{equation}\label{eq:rlex}
 \mathcal R_{\mathrm{lex}}:=\frac{\beta}{\mu}\rho(W)<1,
\end{equation}
and unstable if $\mathcal R_{\mathrm{lex}}>1$.
\end{theorem}
\begin{proof}
The Jacobian of \eqref{eq:nis} at zero is $J=\beta W-\mu I$. By Perron--Frobenius theory, the spectral abscissa of $W$ is its nonnegative eigenvalue $\rho(W)$. Hence the spectral abscissa of $J$ is $\beta\rho(W)-\mu$. Standard linearisation gives local asymptotic stability when this quantity is negative and instability when it is positive. At equality the Jacobian is non-hyperbolic, so this theorem deliberately makes no claim.
\end{proof}

\lingnote{The index $\mathcal R_{\mathrm{lex}}=(\beta/\mu)\,\rho(W)$ plays the role that $R_0$ plays for an epidemic. It multiplies how contagious the word is ($\beta$) and how long speakers keep using it ($1/\mu$) by how densely connected the community is ($\rho(W)$). In the deterministic approximation, values above one make the inactive equilibrium locally unstable, whereas values below one make it locally asymptotically stable. It does not say that an individual lexical item certainly survives or fades in the finite stochastic process. The lesson for the linguist is that the same word can spread in a well-connected community and die in a sparse one: survival is a property of the network, not only of the word.}

The criterion concerns invasion in the approximation \eqref{eq:nis}. For the finite static, unforced linear two-state chain with $\mu>0$, the all-inactive state is reached almost surely in finite time: from every nonabsorbing state a sequence of recoveries has positive probability, so no other closed communicating class exists. This statement is not asserted for arbitrary time-varying rates; the phenomenon of interest above threshold is long-lived finite-horizon or quasi-stationary persistence, the simulations of Section~\ref{sec:insilico} measure finite-horizon persistence, not a quasi-stationary distribution. For time-varying networks, $\rho(W(t))$ at a single time is generally insufficient; growth is governed by the evolution operator of the linearised nonautonomous system.

\begin{remark}[What the vanishing derivative does and does not say]\label{rem:hill}
For the unforced homogeneous two-state closure $p_i\prime=\beta(1-p_i)H_{r,\theta}(\sum_jW_{ij}p_j)-\mu p_i$, the Hill response with $r>1$ has derivative zero at the origin, so the Jacobian at $p=0$ is $-\mu I$ and no spectral invasion criterion of the form \eqref{eq:rlex} exists: invasion from an \emph{infinitesimal prevalence} in the deterministic limit is impossible. This does not transfer to the finite chain. There a single active neighbour $j$ delivers exposure $W_{ij}>0$, so the adoption rate of speaker $i$ is $\beta H_{r,\theta}(W_{ij})>0$ and spread from one seed speaker has positive probability at any $r$. The vanishing derivative concerns this particular nonlinear deterministic closure. Replacing $\mathbb E[H(\sum_jW_{ij}\ind\{X_j=A\})]$ by $H(\sum_jW_{ij}p_j)$ is an additional approximation, not a consequence of independence; even independent rare active neighbours contribute terms of first order in their activation probabilities. Bistability or an effective positive critical mass may occur, but neither is universal: existence depends on the full nonlinear system, baseline adoption, and network structure. We therefore treat critical mass as an estimand or model-specific result, not as a theorem of the general framework. Figure~\ref{fig:contagion} shows the two response shapes that this remark contrasts.
\end{remark}

\section{Community transmission and cross-community diffusion}\label{sec:community}
Let $V_g=\{i:g(i)=g\}$ and define speaker prevalence
\begin{equation}
 P_g(t)=|V_g|^{-1}\sum_{i\in V_g}\ind\{X_i(t)=A\}.
\end{equation}

Within-community and cross-community edges of $W$ distinguish local contact from boundary-crossing contact, but this graph partition is not identical to Labov's technical distinction.
To avoid conceptual confusion, we explicitly distinguish Labov’s sociolinguistic taxonomy from our network-based operationalization. In Labov’s classical framework, \emph{transmission} denotes the faithful intergenerational acquisition of a speech system by children within a community, whereas \emph{diffusion} refers to structural borrowing across community boundaries, typically mediated by adult contacts.

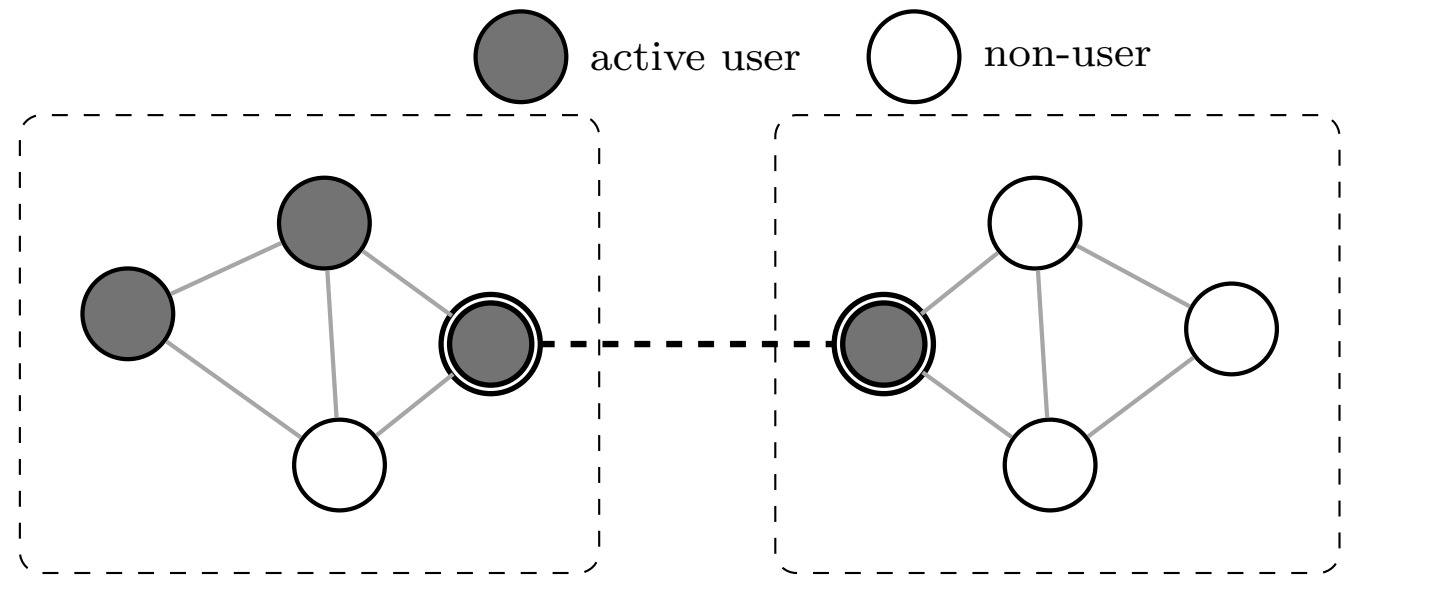
\begin{figure}[t]
\centering
\begin{tikzpicture}[
  >=Latex,
  spk/.style={circle,draw,thick,minimum size=6mm,font=\tiny,inner sep=0pt},
  act/.style={spk,fill=black!55},
  non/.style={spk,fill=white},
  brg/.style={spk,fill=black!55,double,line width=1pt},
  within/.style={thick,gray!70},
  cross/.style={very thick,dashed}
]
\node[act] (a1) at (0,1.4) {};
\node[act] (a2) at (1.3,2.0) {};
\node[non] (a3) at (1.4,0.4) {};
\node[brg] (a4) at (2.4,1.2) {};
\draw[within] (a1)--(a2) (a1)--(a3) (a2)--(a4) (a3)--(a4) (a2)--(a3);
\node[brg] (b1) at (5.0,1.2) {};
\node[non] (b2) at (6.0,2.0) {};
\node[non] (b3) at (6.1,0.4) {};
\node[non] (b4) at (7.3,1.3) {};
\draw[within] (b1)--(b2) (b1)--(b3) (b2)--(b4) (b3)--(b4) (b2)--(b3);
\draw[cross] (a4)--(b1);
\node[draw,dashed,rounded corners,fit=(a1)(a2)(a3)(a4),inner sep=4mm,label={[font=\scriptsize]below:community $g$ (consolidated)}] {};
\node[draw,dashed,rounded corners,fit=(b1)(b2)(b3)(b4),inner sep=4mm,label={[font=\scriptsize]below:community $g'$ (word just arriving)}] {};
\node[act,label={[font=\tiny]right:active user}] at (2.6,3.1) {};
\node[non,label={[font=\tiny]right:non-user}] at (5.2,3.1) {};
\end{tikzpicture}
\caption{Within-community versus cross-community contact. Grey ties inside each dashed box represent within-community contact; the dashed tie represents cross-community contact carried by the two bridging speakers (double circles). The figure illustrates a graph decomposition, not the Labovian transmission/diffusion distinction, which additionally requires age and acquisition type. A word can be fully consolidated in community $g$ (high $P_g$) yet only begin to reach $g'$ through a single bridge.}
\label{fig:transdiff}
\end{figure}

Because our corpus panel lacks speaker age, demographic metadata, and explicitly typed social layers (e.g., $W^{(\text{tr})}$ for child acquisition vs. $W^{(\text{diff})}$ for adult contact), this paper does not implement or test Labovian intergenerational transmission. Instead, our empirical model evaluates within-community versus cross-community graph contact—a structural partition that serves as an operational proxy for geographic reach, but represents a formally weaker distinction than sociolinguistic transmission. All subsequent references to diffusion in this text should be understood strictly in this graph-topological sense. Figure~\ref{fig:pipeline} and Figure~\ref{fig:transdiff} follow the same convention.

\section{Emergence and cultural covariates}\label{sec:emergence}
The count of newly created candidates in community $g$ may be modelled as a non-homogeneous Poisson process with intensity
\begin{equation}\label{eq:emergence}
 \log\nu_g(t)=a_g+u_g(t)^\top b+s(t),
\end{equation}
where $u_g(t)$ contains measured contact, naming pressure, or cultural activity, and $s$ is a smooth temporal component. This process concerns candidate creation, not adoption of an already identified form. Equation~\eqref{eq:emergence} is \emph{not estimated anywhere in this paper}: doing so requires a candidate stream with dated first uses and a model of detection delay, and the corpora used here supply neither. We state it because the framework is incomplete without it and because a confirmatory study must estimate it, and we mark every downstream statement that would depend on it.

To complete the generative link, a candidate event at time $T_{\mathrm{emerge}}$ in community $g$ would initialise the speaker process by placing the coiner in state $A$ and all other speakers in $U$, after which Section~\ref{sec:ctmc} applies; documentation is then the first time the observation model of Section~\ref{sec:observation} yields a positive count. Under that construction first documentation is an observed upper bound on the latent emergence time, $T_{\mathrm{emerge}}\le T_{\mathrm{first\ doc}}$, with the gap governed by detection probability and corpus exposure; left truncation and detection delay must be modelled explicitly, and the first attestation must be a validated lexical use of the form rather than a string match. This documentary lag is well attested empirically: in a four-country press corpus, the metaphorical denominations of the 2007--2012 financial crisis circulate in usage well before they are consigned in dictionaries, and their conceptualisation varies across Romance-language communities, a contrast between terms observed \emph{in vivo} and terminology fixed \emph{in vitro}~\cite{fernandez2015}.

For high-dimensional cultural indicators $x_g(t)$, one may use
\begin{equation}\label{eq:factor}
 x_g(t)=L c_g(t)+\varepsilon_g(t),\qquad c_g(t)=F c_g(t-1)+\zeta_g(t).
\end{equation}
To identify \eqref{eq:factor}, one must fix the factor dimension and impose a rotation convention or equivalent constraints on $L$ and the innovation covariance. Without such restrictions, the factor representation is non-unique (typically up to rotations under the usual covariance normalisation, and more generally under compatible nonsingular reparameterisations). The factor layer is optional, is likewise not estimated here, and should be omitted when data do not support it. Treating neologisms and specialised terms as dynamic, culture-bound units rather than static labels is the premise of recent terminology research~\cite{temmerman2014}; the cultural-covariate layer is one way to encode that dependence in the model.

\section{Observation model}\label{sec:observation}
Let $Y_{wgtc}$ be the token count for word $w$, community $g$, time window $t$, and channel $c$, and let $E_{gtc}>0$ be corpus exposure. The derivation below uses running-token exposure; hours or speaker-sessions require the corresponding unit-specific emission rate. Rescaling exposure by a factor $c$ requires rescaling $u$ and $\phi$ by $1/c$ to preserve the mean.

\subsection{From speaker prevalence to expected counts}\label{subsec:tokenmean}
\begin{figure}[t]
\centering
\resizebox{\linewidth}{!}{%
\begin{tikzpicture}[
  >=Latex,
  fac/.style={draw,rounded corners,thick,minimum height=11mm,text width=20mm,align=center,font=\scriptsize},
  op/.style={font=\large}
]
\node[fac] (pi) {share of text from users\\ $\pi^{\mathrm{tok}}$};
\node[op,right=4mm of pi] (x1) {$\times$};
\node[fac,right=4mm of x1] (u) {how often each one\\ $u$};
\node[op,right=4mm of u] (x2) {$\times$};
\node[fac,right=4mm of x2] (E) {corpus size\\ $E$};
\node[op,right=4mm of E] (x3) {$\times$};
\node[fac,right=4mm of x3] (q) {chance of being recorded\\ $q$};
\node[op,right=4mm of q] (eq) {$=$};
\node[fac,thick,right=4mm of eq] (Y) {expected count\\ $\E[Y]$};
\draw[decorate,decoration={brace,amplitude=5pt,mirror},thick] (pi.south west) -- (u.south east)
  node[midway,below=5pt,font=\scriptsize,align=center]{what speakers do (latent)};
\draw[decorate,decoration={brace,amplitude=5pt,mirror},thick] (E.south west) -- (q.south east)
  node[midway,below=5pt,font=\scriptsize,align=center]{how the corpus filters it};
\end{tikzpicture}}
\caption{Why token frequency is not speaker prevalence. The \emph{expected} count $\E[Y]$ is the product of what speakers do and how the corpus records it; the observed count $Y$ varies around it according to the negative-binomial law \eqref{eq:nb}, so the diagram is a statement about the mean, not an identity between random variables. Media attention can inflate $u$ without any rise in $\pi$; a change in archiving or sampling changes $q$ alone. From aggregate counts only the product $\pi^{\mathrm{tok}}uq$ is recoverable (Proposition~\ref{prop:identify}).}
\label{fig:obsdecomp}
\end{figure}
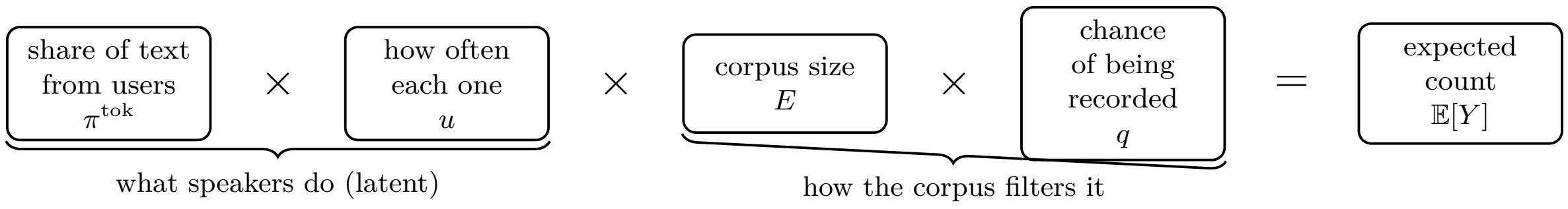

Speaker prevalence and token frequency are different quantities, and the observation mean must be derived rather than asserted; Figure~\ref{fig:obsdecomp} sets out the four factors that separate the one from the other. Let $\pi_{wgt}\in[0,1]$ be the latent proportion of \emph{speakers} who are active users, and let $\tau^{A}_{gt}$ and $\tau^{S}_{gt}$ denote the mean number of running tokens contributed to the corpus by an active and by a non-active speaker respectively, in the window, with both output means positive and word and channel dependence suppressed in the notation. Here $u$ is the token-weighted mean emission rate among active speakers and $q$ the conditional probability that an emitted occurrence is recorded. The share of the sampled text produced by active users is then
\begin{equation}\label{eq:tokprev}
 \pi^{\mathrm{tok}}_{wgt}
 =\frac{\pi_{wgt}\,\tau^{A}_{gt}}{\pi_{wgt}\,\tau^{A}_{gt}+(1-\pi_{wgt})\,\tau^{S}_{gt}} .
\end{equation}
Writing $u_{wgtc}>0$ for the rate at which an active user emits $w$ per token of their own output, and $q_{wgtc}\in(0,1]$ for detection and preservation probability, the expected count is
\begin{equation}\label{eq:nbmean}
 m_{wgtc}=E_{gtc}\,\pi^{\mathrm{tok}}_{wgt}\,u_{wgtc}\,q_{wgtc}.
\end{equation}
\begin{remark}[The sampling assumption]\label{rem:sampling}
Equation \eqref{eq:nbmean} reduces to the familiar $m=E\pi uq$ as an identity for all $0<\pi<1$ exactly when $\tau^{A}_{gt}=\tau^{S}_{gt}$ (the endpoints also agree for unequal output means), that is when active and non-active users contribute text to the corpus in proportion to their numbers. This is an assumption about the sampling frame, not a property of the model, and it is doubtful in press corpora, where the authors who use a new term are plausibly also those who write more about the topic it names. If it fails, $\pi^{\mathrm{tok}}$ and not $\pi$ is what the counts see, and by \eqref{eq:tokprev} the ratio $\pi^{\mathrm{tok}}_{wgt}/\pi_{wgt}=\tau^{A}_{gt}/\big(\pi_{wgt}\tau^{A}_{gt}+(1-\pi_{wgt})\tau^{S}_{gt}\big)$ tends from $1$ at $\pi_{wgt}=1$ to $\tau^{A}_{gt}/\tau^{S}_{gt}$ as $\pi_{wgt}\to0$ (increasing when $\tau^A_{gt}>\tau^S_{gt}$ and decreasing when the inequality is reversed), so at the low prevalences of interest the discrepancy is the full output ratio between active and non-active speakers. We adopt the assumption throughout, state it as such, and note that relaxing it requires speaker-resolved output volumes, which is one more reason the empirical programme of Section~\ref{sec:protocol} needs speaker-level data.
\end{remark}

Conditional on latent paths, covariates and parameters, the fitted likelihood treats distinct available cells as independent. Shared latent paths or parameters still induce marginal dependence. We use dispersions $\psi_c>0$ and the mean--dispersion parameterisation
\begin{align}
 Y_{wgtc}&\sim\operatorname{NegBin}(m_{wgtc},\psi_c),\qquad
 \operatorname{Var}(Y_{wgtc})=m_{wgtc}+m_{wgtc}^2/\psi_c.\label{eq:nb}
\end{align}
For a finite number of token positions, the negative-binomial law is an unbounded count approximation rather than an exact bounded-token sampling law; its suitability must be checked in applications. If $\pi_{wgt}=0$, the model assigns a structural zero; no arbitrary logarithmic offset is needed. Spoken data should be organised at speaker--session level whenever possible. A zero in a short session does not imply that a speaker is a non-user. Historical analyses must additionally model genre composition and documentary survival.

\subsection{What is identified from counts}\label{subsec:identobs}
\begin{proposition}[What aggregate counts identify]\label{prop:identify}
Suppose the sampling assumption of Remark~\ref{rem:sampling} holds, so that
$m_{wgtc}=E_{gtc}\,\pi^{\mathrm{tok}}_{wgt}u_{wgtc}q_{wgtc}$, with $\pi^{\mathrm{tok}}=\pi$, and suppose the use intensities $u>0$ and detection probabilities $q\in(0,1]$ are otherwise free. All rescalings below are restricted to the admissible parameter domain.
\begin{enumerate}
\item[(i)] Within any channel the likelihood depends on $(u,q)$ only through the product $uq$. This is an exact invariance of the specified model, because $u$ and $q$ enter nowhere else. Count observations under this unrestricted factorisation cannot break it without additional constraints or calibration.
\item[(ii)] The map $(\pi^{\mathrm{tok}},u)\mapsto(\pi^{\mathrm{tok}}/s,\,su)$, for any $s>0$ keeping $\pi^{\mathrm{tok}}/s\in[0,1]$, leaves every $m_{wgtc}$ unchanged. With $C>1$ channels and channel-specific free $u_c$ and $q_c$, the same rescaling applied simultaneously in every channel again leaves all channel means unchanged, so additional channels do not by themselves separate prevalence from use intensity.
\item[(iii)] Knowledge of $q$ in one channel does \emph{not} identify prevalence and use intensity separately: it identifies only their product in that channel. Knowledge of the ratios $q_c/q_{c\prime}$ fixes relative detection, but with free channel-specific $u_c$ does not fix relative emission scaling. With shared $u$ it fixes relative emission scaling while leaving an overall scale ambiguity.
\end{enumerate}
\end{proposition}
\begin{proof}
(i) and (ii) are immediate from the multiplicative form of $m_{wgtc}$: each stated map leaves every mean, and hence the likelihood, unchanged. For (iii), fix $q$ in a channel and consider $(\pi^{\mathrm{tok}},u,q)=(0.2,0.1,0.5)$ and $(0.1,0.2,0.5)$; both give $m/E=0.01$, so $q$ is known and the two parameter triples are observationally identical. For the ratio claim, replacing $q_c\mapsto tq_c$ for all $c$ and $u_c\mapsto u_c/t$ leaves every ratio $q_c/q_{c'}$ and every product $u_cq_c$ unchanged, so a known ratio removes no scale.
\end{proof}

The distinction at work here is the standard one between structural non-identifiability, which no amount of data removes because the likelihood is exactly flat along a direction, and practical non-identifiability, in which the likelihood is merely very flat at the sample sizes available; profile likelihoods are the usual diagnostic for telling them apart \cite{raue2009}. Proposition~\ref{prop:identify} concerns the structural case. The constructive question is therefore not whether counts suffice, which they do not, but which additional measurement supplies what. Table~\ref{tab:designs} answers that question design by design.

\begin{table}[t]
\centering\footnotesize
\renewcommand{\arraystretch}{1.2}
\begin{tabularx}{\linewidth}{@{}>{\raggedright\arraybackslash}p{6.1cm}>{\raggedright\arraybackslash}X@{}}
\toprule
\textbf{Information available in addition to counts} & \textbf{What the count mean then identifies}\\
\midrule
none, with prevalence and emission factors unrestricted & only the product $\pi^{\mathrm{tok}}uq$\\
detection $q$ known in one channel & the product $\pi^{\mathrm{tok}}u$ in that channel\\
detection ratios known across channels, use intensity shared & relative channel scaling; an absolute scale ambiguity remains\\
use intensity and detection both known in one channel & $\pi^{\mathrm{tok}}$, and hence $\pi$ under the sampling assumption\\
prevalence measured independently & the product $uq$\\
prevalence and detection measured independently & the use intensity $u$\\
\bottomrule
\end{tabularx}
\caption{What each observation design identifies in the count mean $m=E\,\pi^{\mathrm{tok}}uq$. The first row is the situation of the corpus analysis reported below. Replicated observations of identified speakers can supply the entries lower in the table, but they do not by themselves reveal latent active status: a speaker may be active and emit no token in a given window, so observed use within a window must be distinguished from the latent state and connected to it by an explicit observation model. Calibrating detection through overlapping samples likewise requires assumptions about sampling, record linkage and dependence; having several channels is not sufficient.}
\label{tab:designs}
\end{table}

One distinction in Proposition~\ref{prop:identify} carries through the rest of the paper and is easy to lose. The exchange of $u$ against $q$ in (i) is an exact invariance of the whole model. The rescaling of prevalence against emission in (ii) is an invariance of the \emph{observation layer alone}: an arbitrarily rescaled prevalence path need not satisfy the same nonlinear dynamical equation, and Section~\ref{subsec:emisobs} shows that it does not. In the empirical work below we adopt the convention $\phi_c\equiv1$, which fixes the scale in (ii) by fiat. All rate estimates are then conditional on that convention, as Section~\ref{sec:diagnosis} makes explicit.

\section{An estimable community-level specialisation}\label{sec:estimable}
The full individual process of Sections~\ref{sec:network}--\ref{sec:ctmc} is generative but, without speaker-level contact data, over-parameterised for corpus counts. We therefore formulate the estimable specialisation directly as a continuous-time community ODE and integrate it numerically between observation windows.

Let $B=(B_{gh})$ be a fixed nonnegative community-contact matrix, with its scale fixed before estimation. For $\pi_g(t)\in[0,1]$, define
\begin{equation}\label{eq:commdyn}
 \dot\pi_g(t)=\beta\,[1-\pi_g(t)]\left(\pi_g(t)+\kappa\sum_h B_{gh}\pi_h(t)\right)-\mu\pi_g(t),
 \qquad g=1,\ldots,G.
\end{equation}
Here $\beta,\mu>0$ and $\kappa\ge0$. Counts at observation times $t_k$ follow \eqref{eq:nb} with
\begin{equation}\label{eq:commobs}
 m_{gkc}=E_{gkc}\,\pi_g(t_k)\,\phi_c,\qquad \phi_c=u_cq_c,
\end{equation}
under the sampling assumption of Remark~\ref{rem:sampling}. For window counts this is a point-sampling convention: if $dE_{gc}(s)$ denotes exposure accumulated inside window $I_k$, the corresponding continuous-time mean is $\phi_c\int_{I_k}\pi_g(s)\,dE_{gc}(s)$. Equation~\eqref{eq:commobs} approximates this by $E_{gkc}\phi_c\pi_g(t_k)$; adequacy requires limited within-window variation or a sensitivity check. The free parameters are $\beta$, $\mu$, $\kappa$, the emission scales $\phi_c$ and the dispersions $\psi_c$, together with the initial condition. Initial prevalences must either be estimated with regularisation or anchored by genuinely independent baseline information; they cannot in general be fixed by the sampling frame, and Section~\ref{subsec:seeding} reports what was actually done and what it costs.

With $\kappa=0$ and one community, \eqref{eq:commdyn} is exactly the logistic reduction \eqref{eq:logistic}. Linearisation at $\pi=0$ gives $J=\beta(I+\kappa B)-\mu I$, so the inactive equilibrium is locally asymptotically stable when
\begin{equation}\label{eq:community-threshold}
 \frac{\beta}{\mu}\rho(I+\kappa B)<1
\end{equation}
and unstable when the inequality is reversed. This is the community-level analogue of Theorem~\ref{thm:threshold}, and it inherits the same caveats: it is a statement about \eqref{eq:commdyn}, which is a reduction of the linear two-state replacement, not of the chain of Section~\ref{sec:ctmc}. Because $\beta$ and the scale of $I+\kappa B$ are confounded, the normalisation of $B$ is part of the model definition, and $\kappa$ measures cross-community coupling only relative to that normalisation.

\begin{remark}[The compartmental pattern, and where it stops]\label{rem:compartmental} The shape of this proof, local Lipschitz continuity followed by a bound that rules out blow-up, is the standard one for compartmental language models and it scales well beyond the dimension used here: Area and Jornet \cite{areajornet2025} establish existence, uniqueness and positivity on $[0,\infty)$ for a $(3J+2)$-dimensional system of competing tongues by combining Picard--Lindel\"of with a Gr\"onwall bound on the total population. The contrast worth drawing is not in the dynamics but in what comes after them. There the parameters are explored by simulation over a declared range, and the qualitative conclusions follow from inequalities among them; here they must be recovered from counts, and Sections~\ref{subsec:flatfamily} and \ref{subsec:identreal} locate the point at which that recovery stops being possible.
\end{remark}

\begin{proposition}[Global well-posedness of the community model]\label{prop:global}
For every nonnegative matrix $B$, parameters $\beta,\mu>0$, $\kappa\ge0$, and initial condition $\pi(0)\in[0,1]^G$, system \eqref{eq:commdyn} has a unique solution defined for all $t\ge0$, and $\pi(t)\in[0,1]^G$ for all $t\ge0$. Thus the exact solution requires no clipping. A numerical discretisation must be checked separately for accuracy and preservation of the state domain.
\end{proposition}
\begin{proof}
The right-hand side is polynomial, hence locally Lipschitz, so a unique local solution exists. On the face $\pi_g=0$ the $g$-th component of the vector field equals $\beta\kappa\sum_h B_{gh}\pi_h\ge0$, and on the face $\pi_g=1$ it equals $-\mu<0$; hence the cube $[0,1]^G$ is forward invariant. A trajectory confined to a compact set cannot blow up in finite time, so the local solution extends to all $t\ge0$.
\end{proof}

\begin{remark}\label{rem:normB}
For $B\ge0$ one has $\rho(I+\kappa B)=1+\kappa\rho(B)$, so fixing the normalisation $\rho(B)=1$ makes the community invasion boundary $(\beta/\mu)(1+\kappa)=1$ and renders $\kappa$ comparable across specifications and network scales. We use $\rho(B)=1$ throughout.
\end{remark}

The next proposition is a structural limitation of \eqref{eq:commdyn} on the graph actually used in Section~\ref{sec:diagnosis}, and it is stated here because it constrains what any fit of that specification can possibly express.

\begin{proposition}[Two-trajectory degeneracy on the complete graph]\label{prop:symmetry}
Let $G\ge2$ and $B=(\mathbf1\mathbf1^\top-I)/(G-1)$ be the normalised complete graph on $G$ communities, let $\beta,\mu,\kappa$ be shared across communities, and let the initial condition satisfy $\pi_1(0)=\pi_\star$ and $\pi_g(0)=\pi_0$ for all $g\ge2$. Then $\pi_g(t)=\pi_h(t)$ for all $g,h\ge2$ and all $t\ge0$. Consequently the solution takes at most two distinct values at any time, and if $\pi_\star>\pi_0$ then $\pi_1(t)>\pi_g(t)$ for all $g\ge2$ and all $t\ge0$.
\end{proposition}
\begin{proof}
The vector field is equivariant under any permutation of the community indices that permutes $B$; for the complete graph every permutation does so. The permutation exchanging two indices $g,h\ge2$ fixes the initial condition, so by uniqueness of the solution (Proposition~\ref{prop:global}) it fixes the whole trajectory, giving $\pi_g\equiv\pi_h$. Writing $\pi_1=a$, $\pi_{g\ge2}=b$ and $\bar B$ for the common off-diagonal weight, the reduced planar system is $\dot a=\beta(1-a)\big(a+\kappa\bar B(G-1)b\big)-\mu a$, $\dot b=\beta(1-b)\big(b+\kappa\bar B((G-2)b+a)\big)-\mu b$. Subtracting gives $\dot\Delta=h(t)\Delta$, where $\Delta=a-b$ and $h(t)=\beta[1-a-b-\kappa\bar B\{1+(G-2)b\}]-\mu$. Thus $\Delta(t)=\Delta(0)\exp(\int_0^t h(s)\,ds)$; boundedness of $a,b$ makes the integral finite on finite intervals and preserves the strict sign.
\end{proof}

\begin{remark}\label{rem:symmetryread}
Proposition~\ref{prop:symmetry} says that the specification used on real data below has exactly two latent paths, source and non-source, whatever the data are. It cannot represent one country overtaking another, or a specific multi-country pathway; any such structure in the fitted output is observation noise. Representing distinct country paths requires asymmetric or measured contacts, or independently regularised per-country initial conditions. We verified the degeneracy numerically on the fitted configuration: the four non-source communities agree to machine precision at every observation time.
\end{remark}

\subsection{An exact non-identification family in the low-prevalence regime}\label{subsec:flatfamily}
At small prevalence the quadratic terms in \eqref{eq:commdyn} are negligible and the system is governed by its linearisation
\begin{equation}\label{eq:linear}
 \dot\pi=(aI+bB)\,\pi,\qquad a=\beta-\mu,\qquad b=\beta\kappa .
\end{equation}
Equation~\eqref{eq:linear} defines the linear approximation exactly; its solution need not remain in $[0,1]^G$ at arbitrarily long times. Comparisons with the count model are restricted to horizons on which the relevant states are admissible. The following proposition replaces the informal observation that the likelihood has a flat direction by an exact statement about which parameters the linearised model can and cannot distinguish.

\begin{proposition}[Non-identification of the rate triple]\label{prop:family}
Fix $a\in\R$ and $b\ge0$, an initial condition, and the observation parameters. Every member of the one-parameter family
\begin{equation}\label{eq:family}
 (\beta,\mu,\kappa)=\big(d,\;d-a,\;b/d\big),\qquad d>\max(0,a),
\end{equation}
produces exactly the same linearised system matrix $aI+bB$, hence exactly the same trajectory of \eqref{eq:linear} and the same likelihood. In the full nonlinear system \eqref{eq:commdyn}, writing $F_d$ for the vector field obtained from \eqref{eq:family},
\begin{equation}\label{eq:fdiff}
 F_d(\pi)-F_{d'}(\pi)\;=\;-(d-d')\,\pi\odot\pi ,
\end{equation}
where $\odot$ denotes componentwise multiplication.
\end{proposition}
\begin{proof}
For the triple \eqref{eq:family} we have $\beta-\mu=d-(d-a)=a$ and $\beta\kappa=d\cdot(b/d)=b$, and $\mu>0$ exactly when $d>a$, so the constraints of \eqref{eq:commdyn} are met. Expanding \eqref{eq:commdyn},
\[
 F_d(\pi)=a\pi+bB\pi-d\,\pi\odot\pi-b\,\pi\odot(B\pi),
\]
in which only the coefficient of $\pi\odot\pi$ depends on $d$; subtracting two such fields gives \eqref{eq:fdiff}.
\end{proof}

For completeness, the vector-field identity also gives a finite-horizon trajectory bound. If two nonlinear trajectories with the same initial condition remain in $[0,\varepsilon]^G$ up to time $T$, and $F_d$ has Lipschitz constant $L\ge0$ there in the maximum norm, subtraction of their integral equations and Gronwall's inequality give $\|\pi_d(t)-\pi_{d\prime}(t)\|_\infty\le |d-d\prime|\varepsilon^2(e^{Lt}-1)/L$, interpreted as $|d-d\prime|\varepsilon^2t$ when $L=0$. Thus small curvature is a finite-horizon, parameter-dependent statement, not a uniform guarantee for arbitrarily large rates or long times. The consequence is sharper than a statement about a nearly flat likelihood. In the linear regime the three rates are \emph{exactly} non-identified, and the only information that separates them is carried by the term quadratic in prevalence. Whether that term is informative is an empirical question about the prevalences actually reached, not a property of the model, and Section~\ref{subsec:identreal} answers it for the press panel.

\begin{corollary}[Two observable rates on the complete graph]\label{cor:modes}
Let $G\ge2$ and $B=(\mathbf 1\mathbf 1^\top-I)/(G-1)$, the complete graph normalised so that $\rho(B)=1$. Then $aI+bB$ has eigenvalue
\[
 \gamma_+=a+b \quad\text{on the vector } \mathbf 1, \qquad
 \gamma_-=a-\frac{b}{G-1} \quad\text{on the contrast space } \{v:\mathbf 1^\top v=0\},
\]
so the linearised system has at most two distinct growth rates (two when $b>0$, coincident when $b=0$): a common mode and a single rate for all between-community contrasts. If both are measurable then
\[
 b=\frac{G-1}{G}\,(\gamma_+-\gamma_-),\qquad a=\frac{\gamma_++(G-1)\gamma_-}{G},
\]
and $a$ and $b$ are identified; if only the common mode is excited, only $\gamma_+=a+b$ is observed.
\end{corollary}
\begin{proof}
$\mathbf 1\mathbf 1^\top$ has eigenvalue $G$ on $\mathbf 1$ and $0$ on the contrast space, so $B$ has eigenvalue $1$ on $\mathbf 1$ and $-1/(G-1)$ on contrasts, whence $\rho(B)=1$ and the stated eigenvalues of $aI+bB$. The inversion follows by solving the two linear equations.
\end{proof}

Corollary~\ref{cor:modes} explains Proposition~\ref{prop:symmetry} at the level of the linearised dynamics, and it tells an empirical study what to look for: the coupling strength is recoverable only to the extent that between-country contrasts are excited and measurable. Estimating and reporting $\gamma_+$ and $\gamma_-$, together with their posterior dependence, is therefore more informative than reporting a coupling coefficient.

\subsection{The emission scale in observed coordinates}\label{subsec:emisobs}
The quantity a corpus delivers is not prevalence but an expected relative intensity. Writing $f=\phi\pi$ for a constant common emission scale $\phi$ and multiplying \eqref{eq:commdyn} through by $\phi$ gives
\begin{equation}\label{eq:obscoord}
 \dot f \;=\; a f+bBf-\frac{\beta}{\phi}\,f\odot f-\frac{b}{\phi}\,f\odot(Bf).
\end{equation}
Two things follow that matter for the interpretation of any low-frequency corpus fit.

First, a small observed frequency does not establish a small speaker prevalence. The linear part of \eqref{eq:obscoord} is free of $\phi$, whereas the quadratic part is divided by it, so an emission scale a hundred times smaller corresponds to a prevalence a hundred times larger at the same observed intensity. Statements to the effect that real neologisms necessarily occupy the linear regime are therefore conditional on the chosen emission scale and on the sampling assumption of Remark~\ref{rem:sampling}, and must be reported as such.

Second, the exact nonlinear model does not inherit the observation-layer rescaling of Proposition~\ref{prop:identify}(ii). If $B$ is known, $\phi$ constant, $\kappa>0$ and the trajectory rich enough to identify all four coefficients in \eqref{eq:obscoord}, let $c_{\mathrm{sq}}$ and $c_{\mathrm{cr}}$ denote the estimated coefficients of $f\odot f$ and of $f\odot(Bf)$, both of which are negative. Then
\[
 \phi=-\frac{b}{c_{\mathrm{cr}}},\qquad
 \beta=\frac{b\,c_{\mathrm{sq}}}{c_{\mathrm{cr}}},\qquad
 \mu=\frac{b\,c_{\mathrm{sq}}}{c_{\mathrm{cr}}}-a,\qquad
 \kappa=\frac{c_{\mathrm{cr}}}{c_{\mathrm{sq}}},
\]
each of the four being positive where it must be. A precise sufficient richness condition is available for an ideal continuously observed $f$: let $D(t)$ be the $G\times4$ matrix whose columns are $f(t)$, $Bf(t)$, $f(t)\odot f(t)$ and $f(t)\odot(Bf(t))$. If $\int_0^T D(t)^\top D(t)\,dt$ is positive definite, equality of two vector fields along $f$ forces their four coefficients to coincide, by integrating the squared norm of $D(t)$ times their coefficient difference. The displayed inverse then gives uniqueness of $(\beta,\mu,\kappa,\phi)$. Discrete noisy counts do not supply this ideal trajectory automatically. So in principle the emission scale and the dynamical rates are jointly recoverable from a sufficiently informative nonlinear aggregate trajectory. This is a conditional structural result and not a claim about the present data: it still does not separate use intensity from detection, the required richness fails in synchronised or single-community regimes, and the curvature may be far too weak to estimate. It does show that the non-identification of Proposition~\ref{prop:family} is a property of the low-information regime rather than an inescapable feature of aggregate counts.

\section{In silico validation: does the inference machinery work?}\label{sec:insilico}
Before any corpus is touched, a methodological framework must show that its estimator recovers known parameters and that its model-selection procedure has both power and specificity. This section is a simulation-based characterisation of one estimator; it does not by itself establish practical identifiability in sparse data, and the empirical adequacy of the model for real Spanish is the separate business of Sections~\ref{sec:protocol}--\ref{sec:diagnosis}.

Table~\ref{tab:simconfig} summarises the reported configuration of each experiment, so that the reported numbers are attached to stated designs. Exact reproduction additionally requires the graph, exposure draws, full seed schedule, optimizer settings and code; these materials were not supplied for this review.

\begin{table}[t]
\centering\scriptsize
\renewcommand{\arraystretch}{1.15}
\begin{tabularx}{\linewidth}{@{}p{1.05cm}p{2.5cm}>{\raggedright\arraybackslash}X@{}}
\toprule
Exp. & Object & Configuration\\
\midrule
E1, E7 & parameter recovery & community graph: ring of $G=8$ with two long-range chords; $T=60$ windows; $C=2$ channels; truth $\beta=0.35$, $\mu=0.08$, $\kappa=0.40$, $\phi=(1.2,0.7)$, $\psi=(6,4)$; seeds $\pi_g(0)=10^{-6}$ except $\pi_1(0)=0.05$, $\pi_2(0)=0.03$; exposure $E\sim\mathrm{Unif}(4000,12000)$ per cell; generator: discrete Euler map of \eqref{eq:commdyn} with step $1$ and state clipped to $[10^{-6},1-10^{-6}]$; estimator: BFGS on the log scale, $2$--$3$ random restarts, $\le800$ iterations; $R=120$ independent replicates; simulation exposure uses arbitrary units, with $\phi$ in matching units; master seeds $20260724$ (E1) and $70724$ (E7); bootstrap of RMSE with $2000$ resamples\\
E2 & spectral threshold, finite process & individual-based \emph{two-state} discrete-time SIS on an Erd\H os--R\'enyi graph, $N=500$, mean degree $8$, $\rho(W)=9.16$; synchronous discrete-time updating with per-step infection probability $1-e^{-\beta\,\Delta t\,n_i}$, where $n_i$ is the number of active neighbours, and recovery probability $1-e^{-\mu\Delta t}$, which approximates rather than exactly simulates the continuous-time chain; $\mu=1$, $\Delta t=0.1$, $300$ steps, giving a horizon of $T=30$ time units; $5$ seed speakers drawn uniformly; $180$ Monte Carlo replicates per grid point; $13$ grid points, $\mathcal R_{\mathrm{lex}}\in[0.4,2.4]$\\
E8 & threshold vs network size & as E2 with $N\in\{300,600\}$ at fixed mean degree $8$, giving $\rho(W)=8.83$ and $9.36$; $150$ replicates per point; Wilson $95\%$ bands\\
E3 & identifiability ridge & as E1 with $C=1$, $\phi=1$, $\psi=6$; log-likelihood evaluated on a $60\times60$ grid, $u,q\in[0.3,3.0]$; $u$ and $q$ are unconstrained positive emission factors, not probabilities\\
E4 & coupling ablation & $G=8$, $T=55$, $C=1$, $\beta=0.33$, $\mu=0.08$, $\kappa\in\{0.40,0\}$; single seed community, $\pi_1(0)=0.06$, others $10^{-6}$; leave-one-community-out summed held-out log density, coupled minus no-coupling; $24$ replicates, all finite\\
E5 & forecast calibration & as E1 with $C=1$; sparse regime $E\sim\mathrm{Unif}(0.5,10)$; a single split per dataset, fitting on windows $1$--$30$ and forecasting presence on windows $31$--$60$ with no refitting; $30$ independent datasets, $240$ forecasts each, $7200$ in total\\
E6 & misspecification & data generated with per-community log-normal random effects of s.d.\ $0.30$ on $\beta$ and $\mu$, multiplicative latent-path noise of s.d.\ $0.15$, and generating dispersion $\psi=2$ against a fitted free $\psi$; homogeneous model fitted; recovery on $40$ replicates, calibration on $30$ datasets, ablation on $20$ replicates\\
\bottomrule
\end{tabularx}
\caption{Reported configurations of the in silico experiments; some implementation details remain to be deposited. The generator in E1 and E3--E7 is the discrete Euler map of the community equation, not the Runge--Kutta integration of \eqref{eq:commdyn} used for the real-data fits; Euler has first-order global error as the step tends to zero under the usual regularity conditions, but no step-refinement check is supplied for the unit step used here. Clipping also changes the dynamics. Recovery in these experiments does not establish accuracy of the continuous-time solver. The process simulated in E2 and E8 is the linear two-state replacement of Definition~\ref{def:replacement}, which is the system to which Theorem~\ref{thm:threshold} applies.}
\label{tab:simconfig}
\end{table}

\subsection{Parameter recovery: small observed bias, uneven precision}\label{subsec:recovery}
\begin{figure}[t]
\centering
\includegraphics[width=\linewidth]{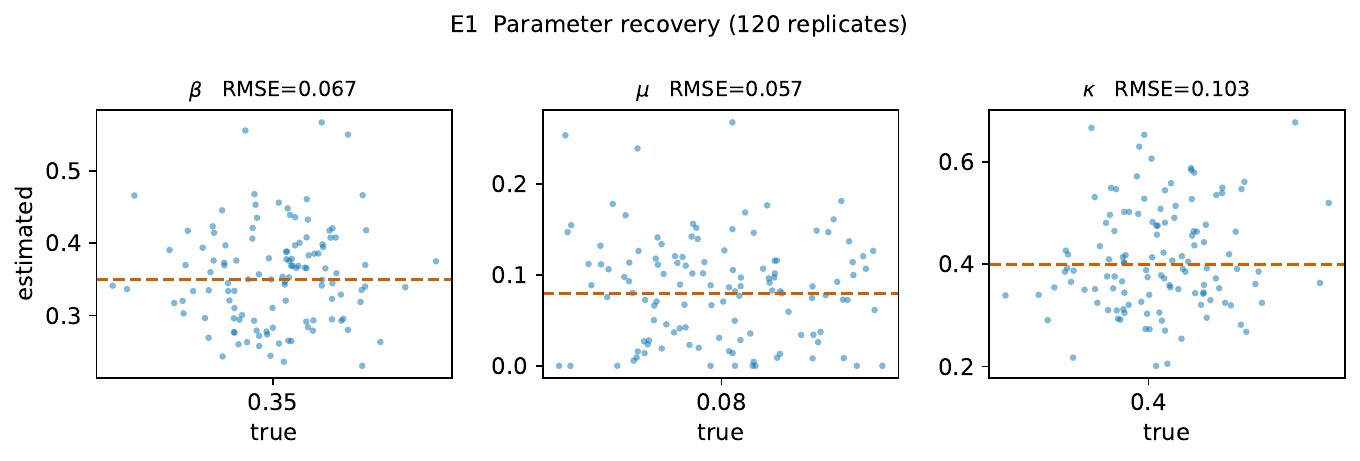}
\caption{Parameter recovery in experiment E1, over $120$ simulated corpora. Points are per-replicate estimates (jittered horizontally); the dashed line is the true value. Transmission $\beta$, abandonment $\mu$ and cross-community coupling $\kappa$ are recovered with small bias. The panel titles give this run's absolute root-mean-square errors, which correspond to relative errors of $19.1\%$, $71.1\%$ and $25.7\%$; the companion run E7 tabulated in Section~\ref{subsec:recovery} gives $17.2\%$, $67.7\%$ and $26.2\%$. The two runs agree on the substantive point, that $\mu$ is recovered without appreciable bias and with very little precision.}
\label{fig:e1}
\end{figure}

\begin{figure}[t]
\centering
\includegraphics[width=0.62\linewidth]{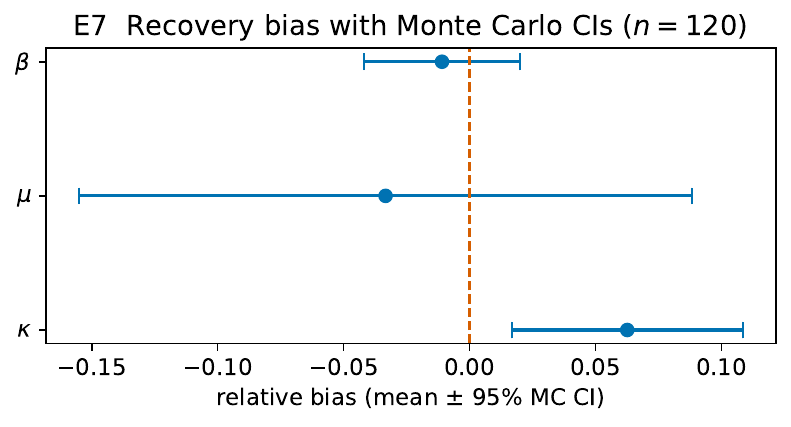}
\caption{Relative bias of the three dynamical parameters with Monte Carlo $95\%$ confidence intervals over the $120$ recovery replicates. Small bias is compatible with poor per-replicate precision, and for $\mu$ that is exactly the situation: the interval straddles zero while the root-mean-square error is about two thirds of the true value (Section~\ref{subsec:recovery}).}
\label{fig:e7bias}
\end{figure}

Two independent recovery experiments were run with the same design and different master seeds, E1 and E7 of Table~\ref{tab:simconfig}; each generated $120$ independently seeded corpora from known parameters and re-estimated them, and in each all $120$ fits converged with no optimiser failures. Table~\ref{tab:recovery} reports E7, which also carries bootstrap intervals; Figure~\ref{fig:e1} displays the per-replicate estimates of E1, whose panel titles give that run's absolute root-mean-square errors and therefore differ slightly from the values tabulated here. Bias is small, but bias is not precision, and the two must be reported together.
\begin{table}[htbp]
\centering\small
\begin{tabular}{@{}lccc@{}}
\toprule
& relative bias & relative RMSE & bootstrap $95\%$ CI for relative RMSE\\
\midrule
$\beta$ & $-1.1\%$ & $17.2\%$ & $[15.4,\,19.1]$\\
$\mu$ & $-3.3\%$ & $67.7\%$ & $[60.7,\,74.6]$\\
$\kappa$ & $+6.3\%$ & $26.2\%$ & $[22.1,\,30.4]$\\
\bottomrule
\end{tabular}
\caption{Parameter recovery in experiment E7, over $120$ independently seeded simulated corpora. Relative bias and relative root-mean-square error are expressed as fractions of the true value; the interval is a bootstrap $95\%$ interval for the relative RMSE over replicates.}
\label{tab:recovery}
\end{table}
The abandonment rate $\mu$ is therefore recovered without appreciable bias but with a root-mean-square error equal to about two thirds of its true value: on any single simulated corpus the estimate of $\mu$ is nearly uninformative. We accordingly describe these experiments as showing small observed average bias, without claiming unbiasedness or reliable single-corpus identification, and we use this phrasing consistently in the abstract and the conclusion. Figure~\ref{fig:e1} shows the per-replicate estimates and Figure~\ref{fig:e7bias} the relative biases with Monte Carlo intervals. Interval calibration is not assessed here, because a Laplace approximation understates uncertainty for the dispersion parameters; it is assessed by replication in Section~\ref{sec:inference}.


\subsection{The spectral index and finite-horizon persistence}
Theorem~\ref{thm:threshold} is a statement about a deterministic approximation to the \emph{linear two-state replacement}. To probe the corresponding finite process we simulate a synchronous discrete-time approximation to that system (the individual-based two-state SIS process of Table~\ref{tab:simconfig}, with mass-action exposure and no dormant compartment) on an Erd\H{o}s--R\'enyi contact network, and sweep $\beta$ at fixed $\mu$. Generator and index therefore correspond, which is what makes the comparison interpretable; nothing here tests the three-state Hill model of Section~\ref{sec:ctmc}.

\begin{figure}[t]
\centering
\includegraphics[width=0.62\linewidth]{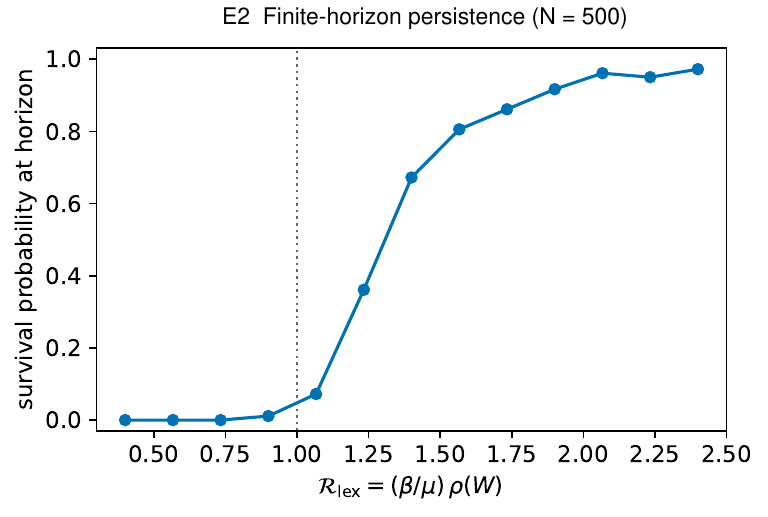}
\caption{Finite-horizon persistence probability of the innovation in the finite two-state individual-based process against $\mathcal R_{\mathrm{lex}}=(\beta/\mu)\rho(W)$. Persistence is essentially zero below the spectral threshold (dotted line at $1$) and rises sharply above it. Horizon, seeding and replicate count are in Table~\ref{tab:simconfig}.}
\label{fig:e2}
\end{figure}

\begin{figure}[t]
\centering
\includegraphics[width=0.62\linewidth]{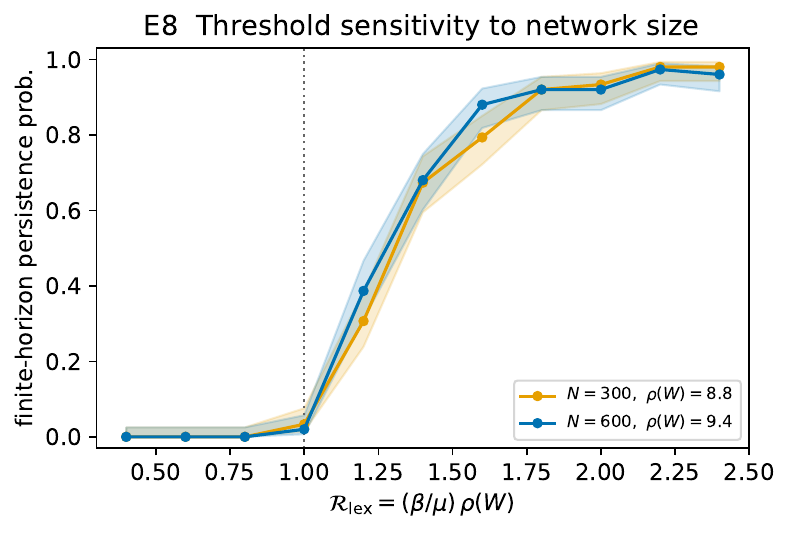}
\caption{Sensitivity of the persistence transition to network size: finite-horizon persistence probability against $\mathcal R_{\mathrm{lex}}$ for $N=300$ and $N=600$ at fixed mean degree $8$, with Wilson $95\%$ bands. Location and shape are essentially unchanged. The comparison holds expected degree fixed and uses one realised graph per size.}
\label{fig:e8}
\end{figure}

The persistence event is the survival of at least one active user at the fixed horizon $T=30$, estimated by Monte Carlo over $180$ independent realisations per grid point. The persistence probability is negligible below $\mathcal R_{\mathrm{lex}}=1$ and rises steeply just above it (Figure~\ref{fig:e2}). The $50\%$ crossing occurs near $\mathcal R_{\mathrm{lex}}\approx1.31$ rather than at $1$. We do not attribute this offset to finite size alone: horizon, seed number and placement, graph topology, and the mean-field closure error all shift it, and the design here varies none of them except network size. What the experiment supports is the weaker and correct statement that the transition is located just above the spectral threshold and is compatible with it. Repeating the sweep at $N=300$ ($\rho(W)=8.83$) and $N=600$ ($\rho(W)=9.36$) at fixed mean degree leaves location and shape essentially unchanged, with Wilson $95\%$ bands (Figure~\ref{fig:e8}), consistent with mean-field threshold theory for epidemic processes on networks \cite{pastorsatorras2015}. Two sizes from one graph family, at fixed expected degree, support a limited sensitivity result, not general robustness to size and topology; the bands reflect Monte Carlo variability at a fixed graph, not variability across independently drawn graphs.


\subsection{Only the product \texorpdfstring{$u\,q$}{u q} is identified from counts}
\begin{figure}[t]
\centering
\includegraphics[width=0.55\linewidth]{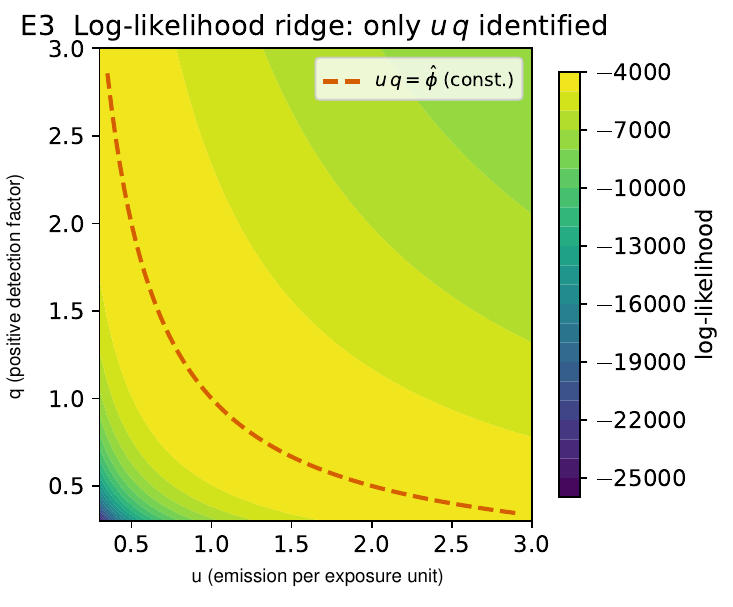}
\caption{Identifiability ridge. Contours of the log-likelihood in the $(u,q)$ plane are level along $u\,q=\hat\phi$ (dashed): only the product is recoverable from counts. The horizontal axis is the emission rate $u$ in the experiment's exposure units and the vertical axis the positive detection factor $q$. In the model $q$ is a probability, and the experiment relaxes it to an unconstrained positive factor purely so that the ridge is visible over a wider range; restricting to $q\le1$ retains an admissible part of the same ridge.}
\label{fig:e3}
\end{figure}

Proposition~\ref{prop:identify} predicts an exactly flat direction, and Figure~\ref{fig:e3} exhibits it: the log-likelihood over the $(u,q)$ plane is constant along the hyperbola $u\,q=\text{const}$. This is a property to be respected in estimation, not a defect to be optimised away.

\subsection{Coupling is favoured mostly, but not only, when present}\label{subsec:ablation}
The decisive test of the coupling term is whether including it improves held-out prediction when the data-generating process truly has diffusion, and, equally, whether it fails to win when the data have none. Using leave-one-community-out prediction, the coupled model beats the no-coupling model in all $24$ usable replicates when coupling is present, with mean gain in summed held-out log predictive density $+2577$ and Monte Carlo standard error $683$ over replicates; the win fraction is $1.00$ with Wilson $95\%$ interval $[0.86,1.00]$. When the same procedure is run on data generated \emph{without} coupling the coupled model does not win on average (mean $-235$, Monte Carlo standard error $50$), but it is still favoured in $3$ of $24$ replicates: a false-selection rate of $12.5\%$ with Wilson $95\%$ interval $[4.3,31.0]$. Specificity is therefore good but far from absolute, and the interval is wide because $24$ replicates is a small experiment.

\begin{figure}[t]
\centering
\includegraphics[width=0.6\linewidth]{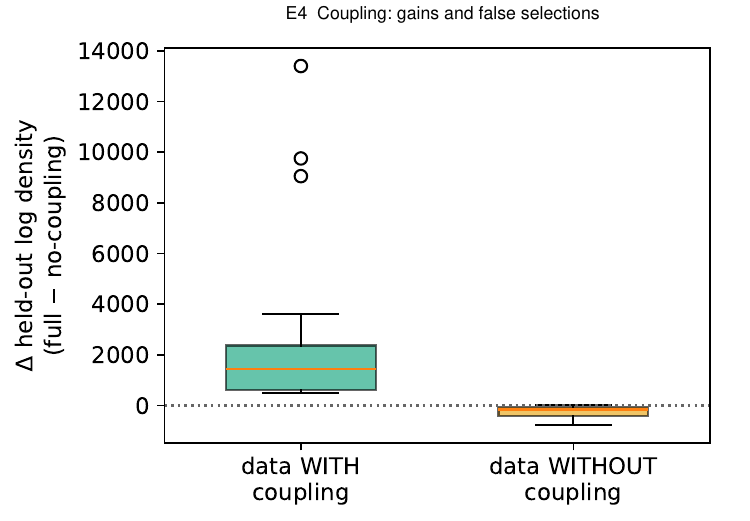}
\caption{Ablation of cross-community coupling. Difference in held-out log predictive density (coupled minus no-coupling) on data generated with coupling (left) and without it (right). The coupled model helps on average only when the generating process contains coupling; it is nevertheless favoured in $3$ of $24$ null replicates, so the null-case win count is not zero (Section~\ref{subsec:ablation}).}
\label{fig:e4}
\end{figure}

Figure~\ref{fig:e4} shows the distribution of the per-replicate score differences in both arms. One design caveat must be recorded, because it works in favour of the coupled model. In a held-out non-source community the imposed seed is $10^{-6}$. Even without coupling it grows logistically when $\beta>\mu$: with the E4 values, the continuous-time logistic solution reaches approximately $0.42$ by time $55$. The discrete generator need not give that exact value, but the absence of coupling does not imply structural silence. Coupling can accelerate early growth from this small seed, so the comparison remains sensitive to the seed convention. The comparison thus measures the value of coupling \emph{given this seeding convention}, and the contribution of the seed convention to the gain has not been isolated. Section~\ref{sec:diagnosis} uses a real-data design in which the compared models use training-country information for the proxy and seed under the reported construction, and the conclusion there is different.

\subsection{Forecast calibration, against a matched baseline}\label{subsec:calib}
\begin{figure}[t]
\centering
\includegraphics[width=\linewidth]{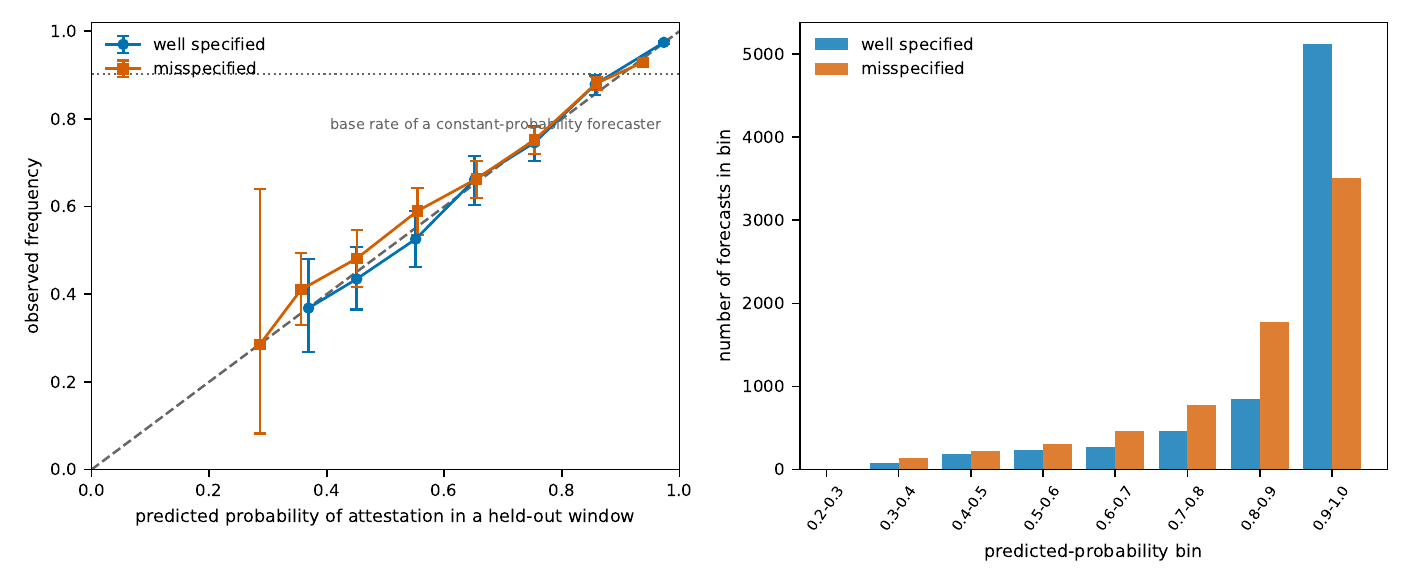}
\caption{Forecast calibration on held-out windows, well specified and misspecified, over $30$ independent datasets each. Left: observed frequency against predicted probability with nominal Wilson $95\%$ intervals that ignore within-dataset dependence. Dataset-level resampling is needed for calibrated uncertainty. The dotted horizontal line shows the well-specified base rate; the misspecified base rate is lower. Right: the number of forecasts in each predicted-probability bin, which is the denominator behind each point on the left.}
\label{fig:calib}
\end{figure}

Fitting on the first thirty windows and forecasting presence over the next thirty, with no refitting, gives a Brier score of $0.067$. The design is a single split repeated over $30$ independently simulated datasets, not a rolling origin within one series, and the $7200$ forecasts are therefore dependent within a dataset through the shared fit and the shared trajectory; the correct resampling unit for uncertainty is the dataset. The supplied figure instead uses ordinary binwise Wilson bars, which do not account for this clustering and must be read as nominal independence-based summaries. A Brier score cannot be read on its own: it mixes calibration with discrimination and depends on how common the event is. The event here (at least one recorded token in a community-year cell) has base rate $0.902$, so a forecaster that always predicts $0.902$ already achieves $0.088$. We therefore report the reliability and resolution summaries associated with the Murphy decomposition \cite{murphy1973} and the skill relative to that matched constant-probability baseline:
\begin{center}\small
\begin{tabular}{@{}lccccc@{}}
\toprule
& base rate & Brier & reliability & resolution & skill score\\
\midrule
well specified (E5) & $0.902$ & $0.067$ & $0.00009$ & $0.0207$ & $0.240$\\
misspecified (E6) & $0.843$ & $0.115$ & $0.00031$ & $0.0175$ & $0.133$\\
\bottomrule
\end{tabular}
\end{center}
For a discrete forecast partition, the Murphy identity is $\mathrm{BS}=\bar y(1-\bar y)+\mathrm{REL}-\mathrm{RES}$. If continuous probabilities are binned, the grouped decomposition need not equal the raw unbinned score; the binning and weighting rules must be supplied with unrounded summaries to reconcile the table exactly. Reliability is the mean squared calibration error and resolution the ability to separate cells that differ in outcome; the skill score is $1$ minus the ratio of the Brier score to that of the constant-probability baseline. Read this way the picture is sharper than a comparison of raw Brier scores. The reported binned point estimates suggest small calibration error under this misspecification: reliability rises from $9\times10^{-5}$ to $3\times10^{-4}$, both negligible against the base variance. The reported point estimates show lower discrimination and forecast value: resolution falls by about $15\%$ and the skill over a constant-probability forecaster falls from $0.24$ to $0.13$, close to a halving. The raw Brier comparison $0.067\to0.115$ overstates the deterioration in calibration and understates the loss of information, because the two experiments have different base rates. Figure~\ref{fig:calib} shows both reliability curves with nominal Wilson intervals and the number of forecasts in each probability bin; the bins are heavily concentrated at high predicted probability, which is itself a limit on what this experiment can establish about the low-probability regime.

\subsection{Recovery under misspecification}\label{subsec:misspec}
We generate data from a process the estimator does not assume: per-community log-normal random effects of standard deviation $0.30$ on both $\beta$ and $\mu$, multiplicative latent-path noise, and generating dispersion $\psi=2$, while dispersion is estimated freely. The latter value is within the fitted negative-binomial family; the heterogeneity and path noise supply the stated misspecification. Fitting the homogeneous model to these data separates cleanly into what is and is not robust. The individual rate parameters are badly biased, as is possible under misspecification; quasi-maximum-likelihood theory does not predict the direction or magnitude of these biases \cite{white1982}: relative bias $+0.45$ (Monte Carlo standard error $0.14$) for $\beta$ and $+1.17$ (standard error $0.44$) for $\mu$, that is, the abandonment rate is overestimated by more than a factor of two on average, while the coupling is nearly unbiased, $+0.03$ (standard error $0.13$). Word-level pooling in Section~\ref{sec:inference} does not by itself correct omitted within-word community effects or latent-path noise. Structural selection also degrades: the coupled model is still favoured in all $20$ replicates when coupling is present (mean $+2606$, standard error $708$) and not favoured on average when it is absent (mean $-168$, standard error $72$), but it wins in $6$ of $20$ null replicates, a false-selection rate of $30\%$ with Wilson $95\%$ interval $[14.5,51.9]$. Six false selections in twenty is substantial, and the interval is wide enough to be compatible with anything from one in seven to one in two. We therefore do not describe structural selection as reliable under misspecification; the defensible statement is that it retains power while losing much of its specificity, at a single misspecification strength that we did not vary.

\section{A protocol for confirmatory validation}\label{sec:protocol}
This section specifies the study that would test the framework. Everything in it is a proposal: no result reported anywhere in this paper follows the protocol below, and Section~\ref{sec:diagnosis} states explicitly which of its rules the analysis that was carried out violates. We keep the two apart deliberately, because the failure to do so is the commonest way that an exploratory corpus analysis comes to be read as a validation.

\subsection{Candidate data sources and sampling frame}
BOBNEO (\textit{Banco de neologismos de los observatorios de neolog\'{\i}a}) \cite{bobneo} is a lexical database developed and maintained by the \textit{Observatori de Neologia} at \textit{Universitat Pompeu Fabra}. It compiles neologisms attested in contemporary usage from a wide range of sources, including newspapers, magazines, audiovisual media, and social media, with data collected systematically since 1989. The database contains records for both Spanish and Catalan and provides linguistic information and metadata such as grammatical category, neologism type, source, date of attestation, and frequency. Neologisms are identified mainly according to a lexicographic exclusion criterion, whereby lexical units absent from the reference dictionaries used by the project are considered candidates for inclusion. BOBNEO also distinguishes different types of neological processes, including semantic neologisms, which makes it a particularly valuable resource for the study and evaluation of methods aimed at detecting emerging word senses.

The neologisms selected for this study were drawn mainly from BOBNEO, with additional items supplied by the authors. They span several morphological and lexical categories, including derivatives formed by prefixation (\textit{micropl\'astico} ``microplastic'') and suffixation (\textit{ochentero} ``eighties-style''; \textit{museificaci\'on} ``museification''), compounds with Greco-Latin roots (\textit{ciberacoso} ``cyberbullying''), lexical compounds (\textit{terraplanista} ``flat-earther''), phrasal compounds (\textit{gemelo digital} ``digital twin''; \textit{techo de cristal} ``glass ceiling''), loanwords (\textit{hater}, \textit{spam}, \textit{esp\'oiler}), acronyms (\textit{covid19} ``COronaVIrus Disease 2019''), and initialisms (\textit{ADSL} ``Asymmetric Digital Subscriber Line''). Semantic neologisms were excluded because the corpora used lack the sense-level information needed to distinguish neological uses from non-neological ones. Certain loanwords exhibit orthographic variation resulting from their adaptation into Spanish (e.g., \textit{spoiler}, \textit{espoiler}, \textit{spolier}, \textit{esp\'oiler}, \textit{sp\'oiler}); the word forms of the different variants were grouped under a single lemma (here, \textit{esp\'oiler}) and their occurrence counts aggregated. The complete list of selected lemmas is provided in Appendix~\ref{app:wordlist}. For each neologism we generated its inflected forms with a large language model and verified them manually. For nouns and adjectives we followed the standard inflectional paradigms; for verbs we generated the full verbal paradigm, noting that neological verbs overwhelmingly follow the regular first-conjugation pattern.

Three items are required for this retrieval to be reproducible and are not yet available: the complete lemma-to-query map, including inflected and orthographic variants and the model and version used to generate them; the manual review rules applied to that map; and a concordance-validation study estimating occurrence-level precision by word and period, with particular attention to first attestations, homographs, proper names, metalinguistic uses and multiword expressions. Appendix~\ref{app:wordlist} lists lemmas only, and cannot by itself reproduce any of the filtered cohorts used below. We report this as an outstanding requirement rather than a completed step, and Section~\ref{subsec:cohortflow} gives the flow counts that a machine-readable cohort table would have to reproduce.

\emph{Two research traditions.} The \emph{lexicographic} tradition relies on stable, reproducible attestations in prestige registers (literary works, the written and broadcast press, official documents, and curated web sources) so that a form can be admitted to a dictionary; BOBNEO and comparable neologism observatories follow this logic \cite{siepmann2015,mcenery2003}. Conventional writing, however, does not capture lexical creation and borrowing in their full complexity \cite{siepmann2015,goody1987,crystal2001}. The \emph{sociolinguistic} tradition instead observes colloquial speech, where much innovation originates and where change in progress appears as a competition among variants \cite{aitchison1991,milroy1992,biber2004}. Change may be led from above, consciously and from prestige groups, or from below, in informal speech; Labov's account of change from above and from below and of the role of local identity is the appropriate source for this distinction \cite{labov2001,labov1994}, and the degree of consolidation of a form depends on how far it diffuses within the community \cite{labov2001}. Social identity theory \cite{tajfel1986} is relevant to why speakers align with a group's usage, but it is not the source of the change-from-above and change-from-below distinction, and we cite it only for the former point. For media-borne innovations the relevant asymmetry is that change from above can enter through broadcast channels directly, which in the model of Section~\ref{sec:ctmc} is the term $M_i$ rather than the peer-exposure term.

\emph{What our evidence can support.} Bottom-up neologisms, which arise in popular oral registers and spread through spoken and social-media networks, are largely invisible to written corpora; top-down neologisms, chiefly technical, political and media-borne terminology, diffuse through the very written and broadcast channels that BOBNEO and written corpora record. Our empirical data are therefore most valid for the top-down, media-borne type. 
CREA \cite{crea} and CORPES XXI \cite{corpes} are linguistic corpora jointly managed by the Royal Spanish Academy and the Association of Academies of the Spanish Language, which represents 24 national language academies. CREA provides a historical baseline for late twentieth-century Spanish, covering 1975--2000 with approximately 160 million words. CORPES XXI is its expanded modern successor, continuously updated, covering 2001 onwards, whose version 1.5 contains over 455 million orthographic forms in 2026 \cite{corpes2026}. The version and extraction date used for the panel must be documented separately. For this study we selected press sources from CREA and CORPES XXI comprising $333{,}885{,}743$ tokens, counting word forms and punctuation marks; the same tokenisation defines both the numerator and the exposure denominator, so the ratio is a relative frequency in a single consistent unit. We then determined the distribution of the selected neologisms by year, country, and news topic to enable geographical and topical stratification. These resources do not directly reveal interpersonal exposure networks. Accordingly, a first validation should target the community-level reduction, with analysis strata defined before outcome inspection by country or region and, when justified, medium or discourse domain. These strata should not be called speech communities unless independent sociolinguistic evidence supports that interpretation.
A further asymmetry is temporal: CREA and CORPES XXI document the spatio-temporal \emph{consolidation} of a form in fine detail, across five decades and the Spanish-speaking countries, but not its \emph{birth} in oral use; the design is strongest for consolidation and weakest for onset. Spontaneous-speech corpora that do carry speaker metadata (sex, age, education, profession, dialect), such as CORLEC \cite{corlec}, C-ORAL-ROM \cite{morenoguirao2006}, MAVIR \cite{mavir} and PRESEEA \cite{preseea}, are too small to observe diffusion at scale, which is why large social-media collections have been used to obtain enough spontaneous data with sociolinguistic attributes \cite{morenosandoval2018}.

In summary, we have not included neologisms originating in spontaneous spoken discourse in this study. Consequently, the empirical design discussed here specifically assesses top-down consolidation via written channels.

\subsection{Cohort construction}
Choose a cutoff date $T_0$ and construct a cohort of forms first documented in the selected database during a fixed recruitment interval $[T_a,T_b]$ before $T_0$. Apply the following rules before fitting:
\begin{enumerate}
 \item retain both successful and unsuccessful candidates, avoiding case selection based on later frequency;
 \item define community, channel, exposure denominator, and operational extinction without using post-$T_0$ outcomes;
 \item audit lexical status: record for each candidate whether it is a new form, a new sense of an existing form, an established multiword expression, or a named referent, and exclude or flag accordingly, since late corpus attestation is not lexical birth. Corpus-driven procedures for detecting emerging forms from frequency trajectories \cite{grieve2017} supply candidates efficiently but do not settle status, so the audit remains a separate manual step;
 \item record the lexicographic exclusion criterion used at the candidate's recruitment date, because later dictionary admission changes its status but not its historical trajectory. The criteria that govern such admission, chiefly frequency and stability of use weighted with formal, semantic, and sociolinguistic dimensions, are themselves an active research question~\cite{bernal2020}, so dictionary admission should be modelled as a downstream outcome and not treated as ground truth for a form's diffusion.
\end{enumerate}
This is a proposed protocol. It is not pre-registered: no dated registration exists for the analyses reported in this paper, and we describe it as a design for future work rather than as a commitment already discharged.

\subsection{Pseudo-prospective validation design}
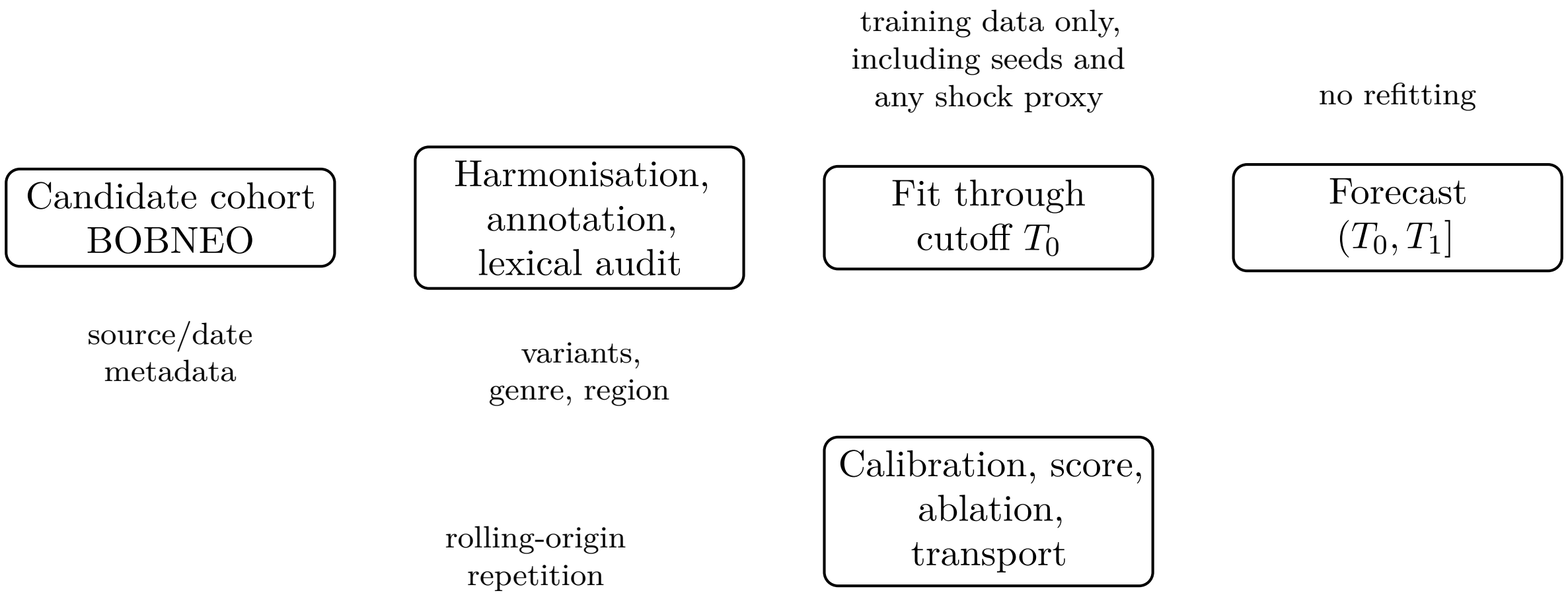
\begin{figure}[t]
\centering
\resizebox{\linewidth}{!}{%
\begin{tikzpicture}[
 >=Latex,
 box/.style={draw,rounded corners,thick,minimum height=10mm,text width=31mm,align=center,font=\small},
 arrow/.style={->,thick},
 note/.style={font=\scriptsize,align=center,text width=31mm}
]
\node[box] (cand) {Candidate cohort\\BOBNEO};
\node[box,right=8mm of cand] (harm) {Harmonisation,\\annotation, lexical audit};
\node[box,right=8mm of harm] (fit) {Fit through\\cutoff $T_0$};
\node[box,right=8mm of fit] (pred) {Forecast\\$(T_0,T_1]$};
\path[arrow] (cand) -- (harm);
\path[arrow] (harm) -- (fit);
\path[arrow] (fit) -- (pred);
\node[box,below=17mm of fit] (score) {Calibration, score,\\ablation, transport};
\path[arrow] (pred) |- (score);
\path[arrow] (score) -| node[note,pos=.22,below] {rolling-origin repetition} (cand);
\node[note,below=4mm of cand] {source/date metadata};
\node[note,below=4mm of harm] {variants, genre, region};
\node[note,above=4mm of fit] {training data only, including seeds and any shock proxy};
\node[note,above=4mm of pred] {no refitting};
\end{tikzpicture}}
\caption{Proposed pseudo-prospective validation workflow. The essential separation is between cohort construction, fitting, and genuinely held-out prediction, and it must apply to every derived quantity, not only to the likelihood.}
\label{fig:validation-workflow}
\end{figure}

Figure~\ref{fig:validation-workflow} sets out the whole design in one diagram. Fit the model only to observations up to $T_0$. Predict, without refitting, community prevalence proxies and token counts over $(T_0,T_1]$. Repeat with rolling origins $T_0^{(1)}<\cdots<T_0^{(K)}$. To test geographical transportability, leave one region out during estimation and predict that region from the remaining communities and measured cross-community exposure. The split must occur by time or community, not by random tokens, because random splitting leaks temporal and speaker information.

The same discipline must extend to every derived quantity, and it is easily lost there. Cohort inclusion, source-community identification, time origin, initial conditions, topic effects, and any global shock proxy must all be constructed from training data only, within each fold. A predictor built from all communities and then used to predict a held-out community is not out of sample, however the model is fitted. Section~\ref{subsec:leak} measures what that violation was worth on the present data.

The primary predictive quantities are: (i) probability of at least one documented use in each future window; (ii) expected token count conditional on corpus exposure; (iii) probability of reaching a new community; and (iv) probability of operational extinction by a fixed horizon. If speaker identifiers are available, active-speaker prevalence and reactivation become primary outcomes; otherwise they remain latent and must not be reported as directly observed.

\subsection{Model comparison and ablation}
Compare the full community-level model with predeclared baselines: constant-rate Poisson or negative binomial, logistic and Bass diffusion curves \cite{bass1969}, a self-exciting count process, homogeneous SIS, and a model without cross-community coupling. Component validity is assessed by ablation:
\begin{enumerate}
 \item remove community coupling to test whether it improves held-out geographical spread;
 \item replace the Hill response by a linear response to test whether reinforcement is supported;
 \item compare a decline-only community model with a model that includes an explicitly defined revival component; note that the community ODE \eqref{eq:commdyn} has abandonment but no dormant compartment, so it cannot ablate speaker-level reactivation directly;
 \item set detection effects to one to quantify the cost of ignoring corpus heterogeneity;
 \item omit cultural covariates or factors to determine whether they add reproducible predictive information.
\end{enumerate}
The full model should be retained only when it improves out-of-sample performance and its posterior predictive checks do not show systematic failure. Use proper scoring rules \cite{gneiting2007}: log predictive density for counts and events, Brier score for future presence and community arrival, and interval coverage for count forecasts. Compare models using paired word-level score differences with uncertainty intervals; token-level tests would overstate precision because observations within a word are dependent.

The in silico ablation of Section~\ref{subsec:ablation} shows that this comparison has power on data of known structure and imperfect specificity, with a null-case false-selection rate of $12.5\%$ $[4.3,31.0]$ rising to $30\%$ $[14.5,51.9]$ under the misspecification studied. A decision threshold must therefore be declared in advance: we suggest requiring the paired word-level mean difference to be positive with a bootstrap $95\%$ interval excluding zero, and reporting the win fraction with its Wilson interval alongside. On the criterion so stated, the analysis reported in Section~\ref{subsec:realbaselines} does not meet it.

\subsection{Criteria for empirical support and falsification}\label{subsec:falsify}
Evidence supporting the method would require: stable calibration across rolling cutoffs; better held-out scores than simpler baselines; improved prediction of new-community arrival when coupling is included; and parameter estimates reproducible across nearby specifications. Failure to outperform a count-only baseline, severe miscalibration, unstable parameter estimates, or no transportability across regions would count against the proposed structure. If only aggregate corpus counts are available, validation is restricted to the observation model and community-level dynamics; it cannot validate individual contagion or causal peer influence. We apply these criteria to our own results in Section~\ref{subsec:realbaselines} and in the conclusion, and they are not met.

\section{The real-data fit as an identification diagnosis}\label{sec:diagnosis}
This section reports the analysis that the available data permit, and uses it to locate the boundary that Sections~\ref{subsec:identobs} and \ref{subsec:flatfamily} predict. We fit the community specialisation \eqref{eq:commdyn}--\eqref{eq:commobs} to a complete national press panel and ask whether cross-country dependence can be distinguished from synchronous adoption. On this design it cannot, and the reasons are structural rather than incidental: the reported nonlinear fits provide little separation of the rates under the fixed emission scale, the positivity of the coupling is a property of its prior, the fitted specification admits only two latent trajectories, and against a constant-rate count baseline the dynamic models lose out of sample on the overwhelming majority of forms. Each of these is quantified below.

\subsection{The panel and the cohort}\label{subsec:cohortflow}
For $466$ candidate forms drawn mainly from BOBNEO and supplemented by the authors, we obtained occurrence counts from the CREA and CORPES XXI press subcorpora on the full grid of $52$ years ($1975$--$2026$), $22$ countries and $6$ thematic domains: $3{,}198{,}624$ cells, with corpus size in tokens per country, year and domain as the exposure denominator and an availability flag equal to one exactly when exposure is positive. Of those cells $1{,}794{,}566$ ($56.1\%$) have positive exposure; the remainder are unavailable and are excluded from the likelihood rather than treated as observed zeros. That figure should not be read as a count of independent sampling units: exposure is a property of a country, year and topic, so the same $3{,}851$ distinct country--year--topic cells recur for each of the $466$ candidate forms, and $3{,}851\times466=1{,}794{,}566$. Total exposure is $333{,}885{,}743$ tokens and the panel contains $143{,}112$ neologism tokens. Coverage of the final years is partial: exposure in $2026$ is $10.7\%$ of the $2010$--$2024$ median and $2025$ about $59\%$ of it. Lower exposure explains why raw counts fall at the end of the record, but it does not by itself make a decline in \emph{relative} frequency documentary, since the offset already absorbs the change in volume. Partial-year coverage may also shift the seasonal and topical composition of what is sampled, which the offset does not absorb. The exact extraction date and results after excluding incomplete final years must be supplied; neither is documented in this version.

The analysis uses the five countries with the largest token totals (Spain, Mexico, Argentina, Colombia and Chile) which carry $69.4\%$ of the neologism tokens and $67.7\%$ of the exposure. The other seventeen countries, from Venezuela ($4.5\%$ of tokens) down to Equatorial Guinea ($0.02\%$), are excluded. This is a coverage-driven, not a linguistic, criterion, and it changes both the population under study and the graph used for geographical hold-out scoring: several excluded countries are exactly the small communities in which confinement or failure would be visible.

Table~\ref{tab:flow} gives the filter cascade in full. The counts below are those reported in the manuscript; the panel and regeneration script remain required for independent verification.

\begin{table}[t]
\centering\footnotesize
\renewcommand{\arraystretch}{1.15}
\begin{tabular}{@{}lcl@{}}
\toprule
Filter stage & forms & role\\
\midrule
Forms queried in the BOBNEO/CREA/CORPES export & $466$ & complete panel\\
with at least one token in the five retained countries & $453$ & country restriction\\
$\ge 2$ tokens in $\ge 3$ retained countries & $248$ & multi-country filter\\
and $\ge 8$ years with a positive count & $225$ & series-length filter\\
and first attestation in $1995$ or later & $144$ & in-window first attestation\\
\addlinespace
relaxed to $\ge 1$ retained country, same year filters & $167$ & selection check\\
no first-attestation filter, $\ge 3$ countries & $225$ & effect of the attestation window\\
\bottomrule
\end{tabular}
\caption{Cohort flow from the complete panel to the fitted cohort, as reported in the source manuscript; independent regeneration remains pending. The $144$-form cohort is used for the hierarchical fit and for the leave-one-country-out comparison of Section~\ref{subsec:realbaselines}.}
\label{tab:flow}
\end{table}

This cohort is not the outcome-independent sample that Section~\ref{sec:protocol} prescribes, and the departures are severe. Requiring attestation in at least three countries conditions on the very cross-country spread that the coupling term is meant to explain, and systematically removes community-confined innovations and failures. Requiring in-window first attestation conditions on a detection event, which depends jointly on prior use and corpus coverage. The retained population is therefore ``forms that reached at least three of the five largest national press markets and were first attested in the selected panel in 1995 or later'', and every estimate below is conditional on that. Section~\ref{subsec:robustfit} reports a partial relaxation of this filter, to forms attested in at least one country; a fully outcome-independent cohort, including forms with no attestation at all, is left to confirmatory work.

The lexical-status audit required by Section~\ref{sec:protocol} has not been carried out on this list. That it is needed is not hypothetical. \emph{Prisi\'on preventiva} sits on the candidate list of Appendix~\ref{app:wordlist} and passes both the multi-country and the series-length filters, failing only the first-attestation cutoff; it is a collocation already present in the 1973 Spanish Penal Code, which establishes that its later appearance in this panel cannot be treated as lexical birth \cite{boe1973}. The reason for any later frequency change is not established by that historical attestation. That a candidate of this kind survives every filter but one, and would enter the cohort under a slightly different recruitment window, shows that the audit must be applied to the whole list rather than to the conspicuous example. Until it is, the cohort should be understood as candidate forms rather than as verified neologisms.

\subsection{Hierarchical specification}\label{subsec:hierspec}
The fit uses \eqref{eq:commdyn}--\eqref{eq:commobs} in a hierarchical Bayesian form in which the cohort of $W$ forms shares population-level rates while retaining word-specific deviations. For word $w$ the dynamical parameters are drawn on the log scale in a non-centred parameterisation,
\begin{align}
 \log\beta_0&\sim\mathcal N(\log 0.4,\,0.6^2), & \log\mu_0&\sim\mathcal N(\log 0.12,\,0.35^2), & \log\kappa_0&\sim\mathcal N(\log 0.5,\,0.8^2),\label{eq:hyper}\\
 \sigma_\beta&\sim\mathrm{HalfNormal}(0.5), & \sigma_\mu&\sim\mathrm{HalfNormal}(0.5), & \sigma_\kappa&\sim\mathrm{HalfNormal}(0.6),\nonumber
\end{align}
with independent standard-normal word effects and $\beta_w=\exp(\log\beta_0+\sigma_\beta z^\beta_w)$ and similarly for $\mu_w,\kappa_w$. The shared negative-binomial dispersion has prior \[\psi\sim\mathrm{LogNormal}(\log 4,\,0.5^2).\] By Proposition~\ref{prop:identify} the observation layer alone identifies only the product of prevalence, use intensity and detectability, so the emission scale is fixed at $\phi\equiv1$: the observation mean is $m_{wgk}=E_{gk}\,\pi_{wg}(t_k)$, exposure is measured in thousands of tokens (so $\phi=1$ means one expected recorded occurrence per thousand tokens contributed by active users), and cells with zero exposure do not enter the likelihood. This fixing is an identification convention, not an estimate, and every rate below is conditional on it. Each form is observed on its own time origin: the series is aligned to the form's first attested year and truncated at $18$ annual observations, which is the window every statement below refers to. The community ODE is integrated between annual observation times by a fixed-step fourth-order Runge--Kutta scheme with two substeps per year, and posterior sampling uses the No-U-Turn sampler \cite{hoffman2014,phan2019} with four chains, $700$ warmup and $700$ retained draws each, and maximum tree depth $8$. The coupling graph $B$ is the complete undirected graph on the five retained countries, normalised to $\rho(B)=1$; with five countries it is a generic any-to-any proxy, not a migration, proximity or communication network, so what it can support is cross-country dependence rather than a particular pathway. For the dynamic fit the six domains are aggregated into a single observation channel, which is an assumption of homogeneous detection across domains that we have not tested.

\subsection{Initial conditions, and what the seeding convention costs}\label{subsec:seeding}
Because $\pi\equiv0$ is invariant for \eqref{eq:commdyn}, a fit that must represent in-window first attestation cannot start every community at zero. We seed each community at $\pi_g(0)=10^{-4}$ and set the source community (the one with the largest count in the first three observed years) to the empirical relative frequency of those years, clipped to $[10^{-3},0.3]$. Since exposure enters in thousands of tokens, these seeds correspond to expected relative frequencies of $0.1$ and $1.0$ tokens per million respectively, with the ceiling at $300$ per million.

That convention is more consequential than it looks, and we quantify it rather than assert that it is harmless. Across the $144$-form cohort the observed first-three-year relative frequency of the source community has median $0.66$ per million, with $5$th and $95$th percentiles $0.10$ and $9.65$ per million. The lower clip therefore binds for $82$ of the $144$ forms, or $56.9\%$: for a majority of the cohort the source community is started at a prevalence higher than the data indicate, in some cases by an order of magnitude. Rare forms are thus given a head start, and the fitted rates absorb the difference. Plausible initial states should instead be estimated with regularisation, or given a prior whose uncertainty propagates into the rates; we did not do so, and this is one of the corrections a confirmatory fit requires.

The seeding also uses information that a genuine hold-out must not: the source community and the time origin are identified from the complete panel, including whichever country is later held out. Section~\ref{subsec:realbaselines} reports a training-only proxy and seed construction, but a complete fold-level audit must also verify time-origin selection and the scoring window.

\subsection{Identifiability at real frequencies}\label{subsec:identreal}
On this panel the fitted latent path of a reference trajectory peaks at about $1.2\times10^{-3}$ under the emission convention $\phi\equiv1$, which by \eqref{eq:obscoord} corresponds to roughly $1.2$ tokens per million. That figure describes one reference trajectory and is not a bound on every fitted word; observed peaks for individual forms are larger, and the two quantities are reconciled by the noise in the counts and by word-specific rates. At prevalences of this order the quadratic term in \eqref{eq:commdyn} contributes little, and the system is close to its linearisation
\begin{equation}\label{eq:linreal}
 \dot\pi\;\approx\;(\beta-\mu)\,\pi+\beta\kappa\,B\pi ,
\end{equation}
so that Proposition~\ref{prop:family} applies: the three rates are exactly non-identified in the linear limit, and only the combinations $a=\beta-\mu$ and $b=\beta\kappa$ can be informed. Whether the residual curvature is enough to separate them is what the fit must answer, and here it is not. Taking $a=-0.152$ and $b=0.240$ (rounded from the displayed marginal-mean reference triple), and integrating \eqref{eq:commdyn} on the fitted configuration over the window of up to $18$ annual observations in Section~\ref{subsec:hierspec}, the triples
\[
 (\beta,\mu,\kappa)=(0.184,\,0.336,\,1.304)\quad\text{and}\quad(0.300,\,0.452,\,0.800)
\]
produce latent paths agreeing to a maximum relative discrepancy of $1.4\times10^{-3}$, that is $0.14\%$, at every community and every time point; comparing the same reference path against every member of the family with $\beta$ from $0.25$ to $1.00$ at fixed $(a,b)$, the worst discrepancy is $0.97\%$. By contrast, moving $\beta$ alone from $0.184$ to $0.300$, holding $\mu$ and $\kappa$ fixed, changes the path by a factor of about $76$, and moving it to $1.00$ by a factor of about $1970$. The stated percentages and factors are retained as reported: independent reproduction requires the precise initial vector, observation grid, integration horizon and unrounded parameters. For an exactly common $b=0.240$, the first triple must use $\kappa=0.240/0.184\simeq1.30435$; the printed $1.304$ gives $b=0.239936$. A single reference trajectory illustrates the flat direction but cannot certify weak curvature for every word in the cohort.

\begin{figure}[t]
\centering
\includegraphics[width=\linewidth]{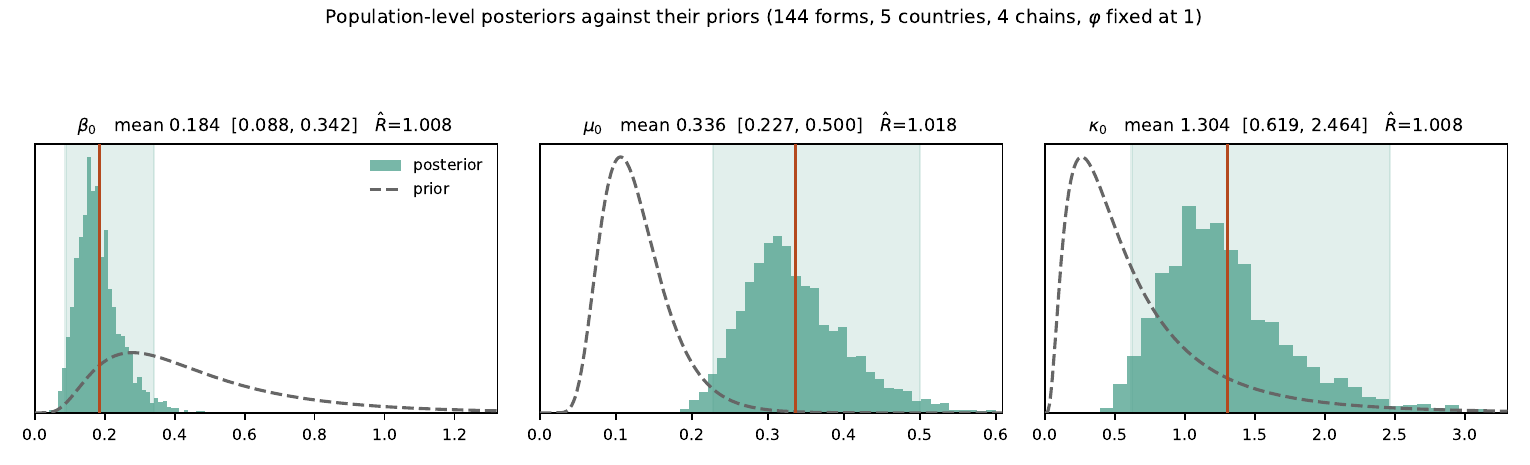}
\caption{Population-level posteriors of run \mbox{R-A}, the run whose values are quoted in the text, shown against their priors ($144$ forms with in-window first attestation, five countries, four chains, $\rho(B)=1$, emission scale fixed at $\phi=1$). Shading marks the posterior $95\%$ interval and the vertical line the posterior mean. The prior for $\kappa_0$ is log-normal and therefore assigns probability one to $\kappa_0>0$ before any data are seen, so the posterior interval excluding zero is guaranteed by the parameterisation; the visible contraction of $\kappa_0$ relative to its prior is $56\%$ on the log scale. The reported maximum $\hat R=1.018$ does not meet the stated $1.01$ criterion, and in any case these are sampling diagnostics, not identification: at these frequencies the leading linear information concerns $\beta-\mu$ and $\beta\kappa$ (Section~\ref{subsec:identreal}), and all values are conditional on the fixed emission scale.}
\label{fig:bobneo-hier}
\end{figure}

The four-chain fit gives posterior means $\beta_0=0.184$ $[0.088,0.342]$, $\mu_0=0.336$ $[0.227,0.500]$ and $\kappa_0=1.304$ $[0.619,2.464]$ (Figure~\ref{fig:bobneo-hier}), with zero divergences and split $\hat R$ at most $1.018$ and effective sample sizes between $232$ and $276$ for the three population rates. These summaries remain provisional: the recommended working threshold is $\hat R<1.01$ \cite{vehtari2021}, which the run does not meet, and an absence of divergences is not by itself evidence of adequate exploration. These are in any case sampling diagnostics. They assess exploration of the specified posterior without certifying that it is adequate; they say nothing about whether the rates are identified, and Proposition~\ref{prop:family} establishes exact non-identification only for the linear approximation, with practical weakness in the nonlinear model requiring its own assessment.

Because $a$ and $b$ are the natural combinations to inspect in the linear approximation, we report them directly rather than inviting the reader to combine the marginal rates. Transforming the posterior draws gives
\[
 a=\beta_0-\mu_0=-0.152\;[-0.204,-0.117],\qquad
 b=\beta_0\kappa_0=\phantom{-}0.214\;[0.178,0.269],
\]
with posterior standard deviations $0.021$ and $0.022$ and a posterior correlation of $-0.80$, so the two are far from independent and should not be quoted separately without that dependence. The posterior probability that $a<0$ is $1.000$: within this specification the uncoupled internal component is decaying. The common-mode rate is instead $\gamma_+=a+b$, whose posterior mean is $0.062$ from these summaries; the contrast-rate mean is $\gamma_-=a-b/4=-0.2055$. Thus negative $a$ does not imply decay of the common mode. Credible intervals and sign probabilities for these modes must be computed from the joint posterior draws; they cannot be inferred from the marginal intervals quoted here. These are transformations of population location parameters on the log scale, not growth rates of an average cohort trajectory: each word has its own $(a_w,b_w)$, and averaging nonlinear trajectories does not give the trajectory at averaged rates. Note also that $b$ must be obtained by transforming the draws and not by multiplying the marginal means, which would give $0.240$ instead of $0.214$; the mean of a product is not the product of the means, and the discrepancy here is about twelve per cent. The triple $(0.184,0.336,1.304)$ used above is a reference triple assembled from the marginal means for the purpose of the trajectory comparison, not a posterior summary.

The coupling deserves separate comment because it is the parameter of interest. It is parameterised as $\kappa_0=\exp(\log\kappa_0)$ with $\log\kappa_0\sim\mathcal N(\log0.5,0.8^2)$, so $\kappa_0>0$ holds \emph{a priori} with probability one and the prior $95\%$ interval is already $[0.10,2.40]$. A posterior interval excluding zero is therefore guaranteed by the parameterisation and is not evidence that coupling is present. Nor is the posterior sharply informative about its magnitude. On the log scale the prior standard deviation is $0.80$ and the posterior standard deviation is $0.350$, a contraction of $56\%$; equivalently the $95\%$ interval narrows from width $3.14$ to $1.37$, a factor of $2.3$. The data therefore move the location of $\kappa_0$ upward and shrink it by rather less than an order of magnitude, which is what \eqref{eq:linreal} predicts when only the product $b=\beta\kappa$ is constrained and $\beta$ is itself weakly determined. Two cautions belong together here. A continuous prior that excludes zero cannot deliver evidence about the exact null $\kappa=0$, so nothing in this posterior speaks to the presence or absence of coupling; a study whose target is presence versus absence needs an explicit null model or a point mass at zero, with a sensitivity analysis for the continuous part. Conversely, the prior does not by itself guarantee that the posterior magnitude will be stable across specifications, and the stability observed in Section~\ref{subsec:robustfit} is a weak empirical fact rather than an artefact of the prior alone. We accordingly report $a$ and $b$ as the empirically meaningful quantities, recommend further sensitivity analysis through priors placed on those combinations, and do not read $\beta_0$, $\mu_0$, $\kappa_0$ as identified transmission, abandonment and coupling rates.

Finally, the cohort is restricted to in-window first attestation. This is a restriction of the fitting pipeline rather than of the differential equation, which accepts any nonzero initial condition and can represent a form already established at the start of the record; what the pipeline lacks is a principled way to set that initial state for such forms without using the outcome. The restriction is therefore avoidable in a future design, and until it is removed the fitted population remains defined partly by the model's implementation rather than by a linguistic question.

\subsection{A leave-one-country-out leakage diagnostic}\label{subsec:leak}
To separate cross-country coupling from a common exogenous event we compare, by leave-one-country-out out-of-sample negative-binomial log score, the family $M_0$ (no coupling), $M_1$ (coupling), $M_2$ (a global word--year shock but no coupling), $M_3$ (coupling and shock) and $M_4$ (directed, lagged coupling), under
\begin{equation}\label{eq:shock}
 \dot\pi_{wg}(t)=\beta_w(1-\pi_{wg})\Big(\pi_{wg}+\kappa_w\!\sum_h B_{gh}\,\pi_{wh}(t-\tau_{gh})+\lambda_w S_w(t)\Big)-\mu_w\pi_{wg},
\end{equation}
Here $\lambda_w\ge0$ is the nonnegative shock-response coefficient, $S_w(t)\ge0$ is a locally bounded proxy for worldwide prevalence, $B\ge0$, and $\tau_{gh}\ge0$. A delay specification requires an admissible continuous prevalence history on $[-\max\tau_{gh},0]$; observed counts are not themselves such a history without an emission convention and interpolation rule. For $M_4$ the lag is fixed at one year and \eqref{eq:shock} is integrated as a delay system with an initial history on $[-1,0]$; the original implementation describes this as observed history, but its construction and availability for a held-out country require documentation before M4 can be certified as out of sample; for $M_0$--$M_3$ the delay is zero and \eqref{eq:shock} is an ordinary differential equation.

Two details determine whether held-out counts enter the proxy or the initial state. The first is the shock proxy: $S_w(t)$ summarises worldwide prevalence, so computing it from all five countries places the held-out country inside the very predictor used to predict it. The second is the initial condition: taking the held-out country's seed from its own first three years hands the model information that a hold-out is supposed to withhold. To measure what these choices are worth we run the comparison under two regimes on identical data. In the \emph{pooled-proxy} regime $S_w$ is computed once from all five countries and each country is seeded from its own record, which is the construction a leave-one-country-out analysis falls into most naturally. In the \emph{training-only} regime, both the proxy and the seeds are constructed inside the fold: $S_w$ is rebuilt inside each fold from the four training countries only, and the held-out country's initial prevalence is the mean of the training countries' seeds, so that the proxy and seed construction excludes held-out counts. The cohort itself was nevertheless selected using all countries; its inclusion and time-window rules must also be audited fold by fold. The reported comparison therefore remains conditional on retrospective cohort selection. The count baselines never see the held-out country under either regime.

\begin{figure}[t]
\centering
\includegraphics[width=0.66\linewidth]{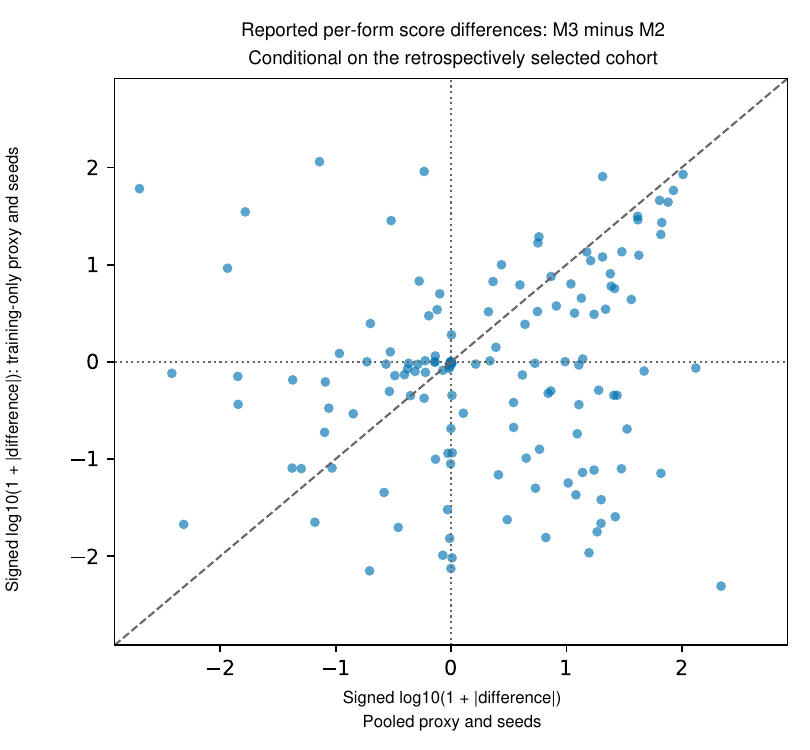}
\caption{Leakage diagnostic. Held-out score difference $M_3-M_2$ for each of the $144$ forms, with the global shock proxy built from all countries (horizontal axis, the pooled-proxy regime) and from the training countries only (vertical axis, the training-only regime, excluding held-out counts from the proxy and seed); the signed logarithmic scale $\mathrm{sgn}(\Delta)\log_{10}(1+|\Delta|)$ is used because the differences span several orders of magnitude. Points off the diagonal are forms whose comparison was affected; points in the upper-left and lower-right quadrants, $60$ of $144$, are forms whose comparison changes sign.}
\label{fig:leak}
\end{figure}

The contrast is large enough to change what the comparison says. Under the pooled-proxy regime $M_3$ beats $M_2$ on $80$ of $144$ forms ($55.6\%$), with paired word-level mean difference $+2.0$ and bootstrap $95\%$ interval $[-8.6,+10.8]$. Under the training-only regime it beats $M_2$ on $56$ of $144$ ($38.9\%$), with mean difference $-4.3$ $[-10.4,+1.3]$. The sign of the per-form difference changes for $60$ of the $144$ forms. Neither regime supports a coupling effect beyond a synchronous shock, but the training-only regime removes two direct sources of contamination; the pooled-proxy regime merely makes the comparison look favourable. Figure~\ref{fig:leak} shows the per-form differences under the two regimes. One qualification belongs with the figure. The two regimes differ in \emph{two} respects at once, the shock proxy and the initial state, so the $60$ sign reversals cannot be attributed to the proxy alone; separating them requires a factorial diagnostic over the four combinations of pooled and training-only proxy with pooled and training-only seeds, which we have not run. What the comparison establishes is the combined effect of removing both sources of information leakage. We report it as a leakage diagnostic and draw no diffusion conclusion from it, and we deliberately do not present these differences as a ranked chart of per-word winners, because a prominent visual conclusion is not repaired by a caveat in its caption.

\subsection{Comparison with simpler baselines}\label{subsec:realbaselines}
The prediction target should be stated before the scores are read, because it is not the one a diffusion study would most want. Each fold withholds an entire country and scores the held-out country's counts over the whole observation period, so the exercise is geographical reconstruction from the remaining four countries, not temporal forecasting of a future the model has not seen. A model earns its score by placing a country correctly given contemporaneous information elsewhere. Temporal forecasting is a separate and stricter test, set out in Section~\ref{subsec:falsify} and not carried out here.

On that target, the same training-only folds allow the comparison that Section~\ref{subsec:falsify} makes decisive: whether the dynamic models out-predict simple count models. They do not. Table~\ref{tab:loco} reports, for the whole $144$-form cohort with five folds per form, the number of forms won, its Wilson $95\%$ interval, and the paired word-level mean difference in held-out log score with a bootstrap $95\%$ interval.

\begin{table}[t]
\centering\footnotesize
\renewcommand{\arraystretch}{1.15}
\begin{tabular}{@{}lcccc@{}}
\toprule
Comparison & forms won & Wilson $95\%$ & mean paired $\Delta$ & bootstrap $95\%$\\
\midrule
$M_1$ (coupling) vs $M_0$ (no coupling) & $28/144$ & $[0.14,0.27]$ & $-0.9$ & $[-2.5,\,+0.5]$\\
$M_3$ (coupling $+$ shock) vs $M_0$ & $41/144$ & $[0.22,0.36]$ & $-25.6$ & $[-34.6,\,-17.1]$\\
$M_3$ vs $M_2$ (shock only) & $56/144$ & $[0.31,0.47]$ & $-4.3$ & $[-10.4,\,+1.3]$\\
$M_4$ (directed, lagged) vs $M_0$ & $45/144$ & $[0.24,0.39]$ & $+0.1$ & $[-1.3,\,+1.4]$\\
\addlinespace
$M_0$ vs constant-rate count baseline & $23/144$ & $[0.11,0.23]$ & $-21.1$ & $[-25.9,\,-16.9]$\\
$M_1$ vs constant-rate count baseline & $21/144$ & $[0.10,0.21]$ & $-22.0$ & $[-26.7,\,-17.9]$\\
$M_3$ vs constant-rate count baseline & $9/144$ & $[0.03,0.11]$ & $-46.7$ & $[-55.8,\,-38.3]$\\
$M_1$ vs log-linear count trend & $21/141$ & $[0.10,0.22]$ & $-16.4$ & $[-20.6,\,-12.6]$\\
$M_3$ vs log-linear count trend & $18/141$ & $[0.08,0.19]$ & $-41.5$ & $[-50.7,\,-32.7]$\\
\addlinespace
log-linear count trend vs constant rate & $65/141$ & $[0.38,0.54]$ & $-5.2$ & $[-8.7,\,-2.0]$\\
\bottomrule
\end{tabular}
\caption{Leakage-free leave-one-country-out comparison on the complete $144$-form in-window cohort, five folds per form. The reported score is the summed held-out negative-binomial log density over the held-out country's series; $\Delta$ is the paired word-level difference, first model minus second, and the bootstrap interval is over forms. Three forms give a non-finite log-trend score and are dropped from the three comparisons that involve that model, whose row-specific denominator is therefore $141$ rather than $144$; non-finite scores from an ordinary count trend are a numerical diagnostic in their own right and are recorded with the results.}
\label{tab:loco}
\end{table}

The numerical comparison remains conditional on the reported implementation. A reproducible specification must supply the fitted mean and dispersion for each baseline, the fitting objective or priors for M0--M4, the definition of $S_w$, the directed graph and history for M4, and whether scores integrate parameter uncertainty or use plug-in estimates. These choices cannot be reconstructed from the aggregate score table alone. Three findings follow from the reported summaries.

First, the comparison does not demonstrate a predictive benefit from cross-country coupling. $M_1$ improves on $M_0$ for $28$ of $144$ forms, a win fraction of $0.19$ $[0.14,0.27]$, and the paired mean interval $-0.9$ $[-2.5,+0.5]$ includes zero; this is not an equivalence test and does not establish an exactly null contribution.

\begin{figure}[t]
\centering
\includegraphics[width=\linewidth]{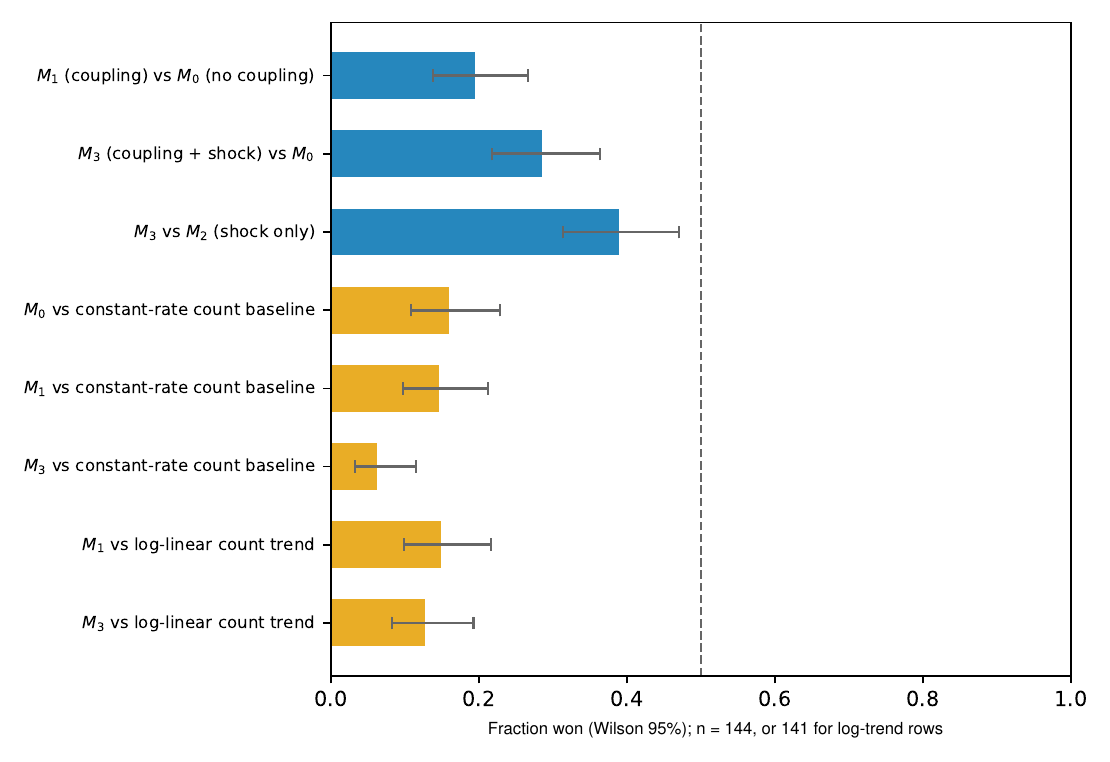}
\caption{Leakage-free leave-one-country-out comparison on the complete $144$-form cohort: fraction of forms for which the first model wins out of sample, with Wilson $95\%$ intervals. The dashed line marks parity. Within the dynamic family (upper bars) coupling has no demonstrated predictive advantage; against the count baselines (lower bars) the dynamic models lose on the large majority of forms.}
\label{fig:realbase}
\end{figure}

Second, the dynamic models are beaten decisively by a constant relative rate with an exposure offset. That baseline wins on $121$ of $144$ forms against $M_0$, $123$ of $144$ against $M_1$ and $135$ of $144$ against $M_3$, and every paired mean difference is negative with a bootstrap interval excluding zero; Figure~\ref{fig:realbase} plots the win fractions of Table~\ref{tab:loco} with their Wilson intervals. Predicting an entirely held-out country from a training-country seed through the ODE is a demanding task, and an exposure-offset constant rate is hard to beat when relative frequencies are this low and this noisy. Adding a shock term makes matters worse, not better, once the shock is built without the held-out country.

Third, and as a control, the log-linear trend does not beat the constant rate either ($65$ of $141$ forms, mean $-5.2$ $[-8.7,-2.0]$). The ranking is therefore not an artefact of the dynamic models being complicated: on this panel the constant-rate model beats M0, M1 and M3 in the displayed comparisons and also beats the log-linear trend on its evaluable subset, which is what one expects when the informative content of the series is its overall level rather than its shape.

Against the falsification criteria of Section~\ref{subsec:falsify} (better held-out scores than simpler baselines, improved prediction when coupling is included) the analysis fails on both counts. We report this as the result.

\subsection{Robustness of the real-data fit}\label{subsec:robustfit}
Two things about the fit invite a check. Because the observation layer alone identifies only the product of prevalence, use intensity and detectability, and the nonlinear fit has weak curvature under the reported convention, the rates can be sensitive to the fixed scale $\phi$; and because the cohort is defined by filters that condition on outcome, the estimates should be recomputed on a relaxed cohort. Table~\ref{tab:robust} reports five configurations, each from a complete recorded run executed on the same machine at production scale, identified by run label so that every estimate is traceable to one execution.

\begin{table}[t]
\centering\footnotesize
\renewcommand{\arraystretch}{1.15}
\begin{tabular}{@{}llccccc@{}}
\toprule
Run & Configuration & $W$ & $\beta_0$ $[95\%]$ & $\mu_0$ $[95\%]$ & $\kappa_0$ $[95\%]$ & $\hat R_{\max}$\\
\midrule
R-A & $\phi=1$ & $144$ & $0.18\,[0.09,0.34]$ & $0.34\,[0.23,0.50]$ & $1.30\,[0.62,2.46]$ & $1.018$\\
R-B & $\phi=2$ & $144$ & $0.16\,[0.08,0.28]$ & $0.32\,[0.23,0.46]$ & $1.19\,[0.59,2.24]$ & $1.008$\\
R-C & $\phi=0.5$ & $144$ & \multicolumn{3}{c}{\emph{did not converge}; $32$ divergences} & $9.19$\\
R-D & $+$ CREA/CORPES channel & $144$ & $0.19\,[0.09,0.33]$ & $0.34\,[0.23,0.49]$ & $1.25\,[0.65,2.35]$ & $1.016$\\
R-E & no $\ge3$-country filter & $167$ & $0.19\,[0.10,0.36]$ & $0.35\,[0.24,0.51]$ & $1.19\,[0.57,2.19]$ & $1.024$\\
\bottomrule
\end{tabular}
\caption{Robustness of the real-data hierarchical fit. All five runs use five countries, four chains, $700+700$ draws, tree depth $8$ and $\rho(B)=1$; \mbox{R-A}--\mbox{R-D} use the $144$-form in-window cohort and \mbox{R-E} the $167$-form cohort without the multi-country filter. All were executed on the same machine and the complete configuration and seed of each is archived under its run label. The pipeline is deterministic given its seed, and we verified this rather than assuming it. Re-executing the $\phi=2$ and $\phi=0.5$ configurations reproduced the reported summaries of their archived runs, in the sense that every reported posterior mean, credible interval, $\hat R$, effective sample size and divergence count was identical, the reproduction extending to the non-convergence of \mbox{R-C}. Re-executing the $\phi=1$ configuration on a different architecture, by contrast, moved the posterior means by at most $1.6\%$ ($\beta_0$ from $0.184$ to $0.181$, a relative change of $1.6\%$; $\mu_0$ from $0.336$ to $0.331$, $1.5\%$; $\kappa_0$ from $1.304$ to $1.312$, $0.6\%$), each well inside the corresponding credible interval. A fixed seed does not guarantee cross-platform equality; the source of these differences has not been isolated by a controlled environment comparison.}
\label{tab:robust}
\end{table}

\emph{Emission scale (\mbox{R-A}--\mbox{R-C}).} The level rates move with $\phi$, confirming that their absolute values are a consequence of the identification convention rather than of the data. The fit fails at $\phi=0.5$, with $\hat R\approx9$ and $32$ divergences; we record this as a computational failure and not as evidence that the scale is inadmissible, since at the observed frequencies $\phi=0.5$ is not implausible and the failure may equally reflect the parameterisation, the seed clipping of Section~\ref{subsec:seeding}, the prior geometry, or integration error, none of which we have separated. What does \emph{not} follow is any conclusion from the stability of $\kappa_0$ from positivity alone: $\kappa_0>0$ is imposed by the prior, but stability of its magnitude is not. The reported similarities are conditional sensitivity results, further limited by the sampling diagnostics in R-A, R-D and R-E.

\emph{The CREA/CORPES boundary (\mbox{R-D}).} The export does not label whether each count comes from CREA (through $2000$) or CORPES XXI (from $2001$); the two do not overlap in time and differ in composition, so a change in corpus composition could masquerade as a temporal increase. Adding a relative pre-$2001$ emission factor $\phi_{\mathrm{CREA}}$ tests this directly. At production scale its posterior is $0.930\,[0.730,1.168]$ with $\hat R=1.004$ and $n_{\mathrm{eff}}=1122$: the interval contains one, and $\Pr(\phi_{\mathrm{CREA}}<1\mid\text{data})=0.74$, so there is no clear pre-$2001$ detection jump at this cohort and scale. The parameter contracts relative to its stated prior (the standard deviation of $\log\phi_{\mathrm{CREA}}$ falls from $0.5$ under the prior to $0.120$, a contraction of $76\%$) and with the channel in the model $\kappa_0$ is essentially unchanged ($1.25$ against $1.30$), so the reported coupling summary changes little in this particular specification.

An interval containing one is nevertheless weak evidence of a negligible discontinuity, and we quantify how weak by reporting how much pre-$2001$ information the cohort actually contains. Of the $10{,}703$ observed cells in the fitted cohort only $585$ ($5.5\%$) fall before $2001$; those cells carry $238$ of the $24{,}446$ tokens, or $0.97\%$; and $100$ of the $144$ forms contribute no pre-$2001$ observation whatsoever. The factor is therefore estimated from about one per cent of the tokens and from the $44$ earliest-emerging forms, which are not a representative subset of the cohort, and a single pooled factor could still average over opposing country- or topic-specific composition shifts that these data cannot resolve. A per-cell corpus label would improve provenance, but with no temporal overlap it would not by itself separate a corpus change from a calendar-period effect; overlap or external calibration is needed.

\emph{The multi-country filter (\mbox{R-E}).} Relaxing the $\ge3$-country requirement to $\ge1$ enlarges the in-window cohort to $167$ forms, including community-confined ones. At production scale this fit completes with zero divergences but does not meet the stated convergence criterion ($\hat R\le1.024$) and the population estimates barely move: $\beta_0=0.19$, $\mu_0=0.35$ and $\kappa_0=1.19\,[0.57,2.19]$ against $1.30\,[0.62,2.46]$ on the filtered cohort, each well inside the other's interval.

The reported estimates change little under this partial filter relaxation, but overlapping credible intervals do not establish equivalence or rule out selection effects, especially while sampling diagnostics remain unresolved. What it does not establish is that the coupling is identified (positivity remains imposed by the log-normal parameterisation exactly as in Section~\ref{subsec:identreal}) or that the enlarged cohort is outcome-independent, since it still requires eight years of positive counts and in-window first attestation. The selection question of Section~\ref{subsec:cohortflow} is therefore narrowed, not settled.

\subsection{What this section establishes}\label{subsec:diagnosis-summary}
\begin{figure}[t]
\centering
\includegraphics[width=\linewidth]{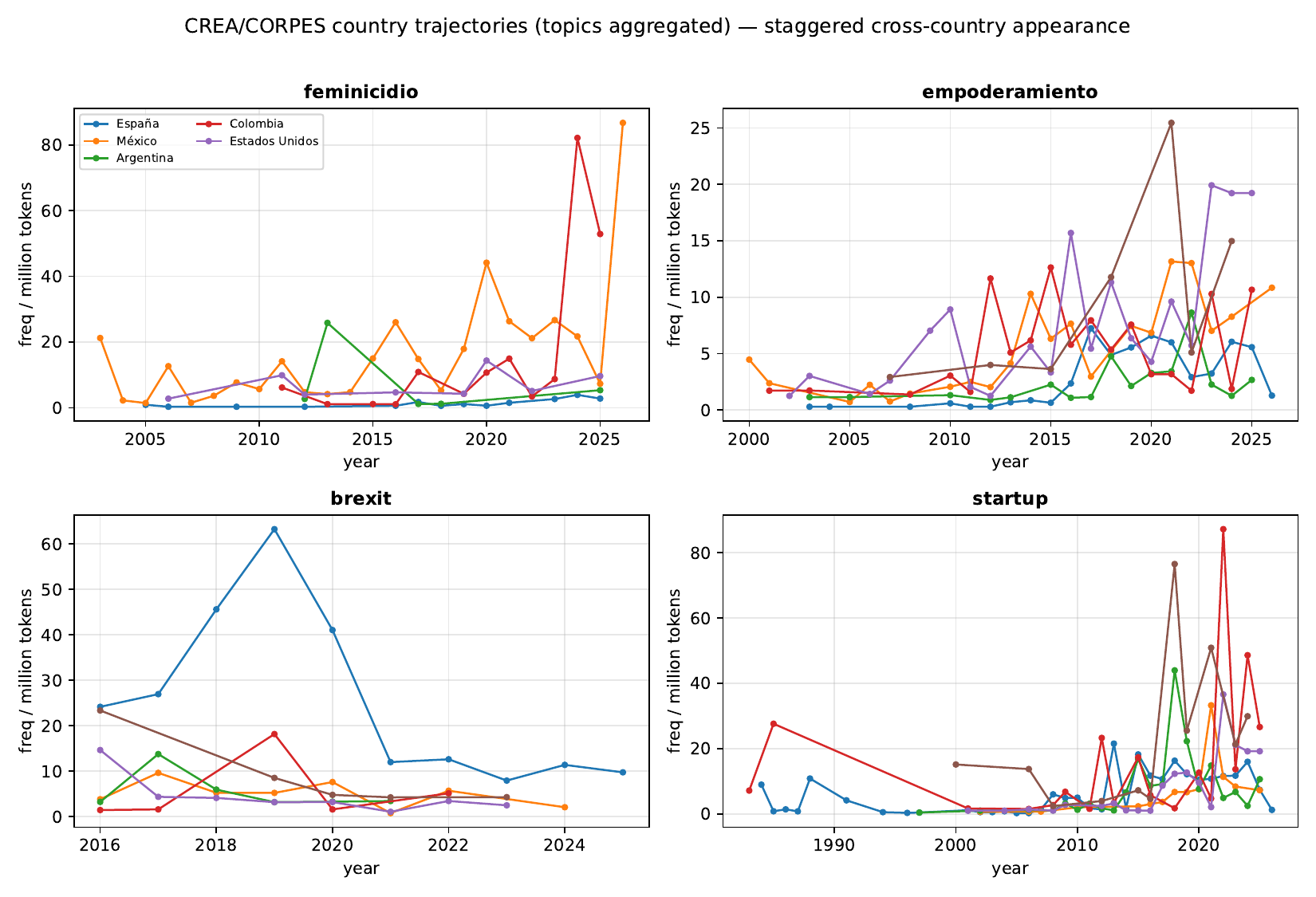}
\caption{Country-level trajectories in CREA/CORPES (frequency per million tokens) for four forms, using observed first attestation rather than apparent visual onset. \emph{Feminicidio} and \emph{empoderamiento} appear earlier in some countries than in others; \emph{brexit} rises and then falls, dominated by Spain. Staggered appearance alone does not distinguish diffusion from a common external shock, and the comparison in Section~\ref{subsec:leak} cannot settle it. The figure is descriptive and its country panel differs from the fitted one: it displays the United States in place of Chile, so one of the five series shown lies outside the five-country specification of Section~\ref{subsec:hierspec}. Years with zero exposure for a country are gaps rather than zeros.}
\label{fig:bobneo-diff}
\end{figure}

Taken together, the reported analyses identify limitations of this particular fit: under the fixed emission scale the leading dynamical information concerns $\beta-\mu$ and $\beta\kappa$, while the nonlinear rate separation is weak; positivity of coupling is imposed by its prior; the symmetric graph with the stated seeds allows at most two latent paths; and the source seed is clipped upward for a majority of forms. Within the retrospectively selected cohort, the reported constant-rate baseline beats M0, M1 and M3 on $84$ to $94\%$ of forms, and adding coupling has no demonstrated score advantage. These empirical statements require the computational and provenance checks specified above. The staggered raw trajectories (Figure~\ref{fig:bobneo-diff}) do not by themselves distinguish diffusion from common external attention.

The partial cohort relaxation and the pre-2001 emission-factor analysis are useful sensitivity summaries, but neither excludes selection effects or a corpus-composition discontinuity. The former retains outcome-based filters and unresolved sampling diagnostics; the latter rests on about one per cent of the cohort tokens and no overlapping corpora. Their limited stability therefore does not establish that the difficulties are intrinsic to aggregate counts in general.

None of this is evidence against diffusion; the present design does not adjudicate it. That delimitation is what the section contributes, and it specifies what a positive test would need: speaker-resolved or social-media data with repeated speakers, so that prevalence can be separated from use intensity, and observed rather than assumed contacts, so that the coupling graph carries information.

\section{What pooling across a cohort does and does not buy}\label{sec:inference}
A hierarchical implementation can share information across words while retaining word-specific parameters; the specification used for the real-data fit is in Section~\ref{subsec:hierspec}. It is often asserted that pooling ``resolves'' the single-word identifiability problem. It does not, and the correct statement is quantitative.

We ran a replicated coverage study rather than a single illustrative fit. In each of $40$ independent replicates, population rates were drawn from a wide design distribution, a cohort of $8$ words on $4$ communities over $10$ windows was simulated, and two models were fitted to the same data: the hierarchical model over the whole cohort, and the identical model with identical priors applied to a single randomly chosen word. Using the same model and priors provides a more controlled pooling comparison. Reproduction still requires the population design distribution, word-effect distributions, emissions, exposure and seed schedule, as well as replicate-level diagnostics; these are not fully specified here.

\begin{table}[t]
\centering\small
\begin{tabular}{@{}lcccc@{}}
\toprule
& \multicolumn{2}{c}{coverage of the $95\%$ credible interval} & \multicolumn{2}{c}{mean interval width}\\
\cmidrule(lr){2-3}\cmidrule(lr){4-5}
& pooled cohort & single word & pooled & single word\\
\midrule
$\beta_0$ & $39/40$, $[0.87,1.00]$ & $40/40$, $[0.91,1.00]$ & $0.359$ & $0.750$\\
$\mu_0$ & $40/40$, $[0.91,1.00]$ & $40/40$, $[0.91,1.00]$ & $0.180$ & $0.235$\\
$\kappa_0$ & $38/40$, $[0.83,0.99]$ & $40/40$, $[0.91,1.00]$ & $0.372$ & $0.645$\\
\bottomrule
\end{tabular}
\caption{Replicated coverage and interval width over $40$ independent cohorts, comparing the hierarchical fit over the whole cohort with the identical model and priors applied to a single randomly chosen word. Coverage counts are followed by Wilson $95\%$ intervals. Pooling narrows the reported intervals; this finite experiment does not establish the absence of a coverage deficit.}
\label{tab:pooling}
\end{table}

Table~\ref{tab:pooling} reports the result. Coverage is at or above the nominal level in both arms, with Wilson intervals wide enough that $40$ replicates cannot distinguish $0.95$ from $1.00$. This is coverage under one fixed simulation design, not simulation-based calibration in the sense of \cite{talts2018}, which checks the whole posterior against draws from the prior and would be the stronger test of the sampler itself. The observed single-word coverage is high and the intervals are wide, but forty replications cannot certify nominal coverage. The reported worst sampling diagnostics also require investigation before interval calibration is considered settled. What pooling buys is precision and computational behaviour. The population $\beta_0$ interval is $2.1$ times narrower and the $\kappa_0$ interval $1.7$ times narrower under pooling, while $\mu_0$, the parameter that Section~\ref{subsec:recovery} showed to be weakly determined, improves by only $1.3$. Sampling is also better behaved: the worst $\hat R$ over all replicates is $1.021$ pooled against $1.033$ unpooled, and the worst divergence count is $10$ against $35$. The log-normal population distribution is itself an identifying restriction: transforming every word along its linear equivalence family need not preserve that distribution. Concentration of population hyperparameters can therefore reflect distributional assumptions and priors as well as trajectory information. Pooling does not remove the unrestricted observation invariances: Proposition~\ref{prop:identify} holds word by word and cohort-wide alike, and no amount of sharing separates $u$ from $q$.

\begin{figure}[t]
\centering
\includegraphics[width=\linewidth]{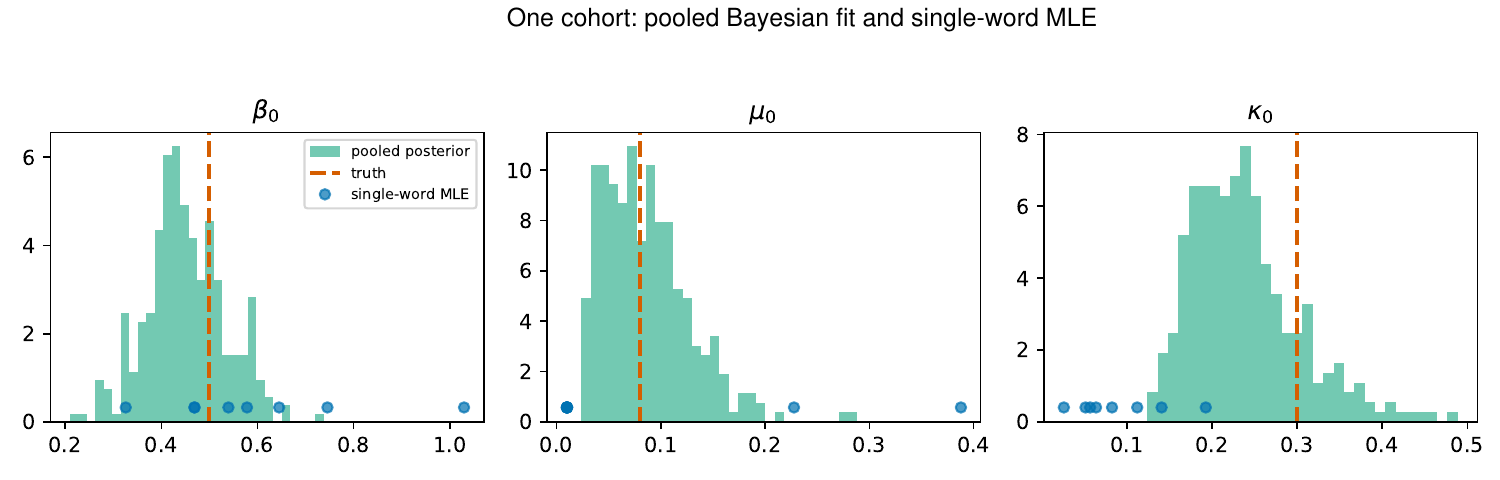}
\caption{Hierarchical Bayesian recovery on one simulated cohort. Pooled posteriors for the population transmission, abandonment, and coupling rates concentrate near the truth (dashed) with $95\%$ credible intervals covering it; single-word maximum-likelihood estimates (points) are scattered and biased, piling at the abandonment boundary and underestimating the coupling. This is one replicate; the replicated comparison against a single-word \emph{Bayesian} fit, which is the like-for-like control, is in Table~\ref{tab:pooling}.}
\label{fig:hier}
\end{figure}

The corresponding maximum-likelihood comparison is starker and is what Figure~\ref{fig:hier} shows: single-word maximum-likelihood fits drive the abandonment rate to its boundary for most words and underestimate the coupling by roughly a factor of three. That contrast is between a pooled Bayesian estimator and an unpooled frequentist one, and it therefore mixes pooling with the effect of the priors; the coverage comparison of Table~\ref{tab:pooling} is the controlled version of the same comparison. This is the intended use of the framework: estimate shared dynamics across a cohort, not structural rates from any single neologism.

\section{Conditional predictions and the observations each would require}\label{sec:predictions}
The framework yields the following conditional predictions. Each is stated with the observable that would test it and the minimum data required, most require measurements absent from the country-count panel of Section~\ref{sec:diagnosis}; the aggregate-versus-community curve comparison is instead limited by the exposure and temporal resolution:
\begin{enumerate}
 \item conditional on initial prevalence and exposure, greater community diversity among early users is associated with broader subsequent reach: requires identified early adopters with community labels;
 \item local cohesion favours persistence, whereas cross-community bridges favour reach: requires an observed contact network with a community partition;
 \item nonlinear reinforcement increases sensitivity to seed size and placement: requires repeated-exposure histories per speaker, because a mechanistic interpretation in terms of repeated contacts requires individual exposure histories;
 \item media exposure can increase token counts without a proportional rise in speaker prevalence: requires a broadcast-volume series alongside a speaker-resolved prevalence estimate;
 \item aggregate sigmoid growth need not imply sigmoid growth within each community: requires community-resolved series with adequate per-community exposure;
 \item \emph{the prediction that distinguishes this framework from a two-state formulation}: among speakers currently not using the form, those who have used it before re-adopt at a different rate from those who never have, at equal current exposure. By Proposition~\ref{prop:lump} a two-state model is the special case in which these two rates coincide, so the comparison is a test with a sharp null. It requires repeated observations of identified speakers with usage histories, which is the minimum datum the dormant compartment demands and which no corpus-count design can supply.
\end{enumerate}
These are hypotheses to be tested, not consequences of Theorems~\ref{thm:wellposed} and \ref{thm:threshold}, and each requires an evaluation and a decision threshold declared before the data are inspected.

\section{Identifiability, causal interpretation, and limitations}\label{sec:limits}
With free emission and contact scales, a single aggregate frequency series cannot identify transmissibility, prevalence, use intensity, detectability, and network scale separately; Proposition~\ref{prop:identify} makes the observational half of this precise and Section~\ref{subsec:identreal} the dynamical half. Homophily further confounds influence with selection, and observing the network does not remove the problem: influence and homophily are generically confounded in observational network data even when the ties are recorded \cite{shalizi2011}, so shared causes and endogenous tie formation require a design (repeated speaker observations, multiple channels, measured contacts, negative controls, credible exogenous variation) and not merely richer data. The framework is generative and predictive; causal interpretations require additional design assumptions that no analysis here supplies.

Network weights and layer coefficients carry the scale invariances of Proposition~\ref{prop:scale}, so a normalisation such as $\rho(W)=1$, fixed layer scales, or an externally calibrated contact unit is required before any rate is reported, and the applicable invariance differs between the Hill and the linear specifications. Sensitive social variables should be minimised and anonymised, with particular care for reidentification in small communities.

\section{Discussion and conclusion}\label{sec:conclusion}
A classical epidemic analogy is productive only after its limitations are made explicit. What propagates is a socially situated practice of use, not a pathogen, and adoption may require reinforcement, carry indexical meaning, and be recorded selectively. The framework developed here formalises these distinctions. It establishes well-posedness of the finite-state process (Theorem~\ref{thm:wellposed}), isolates the assumptions and the small-seed condition behind the logistic S-curve (Proposition~\ref{prop:logistic}), and states a local invasion criterion for a named approximation (Theorem~\ref{thm:threshold}). On that footing it then asks a question that motivates the present model-specific analysis: for this particular mechanism and a corpus that records it only in aggregate, which of the mechanism's parameters survive the passage through the observation layer?

The answer has three parts, and each is exact. Aggregation of the speaker chain is legitimate precisely when the adoption and reactivation intensities agree on the exposures the network can actually deliver, which is a condition on the design and not on the response function alone; demanding it as an identity in the exposure, as occurs for a sufficiently rich class of networks with fixed response coefficients, is what forces the simple response $r=1$ with matched half-saturation $\theta=\theta_D$, unless the contact-driven part of reactivation vanishes altogether (Proposition~\ref{prop:lump} and Corollary~\ref{cor:lumpfun}(i) and (ii)). With unconstrained emission factors, use intensity and detectability are not separated by this count likelihood, whatever its sample size (Proposition~\ref{prop:identify}). And in the linearised community model the three rates collapse onto the two combinations $\beta-\mu$ and $\beta\kappa$, along a one-parameter family whose members are observationally indistinguishable, with the separating information carried by a term quadratic in prevalence whose magnitude is set by the emission scale (Proposition~\ref{prop:family} and Section~\ref{subsec:emisobs}). Together these results say what a corpus panel can be asked to decide, and Corollary~\ref{cor:modes} says what would have to be measured, namely between-community contrasts, for the coupling to become recoverable at all.

The empirical sections then show these bounds operating on real data, and the scope of what they show is worth stating plainly. Print corpora document the spatio-temporal \emph{consolidation} of a form, and media-borne neologisms in particular, whereas the \emph{onset} in speech remains under-observed. Consequently, our empirical design does not claim to track the initial coinage of neologisms. Rather, it specifically evaluates the top-down consolidation and cross-media reach of institutional, political, and media-borne terminology across national written channels, where documentary tracking across decades is fine-grained.

Within that scope the fit of Section~\ref{sec:diagnosis} behaves as the theory predicts: the rate triple is flat along the predicted direction, the coupling posterior excludes zero by construction and its magnitude remains conditional on the prior and model specification, and in leave-one-country-out geographical reconstruction a constant-rate count baseline outperforms M0, M1 and M3 on most forms in the reported comparisons. Measured against the falsification criteria of Section~\ref{subsec:falsify}, which are stated as a design for a confirmatory study and not as a discharged pre-registration, the proposed dynamical structure is not supported on this panel. That is a determinate result about this design, not a verdict on contact-driven diffusion, which these data are not equipped to reach either way.

What the analysis leaves is a specification for the study that could. Repeated speaker observations can help separate prevalence from use intensity when linked to an explicit model of activity and emission; overlapping calibration samples can inform detectability under stated linkage and dependence assumptions; observed rather than assumed contacts, together with genuine between-community contrasts, give the coupling graph content; and a benchmark against the closest generative predecessors on matched data becomes possible with observation units and predictive targets harmonised; a parameter-recovery comparison additionally needs appropriate identification. The framework's practical value is that it names which of these measurements buys which parameter. We therefore offer this article as a mathematical framework with an explicit observation layer, a protocol for falsifiable analysis, an in silico characterisation of its estimator, and a precise account of what aggregate corpus evidence can and cannot decide.

\section*{Supplementary material}
A descriptive illustration for a selection of $37$ forms in the Google Books Spanish corpus, which uses a different sampling frame and no country resolution and therefore plays no part in any estimate reported here, is included among the supplied figure files as \texttt{ngrams\_cohort\_grid.pdf}; its query mapping and provenance still require documentation.

\section*{Data and code availability}
A public archive identifier and the underlying analysis code, panel, posterior draws and score records were not supplied with this review version. The numerical results are retained as reported and have not been independently regenerated from data. A versioned deposit is required before these results can be considered reproducible. It should contain the model engine, simulation and real-data pipelines, fold assignments, complete run configurations and seeds, environment specification, posterior draws, per-form scores, positive-exposure mask, data dictionary, lemma-to-query map, and commands reproducing every figure. The lexical-status audit and concordance validation remain outstanding, as stated in Section~\ref{sec:protocol}; they must not be described as completed deposit contents. Only derived, aggregated counts are intended for sharing; the underlying CREA and CORPES XXI texts remain with their rights holders (\url{https://www.rae.es/banco-de-datos}). The proposed code and derived-data licences are MIT and CC BY 4.0 respectively, subject to author confirmation of redistribution rights. The cohort table must record provenance, first attestation, variants, counts and inclusion flags; until provided, Appendix~\ref{app:wordlist} cannot reproduce the filtered cohorts.

\appendix
\section{List of selected neologisms}\label{app:wordlist}

The candidate list reported here represents an unedited retrieval query set and should not be treated as a manually certified dictionary of emerging forms. In the filtered empirical cohorts evaluated in Section 13, non-neological forms and non-target usages are audited to prevent baseline contamination

The lemmas below are the candidate forms queried against CREA and CORPES XXI. The list is a record of what was searched for, not a verified inventory of neologisms: the lexical-status audit required by Section~\ref{sec:protocol} and discussed in Section~\ref{subsec:cohortflow} has not been carried out on it, and it is known to contain at least one established multiword expression (\textit{prisi\'on preventiva}, already in the 1973 Spanish Penal Code) whose later corpus attestation therefore cannot establish lexical emergence \cite{boe1973}. Spelling and inflectional variants were grouped under a single lemma before counting, so an entry here may correspond to several search strings; the lemma-to-query map, the variant groupings, the lexical-status annotation and the per-cohort inclusion flags must accompany the deposit, and the printed list alone cannot reproduce any of the filtered cohorts of Table~\ref{tab:flow}. Items excluded during preparation must be retained in the deposit with their exclusion reason.

\textit{abertzalismo},
\textit{absurdez},
\textit{achatarramiento},
\textit{achatarrar},
\textit{achuchable},
\textit{acoso callejero},
\textit{acoso digital},
\textit{acoso escolar},
\textit{acoso sexual},
\textit{activo basura},
\textit{acuerdo marco},
\textit{administración electrónica},
\textit{ADSL},
\textit{afectivosexual},
\textit{agenda social},
\textit{agroexportable},
\textit{agroexportación},
\textit{agroexportador},
\textit{agroganadería},
\textit{agroganadero},
\textit{airbag},
\textit{airfryer},
\textit{alfabetización digital},
\textit{altermundismo},
\textit{altermundista},
\textit{amigovio},
\textit{anti cuarentena},
\textit{anti pandemia},
\textit{antibrexit},
\textit{anticovid},
\textit{antiglobalización},
\textit{antivacunas},
\textit{anualización},
\textit{anualizar},
\textit{anumerismo},
\textit{aparcacoches},
\textit{apartahotel},
\textit{aporofobia},
\textit{aprovechategui},
\textit{aserejé},
\textit{astroturístico},
\textit{ataque informático},
\textit{aula virtual},
\textit{autoconducción},
\textit{autoconfinamiento},
\textit{autoconfinarse},
\textit{basura electrónica},
\textit{basura espacial},
\textit{basuraleza},
\textit{baticueva},
\textit{bellezón},
\textit{benchmark},
\textit{bestia parda},
\textit{bicicrocista},
\textit{biocompostable},
\textit{biofirma},
\textit{bitcóin},
\textit{bitcoiner},
\textit{bloguero},
\textit{bluetooth},
\textit{bocachanclas},
\textit{bono basura},
\textit{bono verde},
\textit{brecha digital},
\textit{brexit},
\textit{brexitero},
\textit{bro},
\textit{broker},
\textit{burbuja tecnológica},
\textit{campus virtual},
\textit{capitalismo salvaje},
\textit{cazabobos},
\textit{celiaquismo},
\textit{centenial},
\textit{certificado digital},
\textit{chatbot},
\textit{chatear},
\textit{choriceo},
\textit{choripán},
\textit{choripanero},
\textit{choriplanero},
\textit{ciberacoso},
\textit{ciberatacar},
\textit{cibercafé},
\textit{ciberdelincuente},
\textit{ciberinteligencia},
\textit{cibernauta},
\textit{ciberñoño},
\textit{cibersexo},
\textit{circovirus},
\textit{ciudad inteligente},
\textit{clientelaje},
\textit{clientelismo},
\textit{clientelístico},
\textit{clientelizar},
\textit{cliquear},
\textit{códec},
\textit{código abierto},
\textit{comida basura},
\textit{concertacesión},
\textit{concertístico},
\textit{confinamiento obligatorio},
\textit{confinamiento total},
\textit{confinamiento voluntario},
\textit{congresual},
\textit{conservacionismo},
\textit{conspiranóico},
\textit{conspiranoide},
\textit{contrainteligencia},
\textit{contrato basura},
\textit{copiapega},
\textit{correa de transmisión},
\textit{correo basura},
\textit{cortafuego},
\textit{cortavenas},
\textit{covid persistente},
\textit{covid19},
\textit{coworking},
\textit{cringe},
\textit{criptoactivo},
\textit{criptodivisa},
\textit{criptoeconomía},
\textit{criptoestafa},
\textit{criptoinversión},
\textit{criptomercado},
\textit{criptominería},
\textit{criptomoneda},
\textit{crisis humanitaria},
\textit{crónica rosa},
\textit{crowfunding},
\textit{cuatrolatas},
\textit{cuchareable},
\textit{cultura de la cancelación},
\textit{deepfake},
\textit{demisexual},
\textit{denisovano},
\textit{descarbonización},
\textit{desconexión digital},
\textit{desempoderamiento},
\textit{desempoderar},
\textit{desescalada},
\textit{deslocalización},
\textit{diario digital},
\textit{discurso de odio},
\textit{djembe},
\textit{dólar blue},
\textit{dolarizador},
\textit{dosmilero},
\textit{drag},
\textit{dresscode},
\textit{droga de diseño},
\textit{e-administración},
\textit{early adopter},
\textit{e-bike},
\textit{ebook},
\textit{ecoetiquetado},
\textit{e-commerce},
\textit{economía verde},
\textit{ecopostureo},
\textit{ecotasa},
\textit{ecoturismo},
\textit{efecto 2000},
\textit{electrofunk},
\textit{email},
\textit{emergencia climática},
\textit{emoji},
\textit{emoticono},
\textit{emplatar},
\textit{empoderador},
\textit{empoderamiento},
\textit{empoderante},
\textit{en plan},
\textit{energía verde},
\textit{escape room},
\textit{escrache},
\textit{espóiler},
\textit{eurodiputado},
\textit{euroescéptico},
\textit{euroizquierda},
\textit{eurozona},
\textit{fachosfera},
\textit{fact checking},
\textit{family-friendly},
\textit{fascistoide},
\textit{fast-food},
\textit{femicida},
\textit{feminicidio},
\textit{filonazi},
\textit{filonazismo},
\textit{fintech},
\textit{firma electrónica},
\textit{follamigo},
\textit{fototrampeo},
\textit{friqui},
\textit{frontwoman},
\textit{gamer},
\textit{gaming},
\textit{gauchocracia},
\textit{gel hidroalcohólico},
\textit{gemelo digital},
\textit{gente guapa},
\textit{gentrificación},
\textit{geoetiquetar},
\textit{geointeligencia},
\textit{georeferenciar},
\textit{georreferenciación},
\textit{gestación subrogada},
\textit{gigante asiático},
\textit{glamping},
\textit{glasnost},
\textit{globalización},
\textit{googleo},
\textit{gordofobia},
\textit{gran hermano},
\textit{guerra comercial},
\textit{hackatón},
\textit{hackeo},
\textit{hástag},
\textit{hater},
\textit{headhunting},
\textit{heterocurioso},
\textit{heteroflexible},
\textit{heteroparental},
\textit{hiperventilar},
\textit{hipervincular},
\textit{hispabono},
\textit{homoparental},
\textit{huella de carbono},
\textit{indietrónica},
\textit{indignado},
\textit{influencer},
\textit{infodemia},
\textit{infografista},
\textit{inmunonutrición},
\textit{instamodelo},
\textit{inteligencia ambiental},
\textit{inteligencia artificial},
\textit{inteligencia emocional},
\textit{internet},
\textit{islamofobia},
\textit{janjawid},
\textit{jevy},
\textit{jihadista},
\textit{jiujitsu},
\textit{jugoterapia},
\textit{kale borroka},
\textit{kombucha},
\textit{k-pop},
\textit{lado oscuro},
\textit{lenguaje inclusivo},
\textit{libro electrónico},
\textit{linkear},
\textit{lipotransferencia},
\textit{llave en mano},
\textit{long covid},
\textit{lulismo},
\textit{macrocorrupción},
\textit{macrofestival},
\textit{macroproyecto},
\textit{macrorregión},
\textit{magufo},
\textit{malvinización},
\textit{marea blanca},
\textit{marea negra},
\textit{marea verde},
\textit{masterchef},
\textit{meeting point},
\textit{megahit},
\textit{meme},
\textit{metaverso},
\textit{microbasural},
\textit{micromachismo},
\textit{micromecenazgo},
\textit{microplástico},
\textit{milénico},
\textit{mileurismo},
\textit{mileurista},
\textit{mobbing},
\textit{moda sostenible},
\textit{moneda social},
\textit{monetizar},
\textit{motomami},
\textit{MP3},
\textit{multitasking},
\textit{museificación},
\textit{mutear},
\textit{nanotecnología},
\textit{nativo digital},
\textit{necropolítica},
\textit{negacionismo},
\textit{neobanco},
\textit{networking},
\textit{neuroderecho},
\textit{no binario},
\textit{nómada digital},
\textit{nueva normalidad},
\textit{nueva ola},
\textit{ochentero},
\textit{ojo de halcón},
\textit{okupa},
\textit{packaging},
\textit{paleoartista},
\textit{pandemia del coronavirus},
\textit{pasaporte covid},
\textit{pendrive},
\textit{pera limonera},
\textit{perestroika},
\textit{periódico digital},
\textit{perreo},
\textit{perroflauta},
\textit{photobook},
\textit{picar el bichito},
\textit{pinponear},
\textit{piromusical},
\textit{plan de desconfinamiento},
\textit{pódcast},
\textit{podcastero},
\textit{podemismo},
\textit{podemista},
\textit{podemita},
\textit{podemización},
\textit{podemizar},
\textit{podemizarse},
\textit{poder blando},
\textit{poliamor},
\textit{poliamoroso},
\textit{polimedicar},
\textit{política antidrogas},
\textit{politono},
\textit{populismo},
\textit{posbrexit},
\textit{posconfinamiento},
\textit{poscovid},
\textit{pos-covid-19},
\textit{poshumanismo},
\textit{pospandemia},
\textit{postureo},
\textit{posverdad},
\textit{precarización},
\textit{precovid},
\textit{precuela},
\textit{prensa basura},
\textit{prepandemia},
\textit{prisión preventiva},
\textit{probrexit},
\textit{programa basura},
\textit{programa espía},
\textit{prosumidor},
\textit{puntocom},
\textit{quemagrasas},
\textit{racializar},
\textit{radiobasura},
\textit{raider},
\textit{realidad aumentada},
\textit{realidad virtual},
\textit{red wifi},
\textit{refugiado climático},
\textit{refugio climático},
\textit{reguetón},
\textit{respaldatorio},
\textit{roaming},
\textit{robacarros},
\textit{roboadvisor},
\textit{sanchismo},
\textit{sándwich},
\textit{satisfyer},
\textit{securitario},
\textit{selfi},
\textit{semiconfinamiento},
\textit{sensorizar},
\textit{señorero},
\textit{seropositivo},
\textit{sexteo},
\textit{sexting},
\textit{sin contacto},
\textit{sobrepensar},
\textit{sociedad digital},
\textit{software libre},
\textit{spam},
\textit{squatter},
\textit{stablecoin},
\textit{stalker},
\textit{startup},
\textit{startupero},
\textit{streaming},
\textit{suicidio político},
\textit{súperalimento},
\textit{superfood},
\textit{superinteligencia},
\textit{tablet},
\textit{tabloide},
\textit{tabula rasa},
\textit{tacleador},
\textit{taclear},
\textit{tacleo},
\textit{tacticismo},
\textit{tai chi},
\textit{take away},
\textit{talde},
\textit{talk show},
\textit{tamagochi},
\textit{tamaliza},
\textit{tamayazo},
\textit{tapabocas},
\textit{tardeo},
\textit{tardofranquista},
\textit{tarifa plana},
\textit{tarifazo},
\textit{tarjeta black},
\textit{tarjeta prepago},
\textit{táser},
\textit{tataki},
\textit{tauromafia},
\textit{techo de cristal},
\textit{tecnooligarca},
\textit{teenager},
\textit{tejido social},
\textit{tele basura},
\textit{teletrabajo},
\textit{terraplanismo},
\textit{terraplanista},
\textit{thrasher},
\textit{tirar cohetes},
\textit{todo incluido},
\textit{trabajo remoto},
\textit{transfobia},
\textit{transgénico},
\textit{transición energética},
\textit{troika},
\textit{trolear},
\textit{trumpismo},
\textit{tuit},
\textit{tunelero},
\textit{turismo ecológico},
\textit{turismofobia},
\textit{tutelador},
\textit{uberización},
\textit{uberizar},
\textit{ubersexual},
\textit{ultraconservador},
\textit{ultraortodoxo},
\textit{vaping},
\textit{videocomunicado},
\textit{videoconsola},
\textit{videojuego},
\textit{videollamada},
\textit{vientre de alquiler},
\textit{vinilo},
\textit{virucida},
\textit{wasap},
\textit{wasapear},
\textit{wasp},
\textit{waterproof},
\textit{webcam},
\textit{wellness},
\textit{whatsappitis},
\textit{wireless},
\textit{woke},
\textit{yakuza},
\textit{yapear},
\textit{yapeo},
\textit{yayoflauta},
\textit{youtuber},
\textit{yupi},
\textit{zombificación},
\textit{zoonótico},
\section*{Acknowledgments}
The research carried out by the second and third co-authors has been funded by the UAM-Instituto de Ingeniería del Conocimiento Chair in Computational Linguistics.


\begin{thebibliography}{99}
\bibitem{labov1994} W.~Labov. \emph{Principles of Linguistic Change: Internal Factors}. Blackwell, Oxford, 1994.
\bibitem{labov2001} W.~Labov. \emph{Principles of Linguistic Change: Social Factors}. Blackwell, Oxford, 2001.
\bibitem{labov2010} W.~Labov. \emph{Principles of Linguistic Change: Cognitive and Cultural Factors}. Wiley-Blackwell, Chichester, 2010.
\bibitem{labov2007} W.~Labov. Transmission and diffusion. \emph{Language}, 83(2):344--387, 2007. \doi{10.1353/lan.2007.0082}.
\bibitem{wuerschinger2021} Q.~W\"urschinger. Social networks of lexical innovation: investigating the social dynamics of diffusion of neologisms on Twitter. \emph{Frontiers in Artificial Intelligence}, 4:648583, 2021. \doi{10.3389/frai.2021.648583}.
\bibitem{goel2016} R.~Goel, S.~Soni, N.~Goyal, J.~Paparrizos, H.~Wallach, F.~Diaz, and J.~Eisenstein. The social dynamics of language change in online networks. \emph{Social Informatics}, pages 41--57, 2016. \href{https://doi.org/10.1007/978-3-319-47880-7_3}{doi:10.1007/978-3-319-47880-7\_3}.
\bibitem{zhu2021} J.~Zhu and D.~Jurgens. The structure of online social networks modulates the rate of lexical change. In \emph{Proceedings of NAACL-HLT}, pages 2201--2218, 2021. \doi{10.18653/v1/2021.naacl-main.178}.
\bibitem{laitinen2020} M.~Laitinen, M.~Fatemi, and J.~Lundberg. Size matters: digital social networks and language change. \emph{Frontiers in Artificial Intelligence}, 3:46, 2020. \doi{10.3389/frai.2020.00046}.
\bibitem{tarrade2024} L.~Tarrade, J.-P.~Chevrot, and J.-P.~Magu\'e. How position in the network determines the fate of lexical innovations on Twitter. \emph{PLOS Complex Systems}, 1(1):e0000005, 2024. \doi{10.1371/journal.pcsy.0000005}.
\bibitem{newberry2017} M.~G.~Newberry, C.~A.~Ahern, R.~Clark, and J.~B.~Plotkin. Detecting evolutionary forces in language change. \emph{Nature}, 551(7679):223--226, 2017. \doi{10.1038/nature24455}.
\bibitem{areajornet2025} I.~Area and M.~Jornet. A mathematical analysis for the dynamics of multiple languages. \emph{CSIAM Transactions on Applied Mathematics}, 6(4):842--861, 2025. \doi{10.4208/csiam-am.SO-2024-0056}.
\bibitem{blythe2012} R.~A.~Blythe and W.~Croft. S-curves and the mechanisms of propagation in language change. \emph{Language}, 88(2):269--304, 2012. \doi{10.1353/lan.2012.0027}.
\bibitem{stadler2016} K.~Stadler, R.~A.~Blythe, K.~Smith, and S.~Kirby. Momentum in language change: a model of self-actuating S-shaped curves. \emph{Language Dynamics and Change}, 6(2):171--198, 2016. \doi{10.1163/22105832-00602005}.
\bibitem{ghanbarnejad2014} F.~Ghanbarnejad, M.~Gerlach, J.~M.~Miotto, and E.~G.~Altmann. Extracting information from S-curves of language change. \emph{Journal of the Royal Society Interface}, 11(101):20141044, 2014. \doi{10.1098/rsif.2014.1044}.
\bibitem{amato2018} R.~Amato, L.~Lacasa, A.~D\'iaz-Guilera, and A.~Baronchelli. The dynamics of norm change in the cultural evolution of language. \emph{Proceedings of the National Academy of Sciences}, 115(33):8260--8265, 2018. \doi{10.1073/pnas.1721059115}.
\bibitem{grieve2017} J.~Grieve, A.~Nini, and D.~Guo. Analyzing lexical emergence in Modern American English online. \emph{English Language and Linguistics}, 21(1):99--127, 2017. \doi{10.1017/S1360674316000113}.
\bibitem{morenosandoval2018} A.~Moreno Sandoval and F.~Moreno Fern\'andez. Configuraci\'on ling\"u\'{\i}stica de anglicismos procedentes de Twitter en el espa\~nol estadounidense. \emph{Revista Signos}, 51(98):382--409, 2018. \doi{10.4067/S0718-09342018000300382}.
\bibitem{gneiting2007} T.~Gneiting and A.~E.~Raftery. Strictly proper scoring rules, prediction, and estimation. \emph{Journal of the American Statistical Association}, 102(477):359--378, 2007. \doi{10.1198/016214506000001437}.
\bibitem{white1982} H.~White. Maximum likelihood estimation of misspecified models. \emph{Econometrica}, 50(1):1--25, 1982. \doi{10.2307/1912526}.
\bibitem{pastorsatorras2015} R.~Pastor-Satorras, C.~Castellano, P.~Van Mieghem, and A.~Vespignani. Epidemic processes in complex networks. \emph{Reviews of Modern Physics}, 87(3):925--979, 2015. \doi{10.1103/RevModPhys.87.925}.
\bibitem{hoffman2014} M.~D.~Hoffman and A.~Gelman. The No-U-Turn Sampler: adaptively setting path lengths in Hamiltonian Monte Carlo. \emph{Journal of Machine Learning Research}, 15:1593--1623, 2014.
\bibitem{phan2019} D.~Phan, N.~Pradhan, and M.~Jankowiak. Composable effects for flexible and accelerated probabilistic programming in {NumPyro}. \emph{arXiv:1912.11554}, 2019.
\bibitem{bernal2020} E.~Bernal, J.~Freixa, and S.~Torner. Criterios para la diccionarizaci\'on de neologismos: de la teor\'{\i}a a la pr\'actica. \emph{Revista Signos}, 53(104):592--618, 2020. \doi{10.4067/S0718-09342020000300592}.
\bibitem{fernandez2015} A.~Fern\'andez Rodr\'{\i}guez and I.~Galanes Santos. La crise hypoth\'ecaire et ses d\'enominations. \emph{Babel}, 61(2):265--282, 2015. \doi{10.1075/babel.61.2.06fer}.
\bibitem{temmerman2014} R.~Temmerman and M.~Van Campenhoudt, editors. \emph{Dynamics and Terminology: An Interdisciplinary Perspective on Monolingual and Multilingual Culture-Bound Communication}. John Benjamins, Amsterdam and Philadelphia, 2014. \doi{10.1075/tlrp.16}.
\bibitem{crea} Real Academia Espa\~nola. \emph{CREA: Corpus de Referencia del Espa\~nol Actual}. \url{https://www.rae.es/banco-de-datos/crea}. Accessed 2026.
\bibitem{centola2010} D.~Centola. The spread of behavior in an online social network experiment. \emph{Science}, 329(5996):1194--1197, 2010. \doi{10.1126/science.1185231}.
\bibitem{kiss2017} I.~Z.~Kiss, J.~C.~Miller, and P.~L.~Simon. \emph{Mathematics of Epidemics on Networks}. Springer, Cham, 2017. \doi{10.1007/978-3-319-50806-1}.
\bibitem{lajmanovich1976} A.~Lajmanovich and J.~A.~Yorke. A deterministic model for gonorrhea in a nonhomogeneous population. \emph{Mathematical Biosciences}, 28:221--236, 1976. \doi{10.1016/0025-5564(76)90125-5}.
\bibitem{bass1969} F.~M.~Bass. A new product growth model for consumer durables. \emph{Management Science}, 15(5):215--227, 1969. \doi{10.1287/mnsc.15.5.215}.
\bibitem{bobneo} Observatori de Neologia (OBNEO). \emph{BOBNEO: Banc de dades de neologismes}. Universitat Pompeu Fabra. Data collected since 1989. \url{https://bobneo.upf.edu}. Accessed 2026.
\bibitem{corpes} Real Academia Espa\~nola. \emph{CORPES XXI: Corpus del Espa\~nol del Siglo XXI}. \url{https://www.rae.es/banco-de-datos/corpes-xxi}. Accessed 2026.
\bibitem{sole2026} R.~Sol\'e, G.~Ruffini, F.~Castaldo, M.~Tuccio, L.~F.~Seoane, M.~de Domenico, S.~F.~Elena, D.~C.~Krakauer, and M.~Levin. Large-language models as a cognitive virus. \emph{arXiv preprint} arXiv:2609.03344, 2026. \url{https://arxiv.org/abs/2609.03344}.
\bibitem{aitchison1991} J.~Aitchison. \emph{Language Change: Progress or Decay?} 2nd ed. Cambridge University Press, Cambridge, 1991.
\bibitem{milroy1992} J.~Milroy. \emph{Linguistic Variation and Change}. Blackwell, Oxford, 1992.
\bibitem{tajfel1986} H.~Tajfel and J.~C.~Turner. The social identity theory of intergroup behaviour. In S.~Worchel and W.~G.~Austin (eds.), \emph{Psychology of Intergroup Relations}, pages 7--24. Nelson-Hall, Chicago, 1986.
\bibitem{siepmann2015} D.~Siepmann. Dictionaries and spoken language: a corpus-based review of French dictionaries. \emph{International Journal of Lexicography}, 28(2):139--168, 2015. \doi{10.1093/ijl/ecv006}.
\bibitem{crystal2001} D.~Crystal. \emph{Language and the Internet}. Cambridge University Press, Cambridge, 2001.
\bibitem{goody1987} J.~Goody. \emph{The Interface Between the Written and the Oral}. Cambridge University Press, Cambridge, 1987.
\bibitem{mcenery2003} T.~McEnery. Corpus linguistics. In R.~Mitkov (ed.), \emph{The Oxford Handbook of Computational Linguistics}, pages 448--463. Oxford University Press, Oxford, 2003.
\bibitem{biber2004} D.~Biber, S.~Conrad, and V.~Cortes. If you look at\ldots: lexical bundles in university teaching and textbooks. \emph{Applied Linguistics}, 25(3):371--405, 2004. \doi{10.1093/applin/25.3.371}.
\bibitem{preseea} F.~Moreno-Fernández. Corpus para el estudio del espa~nol en su variaci\'on geogr\'afica y social. El corpus PRESEEA. \emph{Oralia}, 8, 2005 \doi{https://doi.org/10.25115/oralia.v8i1.8333} 
\bibitem{corlec} https://www.lllf.uam.es/ESP/Corlec.html.
\bibitem{mavir} https://leptis.lllf.uam.es/corpus/cgi-bin/mavir.pl.
\bibitem{morenoguirao2006} A.~Moreno-Sandoval and J.~M.~Guirao. Morpho-syntactic tagging of the Spanish C-ORAL-ROM corpus: methodology, tools and evaluation. In Y.~Kawaguchi, S.~Zaima, and T.~Takagaki (eds.), \emph{Spoken Language Corpus and Linguistic Informatics}, pages 199--218. John Benjamins, Amsterdam, 2006.
\bibitem{eisenstein2014} J.~Eisenstein, B.~O'Connor, N.~A.~Smith, and E.~P.~Xing. Diffusion of lexical change in social media. \emph{PLOS ONE}, 9(11):e113114, 2014. \doi{10.1371/journal.pone.0113114}.
\bibitem{ananthasubramaniam2024} A.~Ananthasubramaniam, D.~Jurgens, and D.~Romero. Networks and identity drive the spatial diffusion of linguistic innovation in urban and rural areas. \emph{npj Complexity}, 1:9, 2024. \doi{10.1038/s44260-024-00009-9}.
\bibitem{raue2009} A.~Raue, C.~Kreutz, T.~Maiwald, J.~Bachmann, M.~Schilling, U.~Klingm\"uller, and J.~Timmer. Structural and practical identifiability analysis of partially observed dynamical models by exploiting the profile likelihood. \emph{Bioinformatics}, 25(15):1923--1929, 2009. \doi{10.1093/bioinformatics/btp358}.
\bibitem{vehtari2021} A.~Vehtari, A.~Gelman, D.~Simpson, B.~Carpenter, and P.-C.~B\"urkner. Rank-normalization, folding, and localization: an improved $\widehat R$ for assessing convergence of MCMC. \emph{Bayesian Analysis}, 16(2):667--718, 2021. \doi{10.1214/20-BA1221}.
\bibitem{talts2018} S.~Talts, M.~Betancourt, D.~Simpson, A.~Vehtari, and A.~Gelman. Validating Bayesian inference algorithms with simulation-based calibration. \emph{arXiv preprint} arXiv:1804.06788, 2018. \url{https://arxiv.org/abs/1804.06788}.
\bibitem{shalizi2011} C.~R.~Shalizi and A.~C.~Thomas. Homophily and contagion are generically confounded in observational social network studies. \emph{Sociological Methods \& Research}, 40(2):211--239, 2011. \doi{10.1177/0049124111404820}.
\bibitem{centola2018} D.~Centola, J.~Becker, D.~Brackbill, and A.~Baronchelli. Experimental evidence for tipping points in social convention. \emph{Science}, 360(6393):1116--1119, 2018. \doi{10.1126/science.aas8827}.
\bibitem{ball1993} F.~Ball and G.~F.~Yeo. Lumpability and marginalisability for continuous-time Markov chains. \emph{Journal of Applied Probability}, 30(3):518--528, 1993. \doi{10.2307/3214762}.
\bibitem{murphy1973} A.~H.~Murphy. A new vector partition of the probability score. \emph{Journal of Applied Meteorology}, 12(4):595--600, 1973. \url{https://journals.ametsoc.org/view/journals/apme/12/4/1520-0450_1973_012_0595_anvpot_2_0_co_2.xml}.
\bibitem{boe1973} Bolet\'in Oficial del Estado. Decreto 3096/1973, de 14 de septiembre: C\'odigo Penal, texto refundido. BOE no.~297, 12 December 1973, pp.~24004--24018, article~33. \url{https://www.boe.es/diario_boe/txt.php?id=BOE-A-1973-1715}.
\bibitem{corpes2026} Asociaci\'on de Academias de la Lengua Espa\~nola. El CORPES supera los 455 millones de formas. 2026. \url{https://www.asale.org/noticia/el-corpes-supera-los-455-millones-de-formas}. Accessed 14 September 2026.
\end{thebibliography}
\end{document}